\documentclass[preprint2]{aastex}

\usepackage{natbib}

\usepackage{graphicx}
\usepackage{amsmath}
\usepackage{mathrsfs}
\usepackage{eucal}

\usepackage{color}

\newcommand{\leftmean}{{\textstyle <}}
\newcommand{\rightmean}{{\textstyle >}}

\newcommand{\hm}{h_\text{lw}}
\newcommand{\Tm}{\widetilde{T}}
\newcommand{\DTep}{\Delta T_\text{EP}}

\newcommand{\model}{ESTM} 

 \slugcomment{To be submitted to ...}

\shorttitle{Modeling the surface temperature of Earth-like planets}
\shortauthors{Vladilo et al.}

\begin{document}
 
\title{Modeling the  surface temperature of Earth-like planets}

\author{Giovanni Vladilo\altaffilmark{1,2}, 
Laura Silva\altaffilmark{1},
Giuseppe Murante\altaffilmark{3,1},  
\\
Luca Filippi\altaffilmark{3,4}, 
Antonello Provenzale\altaffilmark{3,5}, 
}

\email{vladilo@oats.inaf.it}

\altaffiltext{1}{INAF - Trieste Astronomical Observatory, Trieste, Italy}
\altaffiltext{2}{Department of Physics, University of Trieste, Trieste, Italy}
\altaffiltext{3}{Institute of Atmospheric Sciences and Climate - CNR, Torino, Italy}
\altaffiltext{4}{DIMEAS, Politecnico di Torino, Torino, Italy}
\altaffiltext{5}{Institute of Geosciences and Earth Resources - CNR, Pisa, Italy}

\begin{abstract}
We introduce a novel Earth-like planet surface temperature model (ESTM) for  habitability studies based on the spatial-temporal distribution of planetary surface temperatures. The ESTM adopts a surface Energy Balance Model complemented by: radiative-convective atmospheric column calculations,  a set of physically-based parameterizations of meridional  transport, and  descriptions of  surface and cloud properties more refined than in standard EBMs. The parameterization is valid for rotating terrestrial planets
with shallow atmospheres and moderate values of axis obliquity ($\epsilon \la 45^\circ$). Comparison with a 3D model of atmospheric dynamics from the literature shows that the  equator-to-pole temperature differences predicted by the two models agree within $\approx 5$\,K when the rotation rate,  insolation, surface pressure and planet radius are varied in the intervals $0.5 \la \Omega/\Omega_\oplus \la 2$, 
$0.75 \la S/S_\circ \la 1.25$, $0.3 \la p/(\mathrm{1\,bar}) \la 10$, and $0.5 \la R/R_\oplus \la 2$, respectively. The ESTM has an extremely low computational cost and can be used when the planetary parameters are scarcely known (as for most exoplanets) and/or whenever many runs for different parameter configurations are needed. Model simulations of  a test-case exoplanet  (Kepler-62e) indicate that an uncertainty in surface pressure within the range expected for terrestrial planets may impact the mean  temperature by $\sim 60\,$K. Within the limits of validity of the ESTM, the impact of surface pressure is larger than that predicted by uncertainties in  rotation rate, axis obliquity,  and ocean fractions. We discuss the possibility of performing a statistical ranking of planetary habitability taking advantage of the flexibility of the ESTM. 
\end{abstract}


\section{Introduction}

The large amount of exoplanet data collected 
with the doppler and  transit  methods
\citep[e.g.][and refs. therein]{Mayor14,Batalha13}
indicate that Earth-size planets are intrinsically more frequent than giant ones, 
in spite of the fact that they are more difficult to detect.  
Small planets are found in a relatively broad range of metallicities \citep{Buchhave12}
and, at variance with giant planets,  their detection rate drops slowly      
with decreasing metallicity \citep{Wang15}. 
These observational results indicate that 
Earth-like planets are quite common around other stars \citep[e.g.][]{Farr14}
and are expected to be detected in large numbers in the  future.  
Their potential similarity to the Earth makes them primary targets in the quest for habitable environments 
ouside the Solar System. 
Unfortunately,  small  planets are quite difficult to characterize with experimental methods
and a significant effort of modelization is required to cast light on their properties. 
The aim of the present work is to model the surface temperature of these planets
as a contribution to the study of their surface habitability.
The capability of an environment to host life depends on many factors,
such as the presence of liquid water, nutrients, energy sources, and shielding from cosmic ionizing
radiation \citep[e.g.][]{Seager13,Guedel14}. 
A knowledge of the surface temperature is essential to apply the liquid water criterion of habitability
and can also be used to assess the potential presence of different life forms 
according to  other types of temperature-dependent biological criteria \citep[e.g.][]{Clarke14}. 
Here we are interested in modeling the latitudinal and seasonal variations of surface temperature,
$T(\varphi,t)$, as a tool to calculate temperature-dependent indices of fractional habitability \citep[e.g.][]{SMS08}. 

Modeling $T(\varphi,t)$ is a difficult task since many of the physical and chemical quantities 
that govern the exoplanet surface properties are currently not measurable. 
A way to cope with this problem is to treat the unknown quantities as free parameters 
and use fast climate calculations
to explore how variations of such parameters affect the surface temperature. 
%
%
General Circulation Models (GCMs)  are not suited for this type of exploratory work
since they require large amounts of computational resources for each single run
 as well as a detailed knowledge of many planetary characteristics. 
Two types of fast climate tools are commonly employed in studies of planetary habitability:
single atmospheric column calculations and energy balance models. 
Atmospheric column calculations treat in detail the physics of vertical energy transport, 
taking into account the influence of atmospheric composition on the radiative transfer \citep[e.g.,][]{Kasting88}.
This is the type of climate tool that is commonly employed in studies of the ``habitable zone''
\citep[e.g.,][]{Kasting88,Kasting93,vonParis13,Kopparapu13,Kopparapu14}.
Energy balance models (EBMs) calculate the zonal and seasonal energy budget of a planet
using a heat diffusion formalism to describe the horizontal transport
and simple analytical functions of the surface temperature to describe the vertical transport \citep[e.g.,][]{North81}.
EBMs have been employed to address the climate impact 
induced by variations of several planet parameters, 
such as axis obliquity, rotation period and stellar insolation
 \citep{SMS08,SMS09,Dressing10,Spiegel10,Forgan12,Forgan14}.
By feeding classic EBMs with multi-parameter functions extracted from atmospheric column calculations 
one can obtain an upgraded type of EBM that 
takes into account the physics of vertical transport \citep{WK97}.
Following a similar approach,
in a previous paper we investigated the impact of surface pressure
on the habitability of Earth-like planets 
by incorporating a physical treatment of the thermal radiation
and a simple scaling law for the meridional transport \citep[][hereafter Paper I]{Vladilo13}.
Here  we  include the transport of the short-wavelength  radiation 
and  we present a physically-based  treatment of the  meridional transport
tested with 3D  experiments. 
In this way we build up an Earth-like planet Surface Temperature Model (\model)  
in which a variety of unknown planetary properties can be treated as free parameters
for a fast calculation of the surface habitability.
The \model\ is presented in the next section. 
In Section 3 we describe  the model calibration and validation.
Examples of model applications are presented in Section 4 and
the conclusions  summarized in Section 5.

\begin{figure*}     
\begin{center}
\includegraphics[width=12.5 cm]{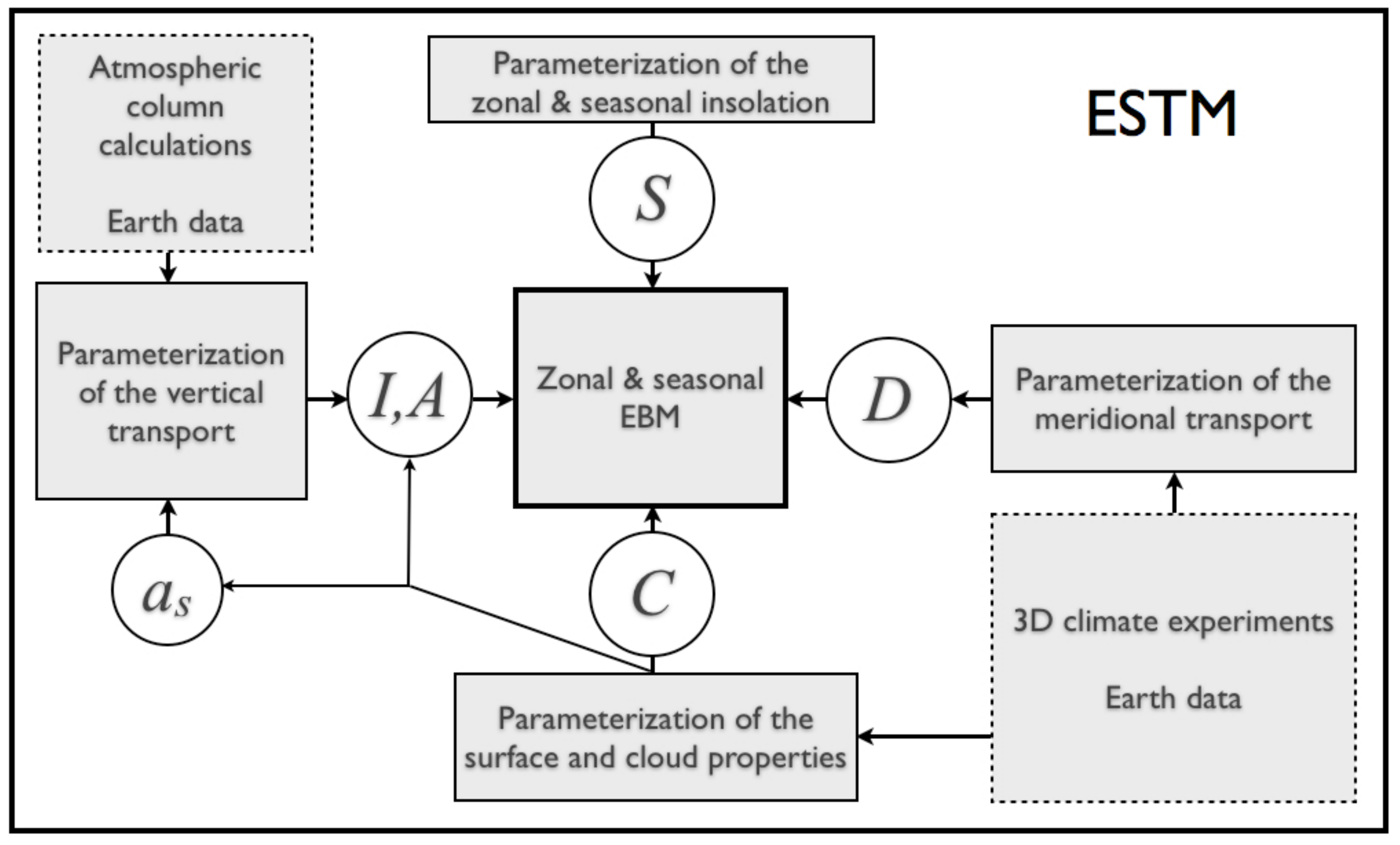} 
\caption{
Scheme of the Earth-like planet Surface Temperature Model (\model). 
The zonal and seasonal  EBM (central box)
is fed by physical quantities (circled symbols) described in \S 2. 
At variance with classic EBMs, the physical quantities are multi-parameter functions
modelled with the aid of atmospheric column calculations ($I$ and $A$) and 3D climate experiments ($D$). 
}
\label{figScheme1}%
\end{center}
\end{figure*}

\section{The model}

The \model\ consists of a set of climate tools and algorithms 
that interchange physical quantities expressed in parametric form.
The core of the model is a zonal and seasonal EBM
fed by multi-parameter physical quantities. 
The parameterization is obtained using physically-based climate tools
that deal with the meridional and vertical energy transport.
The relationship between these  ingredients is shown in the scheme of Fig. \ref{figScheme1}. 
In the following we present the components of the model, starting from the description of the EBM.

In zonal EBMs the surface of the planet is divided into zones delimited
by latitude circles. The surface quantities of interest are averaged in each zone over one rotation period.
In this way, the spatial dependence is determined by a single coordinate, the latitude $\varphi$.
Since the temporal dependence is ``smoothed'' over one rotation period,
 the time, $t$, represents the seasonal evolution during the orbital period.   
The thermal state is described by a single temperature, $T=T(\varphi,t)$,
 representative of the surface conditions.  
By assuming that the heating and cooling rates are  balanced in each  zone,
one obtains an energy balance equation that is used to calculate  $T(\varphi,t)$.
The most common form of EBM equation
\citep{North81,WK97,SMS08,Pierrehumbert10,Gilmore14}  is
\begin{equation}
C  \frac{\partial T}{\partial t} - 
\frac{\partial}{\partial x}
\left[ D \, (1-x^2) \, \frac{ \partial T}{\partial x} \right]
+ I = S \, (1-A) ,
\label{diffusionEq}
\end{equation}
where $x=\sin \varphi$ and all terms are normalized per unit area.
The first term of this equation 
represents the zonal heat storage and describes
the temporal evolution of each zone; $C$ is the zonal heat capacity per unit area \citep{North81}. 
The second term  represents the amount of heat per unit time and unit area
leaving each  zone along the meridional direction  \citep[][Eq. (21)]{North81}.
It is called the ``diffusion term'' because the coefficient $D$ is defined on the basis of the analogy with heat diffusion, i.e.
\begin{equation} 
\Phi  \equiv - D   \frac{\partial T}{\partial \varphi},
\label{eq:DiffA}
\end{equation}
where 
$2 \pi R^2 \Phi \cos \varphi$
is the net rate of energy transport\footnote{
With the adopted definitions of $D$ and $\Phi$ it is easy to show that
${\partial \over \partial x} \left(  \Phi \cos \varphi \right)= 
- {\partial \over \partial x} \left( D \cos \varphi \frac{\partial T}{\partial \varphi} \right)= 
- {\partial \over \partial x} \left[ (1-x^2) D \frac{\partial T}{\partial x} \right]$
represents the latitudinal transport per unit area \citep[see][Eq. 21]{North81}. 
}
across a circle of constant latitude and $R$ is the planet radius \citep[see][]{Pierrehumbert10}.
%
The term $I$ represents the thermal radiation emitted by the zone,   
also called Outgoing Longwave Radiation (OLR).
The  right side of the equation represents the 
fraction of  stellar photons that heat the surface of the zone;
$S$ is the incoming stellar radiation  
and $A$ the planetary albedo at the top of the atmosphere. 
All coefficients of the equation depend, in general, on both time and latitude,
either directly or indirectly, through their dependence on $T$.  

In classic EBMs the coefficients $D$, $I$ and $A$ 
are  expressed in a very simplified form. 
As an example,  
$D$ is often treated as a constant, in spite of the fact that the meridional transport is influenced
by planetary quantities that do not appear  in the formulation (\ref{eq:DiffA}).
The OLR and albedo
are  modelled as simple analytical functions, $I=I(T)$ and $A=A(T)$,
while they should depend not only on $T$, but also on 
other physical/chemical quantities that influence the vertical transport.
This simplified formulation of $D$, $I$ and $A$ prevents important planetary  properties
to appear in the energy balance equation (\ref{diffusionEq}). 
To obtain a physically-based parameterization
we describe
the vertical transport using single-column atmospheric calculations 
and the meridional transport using algorithms  tested with 3D climate experiments.
Thanks to this type of parameterization\footnote{
A simpler parameterization, not tested with 3D climate calculations,
was adopted by \citet{WK97} and in Paper I.}
the \model\ features a  dependence on 
surface pressure, $p$, gravitational acceleration, $g$, planet radius, $R$,
rotation rate, $\Omega$, surface albedo, $a_s$,  stellar zenith distance, $Z$, 
atmospheric chemical composition, and mean radiative properties of the clouds.
By running the simulations described in Appendix A, the \model\ generates
a ``snapshot'' of the surface temperature $T(\varphi,t)$
in a very short computing time,
for any combination of planetary parameters that yield  a stationary solution 
of Eq. (\ref{diffusionEq}). 
We now describe the parameterization of the model.

\subsection{The meridional transport \label{sectMeridionalTransport}}

The heat diffusion analogy (\ref{eq:DiffA}) guarantees the existence of physical solutions and
contributes to the high computational efficiency of EBMs.
In order to keep these advantages and at the same time 
introducing a more realistic treatment of the latitudinal transport, here  we derive  
$\Phi$ and $D$ in terms of planet properties relevant to the physics of the horizontal transport.
To keep the problem simple we focus on the atmospheric transport
(the ocean transport  is discussed below in \S \ref{sectOceanTransport}).
The atmospheric flux can be derived applying basic equations of fluid dynamics 
to the energy content of a parcel of atmospheric gas.
The energy budget of the parcel is expressed in terms of 
the moist static energy (MSE) per unit mass,
\begin{equation}
m=c_p T+ L_v r_v+ gz
\label{eq:mseA}
\end{equation}
where  the terms $c_p T$, $L_v r_v$ and $gz$ measure the sensible heat, the latent heat and the potential energy
content of the parcel at height $z$, respectively; 
$L_v$  is the latent heat of the phase transition between the vapor and the condensed phase;
 $r_v$ the mass mixing ratio of the vapor over dry component; $g$ is the surface gravity acceleration.
The MSE and the velocity of the parcel  are
a function of time, $t$, longitude, $\lambda$, latitude, $\varphi$, and height, $z$.
The latitudinal transport is obtained by integrating the fluid equations   
in longitude and vertically, the height $z$ being replaced by the pressure coordinate, $p=p(z)$.   
Starting from a simplified mass continuity relation valid for the case
in which condensation takes away a minimal atmospheric mass \citep[][\S 9.2.1]{Pierrehumbert10},
one obtains 
the mean zonal flux 
\begin{equation}
\Phi (t,\varphi) 
=  \frac{1}{R} \int_0^{2\pi} d\lambda \int_0^{p} v m \frac{dp'}{g} 
= \frac{1}{R}  \frac{p}{g} \, \overline{  v \, m  }
\label{eq:FluxA}
\end{equation}
where $p$ is the surface atmospheric pressure
and $v$ the meridional velocity component of the parcel. 
The second equality of this expression is valid for a shallow atmosphere,
where $g$ can be considered constant, 
as in the case of the Earth.  

To proceed further, we assume that (\ref{eq:FluxA}) is valid when
the physical quantities are averaged over one rotation period,
since this is the approach used in EBMs. In this case, the
time $t$ represents the seasonal (rather than instantaneous) evolution of the 
system: variability on time scales shorter than one planetary day are averaged out. 
At this point we split the problem in two parts. 
First we derive a relation for $\Phi$ and $D$ valid for the extratropical transport regime. 
Then we introduce a formalism to empirically improve the treatment of the transport inside the Hadley cells.
 
%
%

\subsubsection{Transport in the extratropical region \label{sectEddiesTransport}}

We consider an ideal planet with constant insolation and null axis obliquity, 
in such a way that we can neglect the (seasonal) dependence on $t$.  
We restrict our problem to the atmospheric circulation  typical of fast-rotating terrestrial-type planets, i.e. with
latitudinal transport dominated by eddies in the baroclinic zone. 
A commonly adopted formalism used to treat the eddies consists in dividing
the variables of interest  into a mean component and a perturbation from the mean\footnote{
In general, the ``mean'' and the ``perturbations'' are referred to time and or space variations.
}
, representative of the eddies.
By indicating the mean with an overbar and perturbations with a prime,
we have for instance $v = \overline{v} + v'$ and $m = \overline{m} + m'$.  
It is easy to show that $\overline{vm}= \overline{v}\,\overline{m} + \overline{ v' m'}$. 
When the eddies transport dominates, the term $\overline{v}\,\overline{m}$ can be neglected
so that
\begin{equation}
\Phi \simeq \frac{1}{R}  \frac{p}{g} \, \overline{ v'  m' }  
\label{eq:fluxC}
\end{equation}
and we obtain  
\begin{equation}
D = - \Phi  \left(  \frac{\partial T}{\partial \varphi}  \right)^{-1}
= \frac{1}{R^2}  \frac{p}{g} \, \left(  \frac{\partial T}{\partial y} \right)^{-1} \, \overline{ v'  m' }
\label{eq:DiffB}
\end{equation} 
where $dy=R \, d\varphi$ is the infinitesimal meridional displacement.
To calculate $\overline{ v'  m' }$
we consider the surface value of  moist static energy\footnote
{The MSE is conserved under conditions of dry adiabatic ascent
and is approximately conserved in saturated adiabatic ascent. 
Therefore the MSE is, to some extent, independent of $z$. 
Results obtained by \citet{Lapeyre03} suggest that lower layer values of moist static energy
are most appropriate for diffusive models of energy fluxes.},
%
$m = c_p T + L_v r_v$,
%
%
from which we obtain 
\begin{equation}
\overline{v' m'}=
 c_p \overline{v'T'} + L_v \overline{v'r'_v} ~.
\label{eq:vmfluctA}
\end{equation}
%
%
We express the  mean values of the perturbation products as
\begin{equation}
\overline{ v' T' } = k_\mathrm{S} \, |v'| ~ |T'| 
\label{eq:vTfluctA}
\end{equation}
and 
\begin{equation}
\overline{ v' r_v' } = k_\mathrm{L} |v'|  ~ |r_v'| ,
\label{vpqp}
\end{equation}
where $||$ means a root-mean square magnitude\footnote
{Root-mean square values must be introduced since  
the time mean of the linear perturbations is zero. }
and $k_\mathrm{S}$ and $k_\mathrm{L}$ are correlation coefficients
\citep[e.g.][]{Barry02}. 
%
At this point we need to quantify the perturbations of $T$ and $r_v$, i.e. of the quantities being mixed.
In eddy diffusivity theories these perturbations  
can be written as a mixing length, $\ell_\mathrm{mix}$, times the spatial gradient of the quantity.
We consider the gradient along the meridional  coordinate $y$ 
and we write
\begin{equation}
|T'| = - \ell_\mathrm{mix}    \frac{\partial  T }{\partial y}  
\label{eq:Tfluct}
\end{equation}
and  
\begin{equation}
|r_v'| = - \ell_\mathrm{mix}   \frac{\partial \,  r_v}{\partial y}  
\label{eq:rvfluctA}
\end{equation}
where 
$T$ and $r_v$ are mean zonal quantities; since the mixing is driven by turbulence
we assume that the mixing length is the same for sensitive and latent heat. 
To estimate  $\partial r_v/\partial y$ 
we recall that 
\begin{equation}
r_v  = \frac{\mu_v}{\mu_\mathrm{dry}} \frac{p_v}{p_\mathrm{dry}}
= \frac{\mu_v}{\mu_\mathrm{dry}} \frac{ q \, p_v^* }{p_\mathrm{dry}}
\label{eq:vmr}
\end{equation} 
where 
$\mu_v$ and $p_v$ are the molecular weight and pressure of the vapor,
$\mu_\text{dry}$ and $p_\text{dry}$ the corresponding quantities of the dry air, 
$q$ is the relative humidity and $p_v^*=p_v^*(T)$ is the saturation vapor pressure.  
We assume constant relative humidity and 
we can write
\begin{equation}
 \frac{\partial r_v}{\partial y} =  
\left( \frac{\partial \, r_v}{\partial T} \cdot  \frac{\partial  T}{\partial y} \right)=
\frac{\mu_v}{\mu_\mathrm{dry}} \frac{ q  }{p_\mathrm{dry}}
\frac{\partial  p_v^*}{\partial T}
\,\frac{\partial  T}{\partial y}  ~.
\label{eq:rvy}
\end{equation}
Combining  the expressions from (\ref{eq:vmfluctA}) to (\ref{eq:rvy}) we obtain
\begin{eqnarray}
\overline{v' m'}= -   \ell_\mathrm{mix} \, |v'| \, \frac{\partial  T}{\partial y} 
\left( k_\mathrm{S} c_p + 
k_\mathrm{L}  L_v \frac{ \mu_v }{\mu_\mathrm{dry}   } \frac{q}{p_\mathrm{dry}}
\frac{\partial p_v^*}{\partial T}  \right)
\label{eq:vmfluctC}
\end{eqnarray}
and inserting this  
in (\ref{eq:DiffB})
we derive
\begin{eqnarray}
D \simeq \frac{1}{R^2} \frac{p}{g}  \ell_\mathrm{mix} \, |v'| \,  
\left( k_\mathrm{S} c_p + 
k_\mathrm{L}  L_v \frac{ \mu_v }{\mu_\mathrm{dry}   } \frac{q}{p_\mathrm{dry}}
\frac{\partial p_v^*}{\partial T}  \right) ~.
\label{eq:DtermA}
\end{eqnarray}
At this point, we need an analytical expression for $\ell_\mathrm{mix} \, |v'|$.  
Among a large number of analytical treatments of the baroclinic circulation 
\citep[e.g.,][]{Green70,Stone72,Gierasch73,Held99},
here we adopt a formalism proposed by \citet[][]{Barry02}
which gives the best agreement with GCM experiments.

According to \citet[][]{Barry02}, the baroclinic zone works as  a diabatic heat engine that
obtains and dissipates energy in the process of transporting heat from a warm to a cold region.
If we call $T_w$ and $T_c$
the temperatures of the warm and cold regions, 
the maximum possible thermodynamic efficiency of the engine is  $\delta T/T_w$, where $\delta T = T_w-T_c$. 
The energy received by the atmosphere per unit time and unit mass, $Q$,
represents the diabatic forcing of the engine. 
The rate of generation (and dissipation) of eddy kinetic energy per unit mass is given by
\begin{equation}
\varepsilon = \eta \, \left( \frac{\delta T}{T_w} \right) \, Q 
\label{eq:diabaticForcing}
\end{equation}
where $\eta$ is an efficiency factor representing the fraction of the generated kinetic energy used by heat transporting eddies. 
Assuming that the average properties of the flow depend only on the  length scale and the dissipation rate 
per unit mass\footnote
{If the eddies exist in an inertial range, the average properties of the flow will
depend only on the dissipation rate and the length scale \citep{Barry02}.},
dimensional arguments yield the velocity scaling law
\begin{equation}
|v'| \propto \left( \varepsilon \, \ell_\mathrm{mix} \right)^{1/3} ~.
\label{eq:vflucA}
\end{equation}
As far as the mixing length is concerned,     
the Rhines scale is adopted 
\begin{equation}
\ell_\mathrm{mix}= \left( \frac{2 |v'| }{\beta} \right)^{1/2} 
\label{eq:rhines}
\end{equation}
where $\beta=\partial f/\partial y$ is the gradient of the Coriolis parameter, 
$f=2 \Omega \sin \varphi$, and $\Omega$ the angular rotation rate of the planet.
The study of \citet{Barry02} suggests that, among other types of length scales 
considered in literature, the Rhines scale yields the best correlations in 3D atmospheric experiments.
The adoption of the Rhines scale is also supported by a study of moist transport 
performed with GCM experiments \citep{Frierson07}.
The Rhines scale must be calculated at the latitude $\varphi_\mathrm{m}$ 
of maximum kinetic energy, i.e. for $\beta=(2 \Omega \cos \varphi_\mathrm{m})/R $.
From the above expressions 
we obtain 
\begin{equation}
\ell_\mathrm{mix} \, |v'|  =
\left( \eta \frac{ \delta T }{T_w }  \, Q \right)^{3/5} 
\left( \frac{R}{\Omega \cos \varphi_\mathrm{m}} \right)^{4/5} ~.
\label{eq:ellvfluc}
\end{equation}
%
%
Inserting this  in (\ref{eq:DtermA}) we obtain
\begin{equation}
D = D_\mathrm{dry} ( 1 + \Lambda)
\label{eq:DtermB}
\end{equation}
where
\begin{eqnarray}
\lefteqn{
D_\mathrm{dry} = k_\mathrm{S}  c_p \, \eta^{3/5} (\cos \varphi_\mathrm{m})^{-4/5} \times}
  \nonumber \\
&  &   {}
\times 
\, R^{-6/5}  \, \frac{p}{g} \,  \Omega^{-4/5}   
\left( \frac{ \delta T }{T_w }   Q \right)^{3/5} 
\label{eq:Ddry}
\end{eqnarray}
is the dry component of the  atmospheric eddies transport  
and
\begin{equation}
\Lambda =  
\frac{k_\mathrm{L}  L_v}{k_\mathrm{S}  c_p } \frac{ \mu_v }{\mu_\mathrm{dry}   } 
\frac{q}{p_\mathrm{dry}}
\frac{\partial  p_v^* }{\partial  T}
\label{eq:Lambda}
\end{equation}
is the ratio of the moist over dry components. 

For the practical implementation of the analytical expressions
(\ref{eq:DtermB}), (\ref{eq:Ddry}) and (\ref{eq:Lambda}) in the EBM code,
we proceed as follows. 
The maximum thermodynamic efficiency $\delta T/T_w$ is calculated by taking 
$T_w=\overline{T}(\varphi_1)$ and $T_c=\overline{T}(\varphi_2)$
where $\varphi_1$ and $\varphi_2$ are the borders of the mid-latitude region and
overbars indicate zonal annual means. 
Following \citet{Barry02}, we adopt $\varphi_1=28^\circ$ and $\varphi_2=68^\circ$,
after testing that the model predictions are virtually unaffected by the exact choice of these values\footnote
{Also the GCM experiments by \citet{Barry02} indicate that the results
are not sensitive to the choice of $\varphi_1$ and $\varphi_2$.}. 
We  estimate the diabatic forcing term (W/kg) as $Q \simeq \left\{ \mathrm{ASR} \right\} /(p/g)$,
where $\left\{ \mathrm{ASR}\right\} = \left\{  S(1-A) \right\}$ is the absorbed stellar radiation 
(W/m$^2$) averaged over one orbital period
in  the latitude range ($\varphi_1$, $\varphi_2$) and $p/g$ the atmospheric columnar mass (kg/m$^2$).
We neglect the contribution of surface fluxes of sensible heat
since they cannot be estimated in the framework of the EBM model.
This approximation is not critical because 
these fluxes yield a negligible contribution to $Q$ according to \citet{Barry02}. 
Treating $k_\mathrm{L}$, $k_\mathrm{S}$, $\eta$, and $\varphi_\mathrm{m}$ as constants\footnote
{Numerical experiments performed with simplified GCMs suggest that the correlation coefficients
$k_\mathrm{L}$, $k_\mathrm{S}$ and the efficiency factor $\eta$
can be treated as constants with good approximation \citep{Barry02,Frierson07}.},
we obtain from Eq. (\ref{eq:Ddry}) a  scaling law for the dry term of the transport
\begin{eqnarray}
\lefteqn{
\mathcal{S}_\mathrm{dry} \propto c_p \, R^{-6/5}  \left( \frac{p}{g} \right)^{2/5}   \Omega^{-4/5} \times  {} }
\nonumber\\
& & {} \times 
\left( \frac{ \delta T }{T_w }  \left\{ \mathrm{ASR} \right\}  \right)^{3/5}  .
\label{eq:SLdry}
\end{eqnarray}
We  estimate the temperature gradient of saturated vapor pressure  
as $\partial  p_v^* /\partial  T \simeq \delta p_v^* / \delta T$,
with $\delta p_v^* = \left[ p_v^*(T_w) -p_v^*(T_c) \right]$. 
Since $k_\mathrm{L}$, $k_\mathrm{S}$ and $L_v$ are constants,
we obtain from Eq.  (\ref{eq:Lambda}) a scaling law for the ratio of the moist over dry components
\begin{equation}
\mathcal{S}_\mathrm{md} \propto
\frac{q}{c_p \, \mu_\mathrm{dry} \, p_\mathrm{dry}}
\frac{\delta  p_v^* }{\delta  T} ~.
\label{eq:SLdm}
\end{equation}
Finally, by applying Eq. (\ref{eq:DtermB}) and 
the scaling laws  (\ref{eq:SLdry}) and (\ref{eq:SLdm}) 
to a generic terrestrial planet
and to the Earth,
indicated by the subscript $\circ$, we obtain 
\begin{equation}
\frac{D}{D_\circ} = \frac{ \mathcal{S}_\mathrm{dry} }{ \mathcal{S}_{\mathrm{dry},\circ}  } \, 
\left[ \frac{1+\Lambda_\circ \cdot
\left( \mathcal{S}_\mathrm{md} / \mathcal{S}_{\mathrm{md},\circ} \right) }
{1+\Lambda_\circ} \right] ~.
\label{eq:DtermF}
\end{equation}
With the above expressions we  calculate $D$ treating
  $R$, $\Omega$, $p$, $g$, 
$c_p$, $\mu_\mathrm{dry}$, $p_\mathrm{dry}$, $q$ as parameters that can vary
from planet to planet, in spite of being constant in each planet. 
The ratio of moist over dry eddie transport of the Earth is set to $\Lambda_\circ=0.7$ \citep[e.g.][]{KS14}. 
For the sake of self-consistency, we adopt the parameters 
$(\delta_T)_\circ$, $(T_w)_\circ$, $\left\{ \mathrm{ASR} \right\}_\circ$ 
and $(\delta  p_v^*)_\circ$ obtained from the Earth's reference model.
Since these parameters vary in the course of the simulation, we perform the calibration of the Earth model in two steps.
First we calibrate the model excluding  
the ratios\footnote{Excluding these ratios is equivalent to setting them equal to unity
in the Earth's model, as they should be by definition.}
$\delta_T/(\delta_T)_\circ$, $T_w/T_{w,\circ}$, 
$\left\{ \mathrm{ASR} \right\}/\left\{ \mathrm{ASR} \right\}_\circ$ and $\delta  p_v^*/(\delta  p_v^*)_\circ$
from the scaling laws of Eq. (\ref{eq:DtermF}). 
Then we reintroduce these ratios in the scaling law adopting 
for $(\delta_T)_\circ$, $(T_w)_\circ$ and $\left\{ \mathrm{ASR} \right\}_\circ$ 
the values $(\delta_T)$, $(T_w)$, $\left\{ \mathrm{ASR} \right\}$ and $(\delta  p_v^*)$
obtained in the first step.
The second step is repeated a few times, until convergence of the parameters 
$(\delta_T)_\circ$, $(T_w)_\circ$ and $\left\{ \mathrm{ASR} \right\}_\circ$ 
and $(\delta  p_v^*)_\circ$ is achieved.

\subsubsection{Transport in the Hadley Cell \label{sectHadleyCell}}

The derivation performed above ignores the existence of the Hadley Cells, since they do not contribute
to the extratropical meridional transport. However, the Hadley circulation is extremely efficient 
in smoothing temperature gradients inside the tropical region. This aspect cannot be completeley
ignored in our treatment, since our goal is to estimate the planet surface temperature distribution. 
Unfortunately, the diffusion formalism of Eq. (\ref{eq:DiffA}) 
is inappropriate inside the Hadley Cells and the only way we have to improve the description of the
tropical temperature distribution is to correct the formalism with some empirical expression.
We summarize the approach that we follow to cope with this problem. 

The global pattern of atmospheric circulation is influenced, among other factors, by the 
seasonal variation of the zenith distance of the star.  In the case of
the Earth, a well known example of this type of influence is the seasonal shift of the Intertropical Convergence Zone
(ITCZ), that moves to higher latitudes in the summer hemisphere. 
The ITCZ is, in practice, a tracer of the thermal equator at the center of the system of the two Hadley Cells,
where we want to improve the uniformity of the temperature distribution. 
A way to do this is to enhance the transport coefficient $D$ in correspondence with such thermal equator.
To incorporate this feature in our model, we scale $D$ 
according to mean diurnal value of $\mu(\varphi,t)=\cos Z$, where $Z$ is the stellar zenith distance.  
In practice, we multiply $D$ by a dimensionless modulating factor, $\zeta(\varphi,t)$, that scales linearly with $\mu(\varphi,t)$,
i.e. $\zeta(\varphi,t)=c_o+c_1 \mu(\varphi,t)$.
We normalize this factor in such a way that its mean global annual value is $\widetilde{\zeta}(\varphi,t)=1$. 
Thanks to the normalization condition, it is possible to calculate the parameters $c_0$ and $c_1$
in terms of a single parameter, $\mathcal{R}= \max \left\{ \zeta(\varphi,t)\right\}/ \min \left\{ \zeta(\varphi,t)\right\}$,
which represents the ratio between the maximum and minimum values of $\zeta$ at any latitude and orbital phase
\citep[see][\S A.2.1]{Vladilo13}. 
With the adoption of the modulation term, the complete expression for the transport coefficient becomes
\begin{eqnarray}
\frac{D}{D_\circ} = \zeta(\varphi,t) \, \frac{ \mathcal{S}_\mathrm{dry} }{ \mathcal{S}_{\mathrm{dry},\circ}  } \, 
\left[ \frac{1+\Lambda_\circ \cdot
\left( \mathcal{S}_\mathrm{md} / \mathcal{S}_{\mathrm{md},\circ} \right) }
{1+\Lambda_\circ} \right] .
\label{eq:DtermG}
\end{eqnarray}
The mean global annual value of this expression equals Eq. (\ref{eq:DtermF}) 
thanks to the normalization condition $\widetilde{\zeta}(\varphi,t)=1$. 
This formalism introduces a dependence on $t$ and on the axis obliquity\footnote
{The mean diurnal value of $\mu(\varphi,t)$ is a function of the axis obliquity.}
in the transport coefficient. 

\begin{figure*}[ht]    
\begin{center} 
\includegraphics[width=7.5cm]{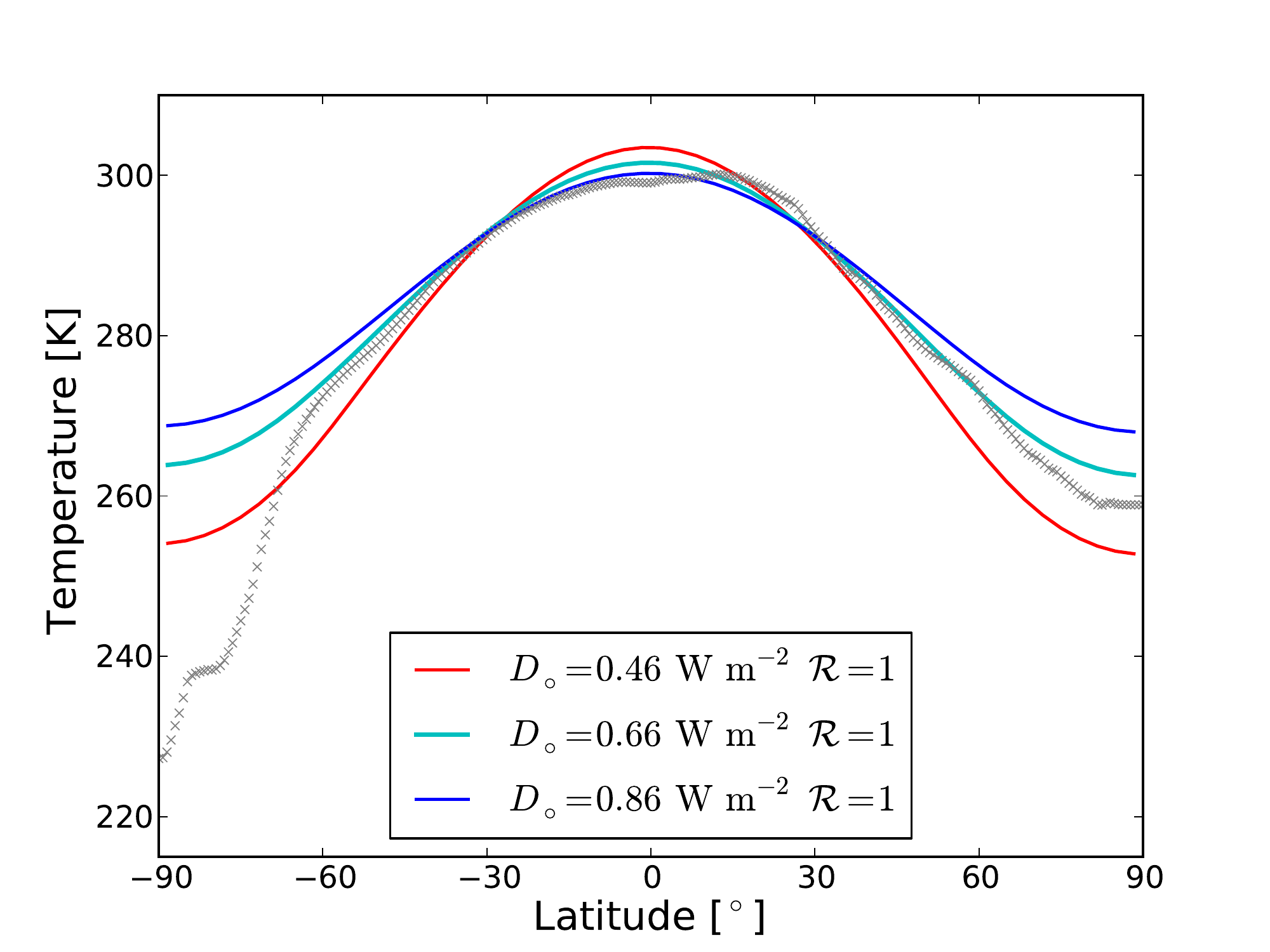}
\includegraphics[width=7.5cm]{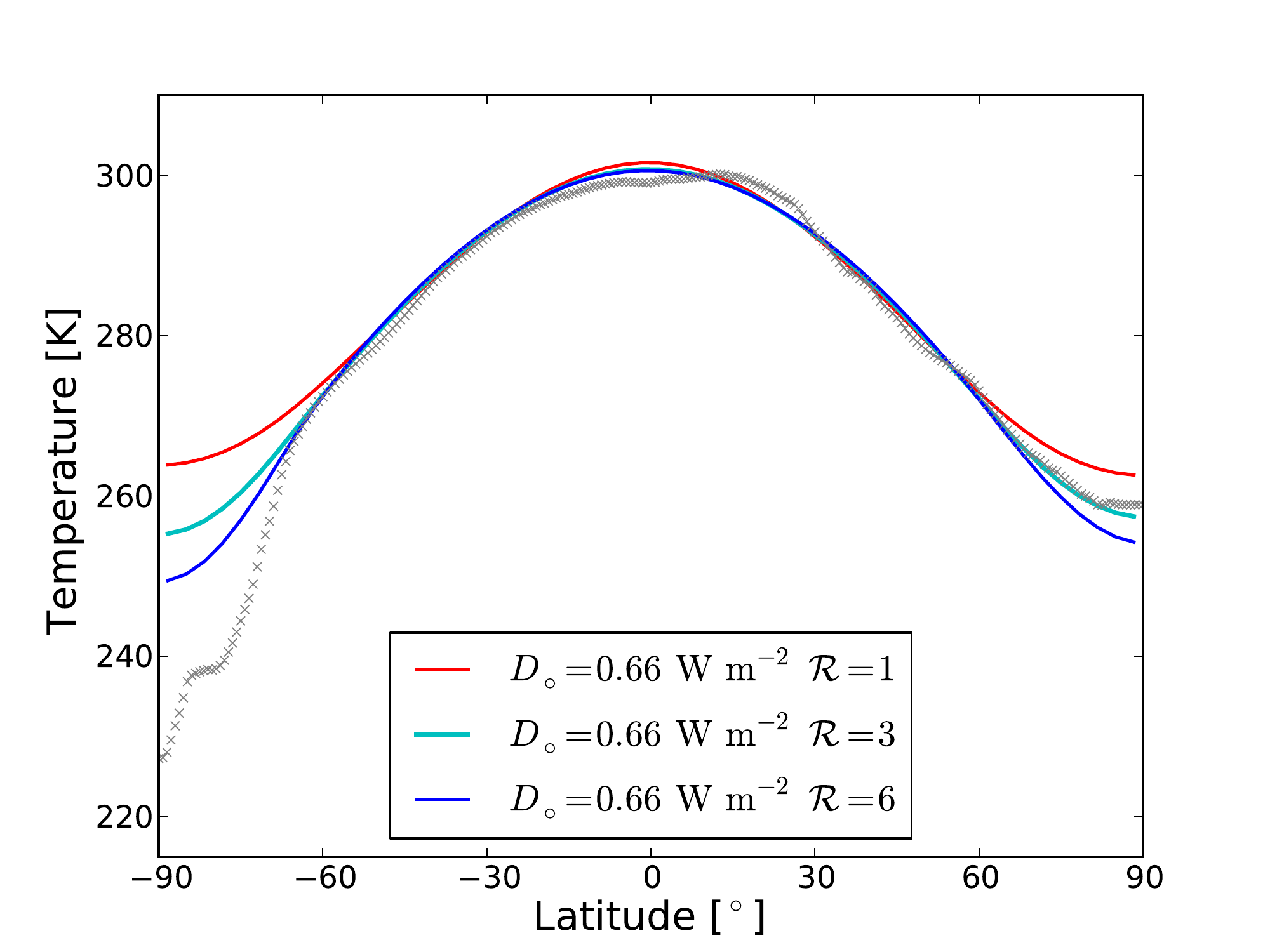}    %
\caption{
Annual temperature-latitude profile of the Earth model obtained by varying the transport parameters $D_\circ$ and $\mathcal{R}$.
Left panel: variation of $D_\circ$ at $\mathcal{R}=1$ (i.e. without seasonal modulation of the transport coefficient).
Right panel: variation of the strength of seasonal modulation, $\mathcal{R}$, at constant $D_\circ$. 
See \S \ref{sectHadleyCell}. 
}
\label{compareDR}%
\end{center}
\end{figure*}

Empirical support for the adoption of the modulation term $\zeta(\varphi,t)$ comes  
from the improved match between the observed and predicted temperature-latitude profile of the Earth.
In the left panel of Fig. \ref{compareDR} we show that it is not possible to accurately match 
the Earth profile by varying $D_\circ$ at constant $\zeta(\varphi,t)=1$ (i.e. $\mathcal{R}=1$). 
This is because the whole profile becomes flatter with increasing $D_\circ$
and the values of $D_\circ$ sufficiently high to provide the desired smooth temperature distribution
inside the tropics yield a profile which is too flat in the polar regions.
This problem can be solved with the introduction of the modulation factor $\zeta$. 
In the right panel of Fig. \ref{compareDR} we show that by increasing  $\mathcal{R}$
the profile declines faster at the poles while becoming slightly flatter at the equator. 
This behavior is different from that induced by changes of $D_\circ$
and provides an extra degree of freedom to match the observed profile.  
For the time being, the parameter $\mathcal{R}$ can be tuned to fit the Earth model,
but cannot be validated with other planets. The validation of $\mathcal{R}$
in rocky planets different from the Earth could be addressed by future GCM calculations.
Meantime, the uncertainty related to the choice of this parameter in other planets
can be estimated by repeating the climate simulations for different values of $\mathcal{R}$. 
Given the lack of  solid theoretical support for the adoption of the $\zeta(\varphi,t)$ formalism,
it is safe to use the smallest possible value of $\mathcal{R}$ (i.e. closest to unity)
that allows the Earth profile to be reproduced. With the upgraded calibration of the Earth
model presented here (Appendix B) we have been able to adopt a lower value ($\mathcal{R}=2.2$, Table \ref{tabFiducialPar}) 
than in Paper I ($\mathcal{R}=6$, Table 2 in \citet{Vladilo13}). 

\subsubsection{Ocean transport \label{sectOceanTransport}}

The algorithm that describes the energy transport has been derived 
assuming that most of the meridional transport is performed by the atmosphere rather than the ocean 
(\S \ref{sectMeridionalTransport}). 
This is a reasonable assumption in the Earth climate regime,
where the atmosphere contributes  78\% of the total transport
in the Northern Hemisphere and 92\% in the Southern Hemisphere 
at the latitude of maximum poleward transport \citep{Trenberth01}. 
In order to assess the importance of the ocean contribution in different planetary regimes 
one needs to run GCM simulations featuring the ocean component. 
This is a difficult task because the ocean circulation is extremely dependent
on the {\em detailed} distribution of the continents  and because the time scale of ocean response 
is much longer than that of the atmosphere. 
As a result, one should run GCMs with a detailed description of the geography for a large number of orbits
in order to include the ocean transport in the modellization of exoplanets.    
With this type of climate simulation it would be impossible to perform an exploratory
study of exoplanet surface temperature, which is the aim of our model. 
Not to mention the fact that the choice of a {\em detailed} 
description of the continental distribution in exoplanets is completely arbitrary. 
It is therefore desirable to find simplified algorithms able to include the ocean transport
in zonal models, such as the \model. To this end, 
one should perform 3D numerical experiments 
aimed at investigating how the energy transport is partitioned  between the atmosphere and the ocean
in a variety of planetary conditions. Preliminary work of this type suggests that the energy transport of
wind-driven ocean gyres\footnote{
In oceanography the term gyre refers to major ocean circulation systems driven
by the wind surface stress.}
vary in a roughly similar fashion to the energy transport of the atmosphere 
as external parameters vary \citep{Vallis09}. 
The existence of mechanisms of compensation that regulate the relative
contribution of the atmosphere and the ocean to the {\em total} transport 
\citep{Bjerknes64,Shaffrey06,vanderSwaluw07,Lucarini11} 
may also help build a simplified description of the atmosphere/ocean transport. 
In the case of the Earth, we note that the {\em total} transport is remarkably similar
in the Southern and Northern hemisphere (see Fig. \ref{annualLatProfiles})
in spite of significant differences between the two hemispheres
in terms of the relative contribution of the ocean and atmosphere \citep[e.g.][Fig. 7]{Trenberth01}.

\subsection{The vertical transport \label{sectCRM}}

%
The outgoing longwave radiation and the top-of-atmosphere albedo are parametrized
using single atmospheric column calculations. 
In the present  version  of the \model,  
the single column calculations are performed 
with standard radiation codes developed at the National Center for Atmospheric Research (NCAR), as part of the Community Climate Model (CCM) project NCAR-CCM \citep{Kiehl98ccm3}.   
To access these codes we use the set of routines CliMT 
\citep{Pierrehumbert10,Caballero12}.


The CCM code employs an Earth-like atmospheric composition, with the possibility
to change the amount of non-condensable greenhouse gases
(i.e. CO$_2$ and CH$_4$). 
We adopt $p\mathrm{CO}_2=380$\,ppmv 
 and $p\mathrm{CH}_4 =1.8$\,ppmv
as the reference values for the Earth's model.
These values can be changed as long as they remain in trace abundances, as in the case of the Earth.
The relative humidity, $q$, is fixed to limit the huge amount of calculations and the dimensions of the tables described below.
%
%
We adopt $q=0.6$, a value 
consistent with the global relative humidity measured on Earth.
A low effective humidity ($q\sim0.6$) is predicted self-consistently by 3-D dynamic climate models
as a result of subsidence in the Hadley circulation \citep[e.g.][]{Ishiwatari02}.
Adoption of saturated water vapour pressure  ($q=1$) tends to understimate the OLR
at high temperatures, leading to excessive heating of the planet. 
%

\subsubsection{Outgoing long-wavelength radiation}

We use a column radiation model scheme for a cloud-free atmosphere to calculate the Outgoing Long-wavelength Radiation (OLR), i.e. the thermal infrared emission that cools the planet. 
The OLR calculations are repeated a large number of times in order to cover 
a broad interval of surface temperature, $T$, background pressure, $p$, gravity acceleration, $g$, and partial pressure of non-condensable greenhouse gases. 
The results of these calculations are stored in tables
OLR=OLR$(T,p,g,p\mathrm{CO}_2,p\mathrm{CH}_4)$. 
In the course of the simulation, these tables are interpolated at the zonal and instantaneous value 
of $T=T(\varphi,t)$. 
The long-wavelength forcing of the clouds is subtracted at this stage, 
taking into account the zonal cloud coverage, as we explain in  \S \ref{sectClouds}.
The total CPU time required to cover the parameter space 
$(T,p,g,p\mathrm{CO}_2,p\mathrm{CH}_4)$ is relatively large. However, once the tables are built up, the simulations
are extremely fast. 

%
%

\subsubsection{Incoming short-wavelength radiation \label{sectTOAalbedo}}

The top-of-atmosphere albedo, $A$, is calculated with 
the CCM code, to take into account the transfer of short-wavelength
stellar photons in the planet atmosphere.  
In each atmospheric column we calculate the fraction of stellar photons that
is reflected back in space for different values of  $T$, $p$,  $g$, $p\mathrm{CO_2}$,
surface albedo, $a_s$, and zenith distance of the star, $Z$.  
In practice, for each set of values ($g$, $p\mathrm{CO}_2$, $p\mathrm{CH}_4$),
we calculate the temperature and pressure dependence of $A$.
Then, for each set of values $(T,p,g,p\mathrm{CO_2},p\mathrm{CH}_4)$ 
the calculations are repeated to cover the complete intervals of surface albedo, $0 < a_\mathrm{s} < 1$, and  zenith distance, $0^\circ < Z < 90^\circ$. The results of these  calculations are stored in multidimensional tables. 
In the course of the \model\ simulations 
these tables are interpolated to calculate $A$ as a function 
of the zonal and instantaneous values of $(T,p,g,p\mathrm{CO_2},a_\mathrm{s},Z)$. 
Each single column calculation of $A$ is relatively fast, compared to the corresponding
calculation of $I$. However, due to the necessity of covering a larger parameter space, 
the preparation of the tables $A=A(T,p,g,p\mathrm{CO_2},a_\mathrm{s},Z)$ requires
a comparable CPU time. 

\subsubsection{Caveats \label{sectCCMcaveats}}

The CCM calculations that we use include pressure broadening \citep[][and refs. therein]{Kiehl98ccm3}, 
but not collision-induced absorption. 
As a result, the model  may underestimate the atmospheric absorption at the highest values of pressure. 
To avoid physical conditions not considered in the  calculations
we limit the  surface pressure at $p \la  10$\,bar.

The  calculations 
are valid for a solar-type spectral  distribution.
The spectral type of the central star affects the vertical transport 
because of the wavelength dependence of the atmospheric albedo \citep[e.g.,][]{Selsis07}. 
The present version of the \model\ should be applied to planets orbiting stars with 
spectral distributions not very different from the solar one.

\subsection{Surface and cloud properties \label{sectSurfaceProperties}} 

\subsubsection{Zonal coverage of oceans, lands, ice and clouds \label{sectCoverage}}
 
The zonal coverage of oceans is a free parameter, $f_o$,  
that also determines the fraction of continents, $f_l=1-f_o$. 
In this way, the planet geography is specified in a schematic way 
by assigning a set of $f_o$ values, one for each zone.   
The zonal coverage of ice and clouds is  parametrized using algorithms calibrated with Earth experimental data.
Following WK97, 
the zonal coverage of ice is a function of the mean diurnal temperature,  
\begin{eqnarray}
f_i (T)=  \max 
\left\{ 0, 
 \left[ 1 - e^{ (T-273.15\,\mathrm{K} )/ 10\,\mathrm{K} }
 \right]  \right\} .
\label{fice}
\end{eqnarray} 
One problem with this formulation is that the ice melts completely 
and instantaneously as soon as  $T > 273.15$\,K.
To minimize this effect, we introduced an algorithm
that mimics the formation of permanent ice when
a latitude zone is below the freezing point for more than half the orbital period.
In this case, we adopt a constant ice coverage for the full orbit, $f_i=f_i (\overline{T})$,
where $\overline{T}$ is the mean {\em annual} zonal temperature.
%
%

As far as the clouds are concerned, we adopt
specific values of zonal coverage for clouds over oceans and continents. 
The dependence of the cloud coverage on the type of underlying surface 
has long been known \citep[e.g.][]{Kondratev69}
and has been quantified in recent studies \citep[e.g.][]{Sanroma12,Stubenrauch13}.
%
Based on the results obtained by \citet{Sanroma12},
we adopt $0.70$ and $0.60$ for the cloud coverage over oceans and lands, respectively. 
In this way, the reference Earth model (Appendix B)  
predicts a
mean annual global cloud coverage $\leftmean f_{c,\circ} \rightmean =0.67$,
in excellent agreement with  most recent Earth data \citep{Stubenrauch13}.
With our formalism the cloud coverage is automatically adjusted for planets with 
cloud properties similar to those of the Earth, but different 
fractions of continents and oceans. 
Since the coverage of ice, $f_i$, depends on the temperature, 
the model simulates the  feedback between temperature and albedo.

\subsubsection{Cloud radiative properties \label{sectClouds}}

The albedo and infrared absorption of the clouds have cooling and warming effects
of the planet surface, respectively.
Even with specifically designed 3D models it is hard to predict which of these two opposite effects dominate.
The single-column radiative calculations used in studies of habitability  
usually assume cloud-free radiative transfer and tune  the 
results by playing with the albedo \citep{Kasting88,Kasting93,Kopparapu13,Kopparapu14}.
The approach that we adopt with the \model\ is to
parametrize the albedo and the long-wavelength forcing of the clouds 
assuming that their global properties are similar to those measured in the present-day Earth.
Following WK97, we express the albedo of the clouds as 
\begin{equation}
a_c =  \alpha +  \beta  Z   
\label{cloudAlbedo}
\end{equation} 
where the parameters $\alpha$ and $\beta$ are tuned to fit
Earth experimental data of cloud albedo as a function
of stellar zenith distance \citep{Cess76}.  For clouds over ice, we adopt the same albedo of frozen surfaces
(see Table \ref{tabFiducialPar}). 
To take into account the long wavelength forcing of the clouds,
we subtract 
$\leftmean \mathrm{OLR} \rightmean_\mathrm{cl,\circ} \, (f_c/\leftmean f_{c,\circ} \rightmean)$  
from the clear-sky OLR obtained from the radiative calculations, where
$\leftmean \mathrm{OLR} \rightmean_\mathrm{cl,\circ}= 26.4$ W m$^{-2}$ is
the mean global long wavelength forcing of the clouds on Earth \citep{Stephens12},
$f_c$ is the mean cloud coverage in each latitude zone, and 
$\leftmean f_{c,\circ} \rightmean =0.67$ the mean global cloud coverage of the 
reference Earth model. 

%

The fact that the \model\ accounts for the mean radiative properties of the clouds is an improvement over classic EBMs,
but one should be aware that 
the adopted parameterization is only valid  
for planets with  global cloud  properties similar to those of the Earth.  
This is a critical point
because the cloud radiative properties may change with planetary conditions,
as suggested by 3D simulations of terrestrial planets  
\citep[e.g.][]{Leconte13,Yang13}. 
To some extent, we can simulate this situation by changing the \model\  cloud-forcing parameters.
An example of this exercise is provided in Fig. \ref{mapsOLRclouds}.
If the predictions of 3D experiments become more robust, it could be possible
in the future to introduce a new \model\ recipe for expressing the cloud forcing as a function of 
relevant planetary parameters.

\subsubsection{The surface albedo \label{sectionAlbedo}}

The mean surface albedo of each latitude zone
is calculated by averaging the albedo of each type of surface present in the zone,
weighted according to its zonal coverage. 
%
%
For the surface albedo of continents and ice we adopt the fiducial values
listed in Table \ref{tabFiducialPar}. The albedo of the oceans  
is calculated as a function of the stellar zenith distance, $Z$,
using an expression calibrated with  experimental data
\citep{Briegleb86,Enomoto07}
%
%
\begin{eqnarray}
\lefteqn{
a_o = { 0.026 \over (1.1 \, \mu^{1.7} + 0.065)} + {} }  \nonumber\\
& &  {}  + 0.15 (\mu-0.1) \, (\mu-0.5) \, (\mu-1.0)   ~,
\end{eqnarray}
where $\mu = \cos Z$.  
Also clouds are treated as surface features, with zonal coverage and albedo parametrized 
as explained above (\S  \ref{sectClouds}). 
%


\subsubsection{Thermal capacity of the surface \label{sectThermalCapacity}}

The term $C$  is calculated by averaging the thermal capacity per unit area 
of each type of surface present in the corresponding zone
according to its zonal coverage (\S \ref{sectCoverage}).
The parameters 
used in these calculations 
are representative of the thermal capacities of oceans and solid surface (Table \ref{tabFiducialPar}).
For the reference Earth model the ocean contribution is calculated 
assuming a 50 m, wind-mixed ocean layer\footnote{
The thermal capacity of the ocean can be changed to simulate idealized aquaplanets 
with a thin layer of surface water  (e.g., \S \ref{sectValidation}).}
\citep{WK97,Pierrehumbert10}.
The atmospheric contribution is calculated as
\begin{equation}
\left( \frac{ C_\mathrm{atm}}{C_{\mathrm{atm},\circ} } \right)
= \left( \frac{ c_p}{c_{p,\circ} } \right) \,
 \left( \frac{ p}{p_\circ} \right) \,  \left( \frac{ g_\circ}{g} \right)~~~,
 \label{Catm}
\end{equation}
where $c_p$ and $p$ are the specific heat capacity and total pressure of the atmosphere,
respectively \citep{Pierrehumbert10}.
The atmospheric term enters as an additive contribution
to the ocean and solid surface terms.
Its impact on these parameters is generally small, the ocean contribution being the dominant one.  

The strong thermal inertia of the oceans implies that
the mean zonal $C$  has an ``ocean-like'' value even when 
the zonal fraction of lands is comparable to that of the oceans \citep{WK97}.
This weak point of the longitudinally-averaged model can be by-passed
 by adopting an idealized orography with continents covering all longitudes
(see \S \ref{sectOceanLand}). 

\subsection{The insolation term $S$}

The zonal, instantaneous stellar radiation $S=S(\varphi,t)$ is calculated  
from the  stellar luminosity, the keplerian orbital parameters and the inclination of the planet rotation axis. 
The model calculates $S$ also for eccentric orbits.
Details on the implementation of $S$ can be found in Paper I \citep[][\S A.5]{Vladilo13}.  
At variance with that paper, the \model\ takes also
into account the vertical transport of short-wavelength  
photons (see \S \ref{sectTOAalbedo}).

\subsection{Limitations of the model}

In spite of the above-mentioned improvements over classic EBMs, the adoption of 
the zonal energy balance formalism at the core of the \model\ leads to well known limitations intrinsic to EBMs. 
One  is that zonally averaged models cannot be applied to tidally-locked
planets that always expose the same side to their central star:
such cases require specifically designed models 
\citep[e.g.,][]{Kite11,Menou13,Mills13,Yang13}.  
Also, it should be clear that 
the \model\ does not track climatic effects that develop in the vertical direction,
even though the atmospheric response is adjusted according to latitudinal and seasonal
variations of $T$ and $Z$.
%
In spite of these limitations, the EBM at the core of our climate tools
provides the flexibility that is required when many runs are needed 
or when one wants to compare the impact of different parameters unconstrained by the observations.
At present time this is still unfeasible with GCMs and even with Intermediate Complexity Models.
While GCMs are invaluable tools for climate change 
studies on Earth, they are heavily parameterized on current Earth conditions, and their use in significantly different conditions raises concern.  In particular, the paper by \citet{Stevens13} caused serious worries about the use of 
GCMs in ``unconstrained'' situations such as those encountered in habitability studies.

\section{Model calibration and validation}

The \model\ is implemented in two stages.
First, a reference Earth model is built up by tuning the parameters  
to match the present Earth climate properties (see  Appendix B).
Then we use results obtained from 3D climate experiments to tune parameters or
validate algorithms that are meant to be applied in Earth-like planets. 
Here we present a test of validation  of the algorithms that describe the meridional transport.
This test is a concrete example of how results obtained by GCMs can be used to validate  the model.

\subsection{Validation of the meridional transport \label{sectValidation}}

\begin{figure*}
\begin{center} 
\includegraphics[width=7.cm]{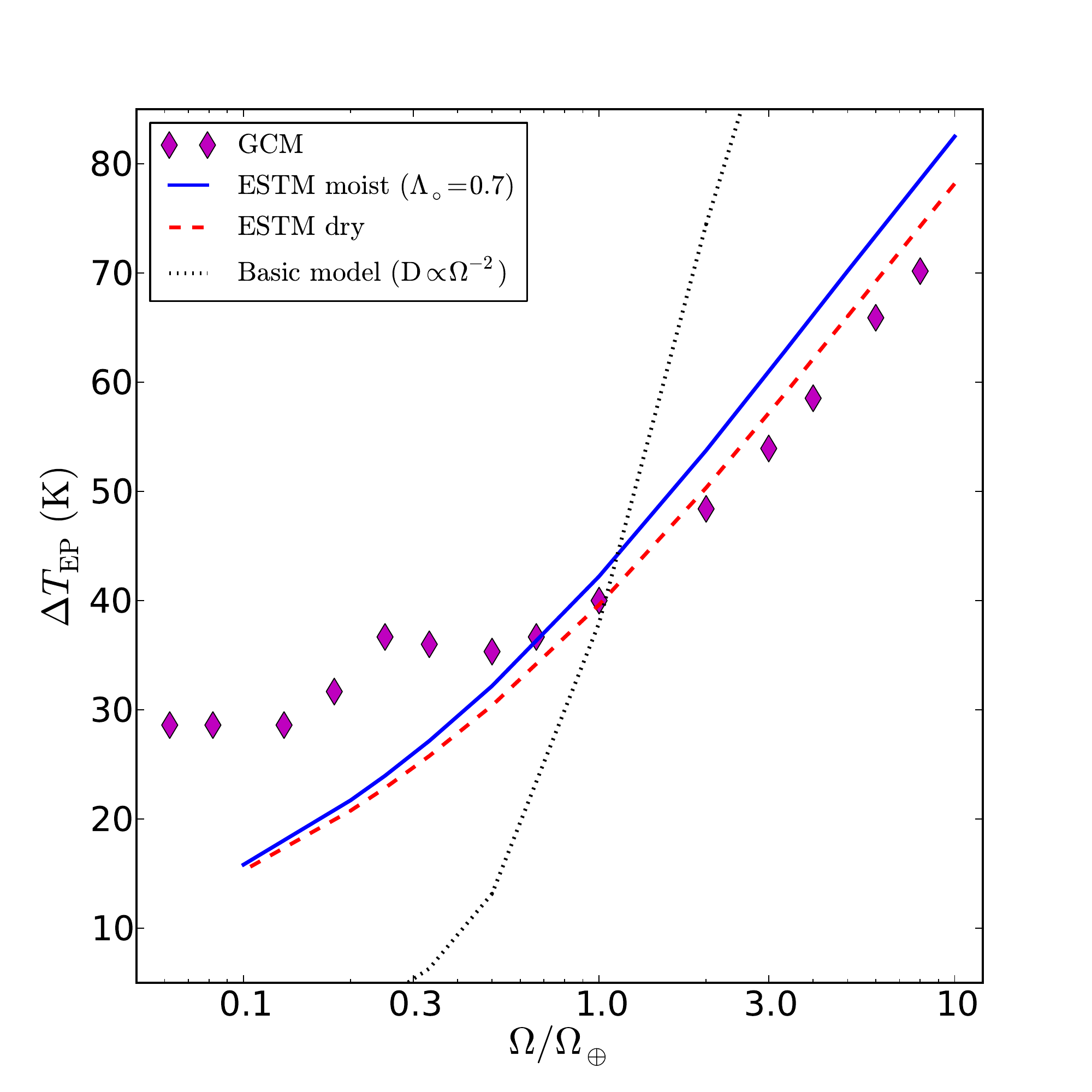}
\includegraphics[width=7.cm]{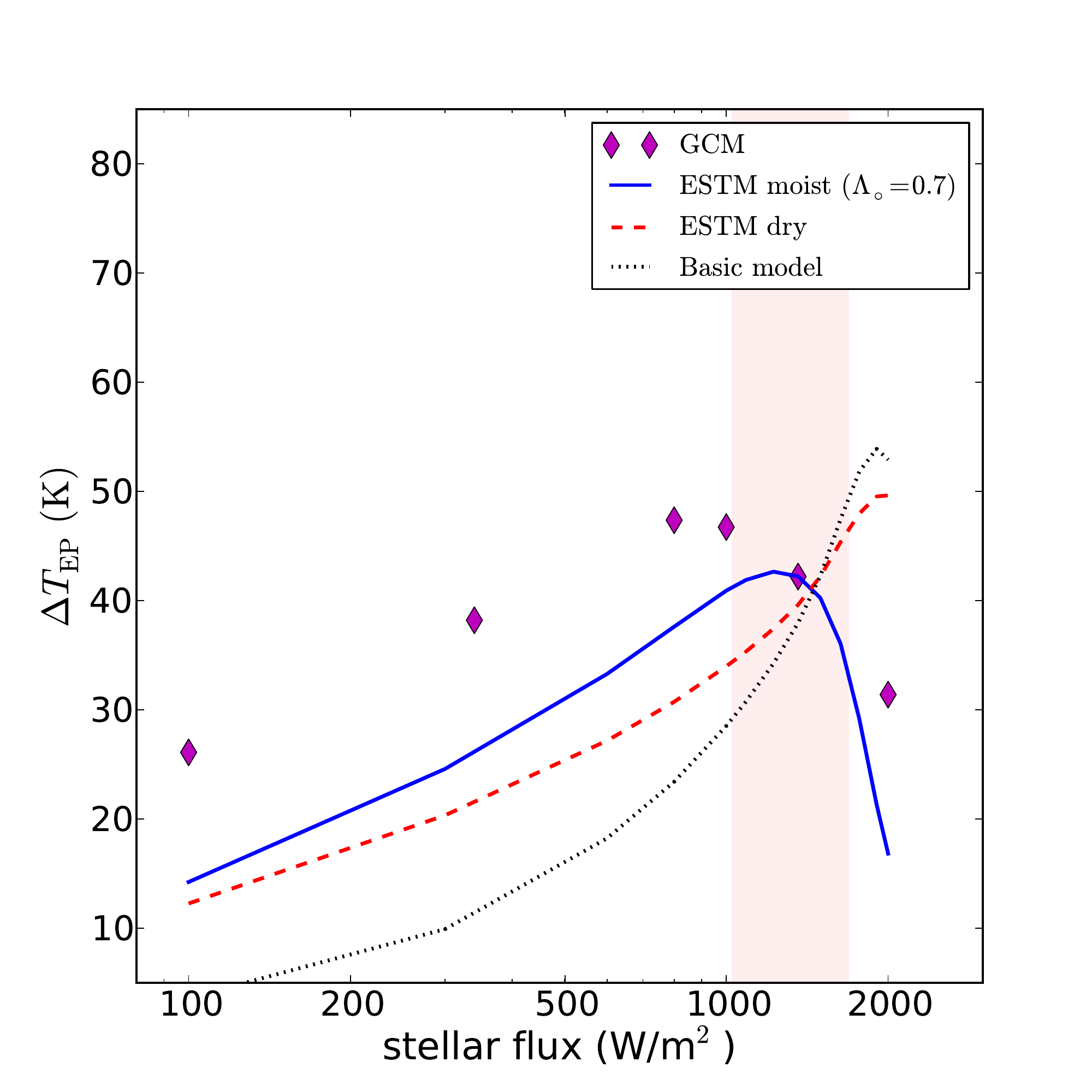}
\includegraphics[width=7.cm]{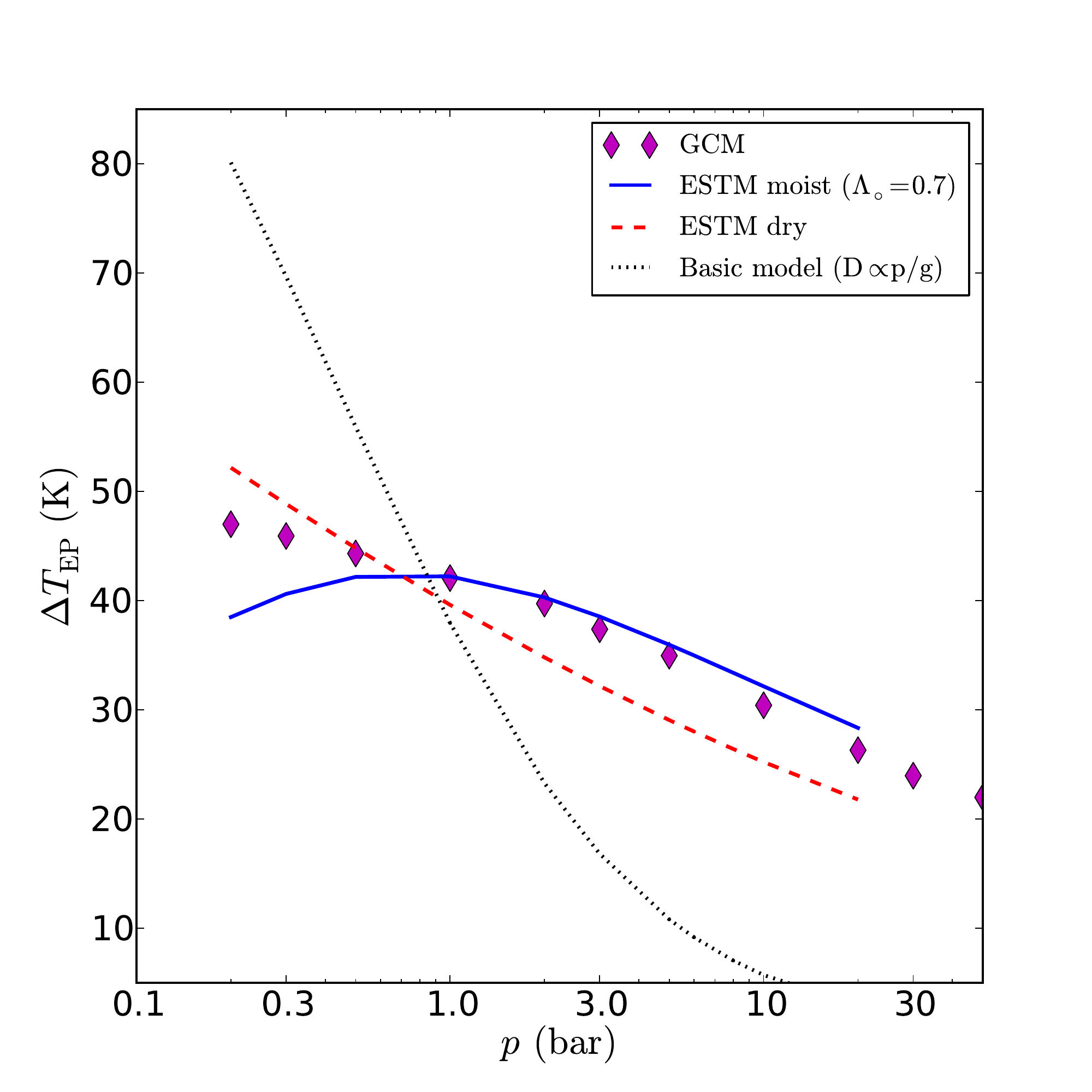}
\includegraphics[width=7.cm]{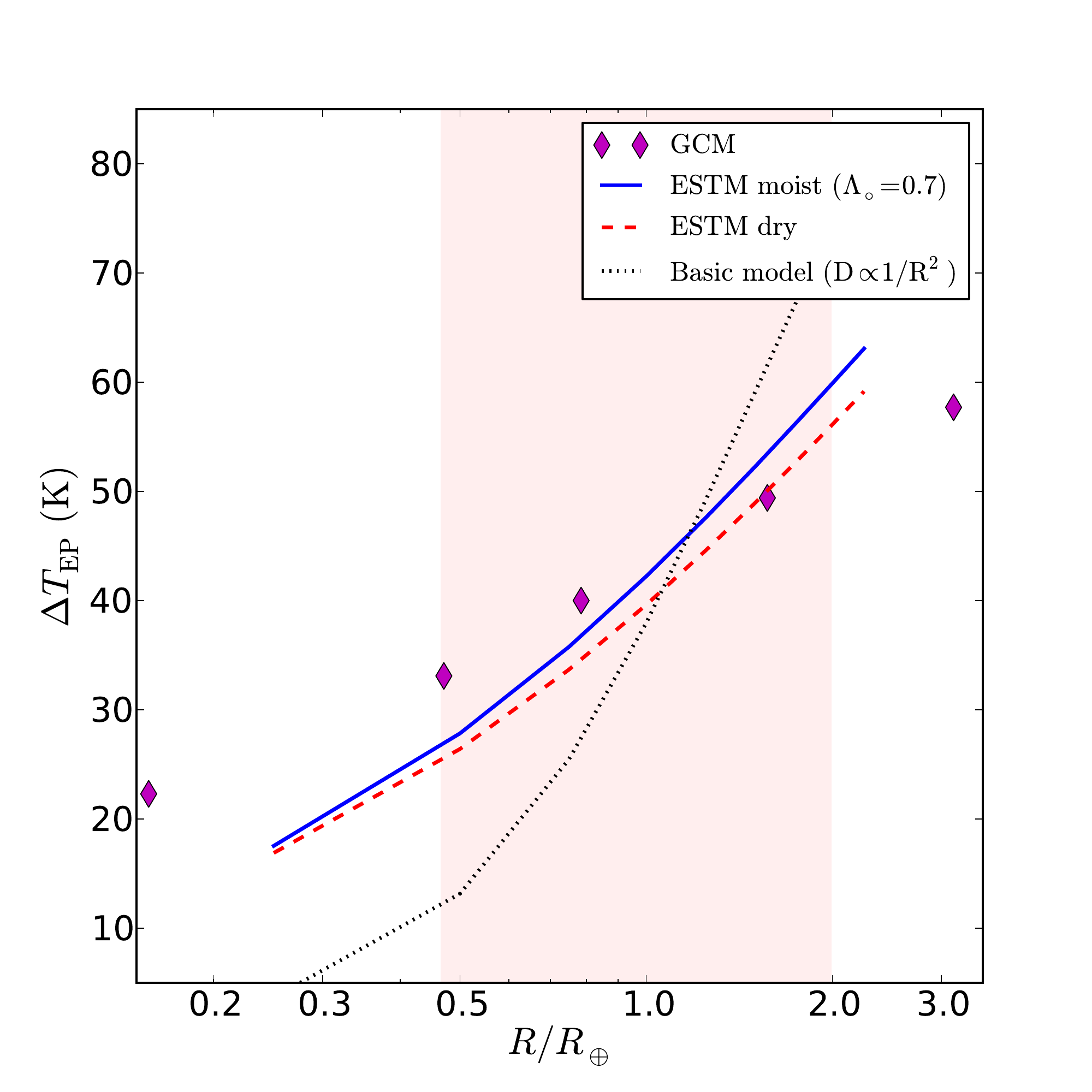}
\caption{
Predicted values of equator-to-pole temperature difference, $\Delta T_\mathrm{EP}$, 
versus planet rotation rate (top left),  stellar flux (top right),
surface atmospheric pressure (bottom left), and planet radius (bottom right)
obtained from different climate simulations of an Earth-like
aquaplanet with axis obliquity $\epsilon=0^\circ$, fixed albedo ($A=0.35$), and no ice. 
Magenta diamonds: 3D GCM simulations \citep{KS14}.
Solid and dashed curves: \model\
(moist and dry transport, respectively).
Dotted curves: ESTM dry model without diabatic forcing term
and with different prescriptions for $D$ 
(see legend in each panel and \S \ref{sectValidation}.)
The shaded area in the top right panel brackets a range of $\pm 25$\% the present-day value
of Earth's insolation. The shaded area in the bottom right panel brackets the range of 
masses 0.1 to 8 M$_\oplus$ for rocky planets with mean density $\rho=\rho_\oplus$. }
\label{figKS14validation}%
\end{center}
\end{figure*}

To perform this test we used 
a  study of the atmospheric dynamics of terrestrial exoplanets performed by \citet{KS14}.
These authors employed a moist atmospheric general circulation model 
to test the response of the atmospheric dynamics over a wide range of planet parameters.
Specifically, they used an idealized aquaplanet with
surface covered by a uniform slab of water 1 m thick;
only vapor-liquid phase change was considered;
the albedo was fixed at 0.35 and insolation was imposed equally between hemispheres;
the remaining parameters were set to mimic an Earth-like climate.
 To validate the \model\ with the results found by \citet{KS14}
we modified the Earth reference model as follows. 
The axis obliquity  was set to zero; the temperature-ice feedback was excluded;
the albedo was fixed  at $A=0.35$;
the fraction of oceans was set to 1, adopting a thermal capacity
corresponding to a mixing layer 1 m thick.   
With this idealized planet model 
we performed
several sets of simulations, varying the planet rotation rate, surface flux,
radius, and surface pressure. To validate the \model\ we analyze 
the mean annual equator-to-pole temperature difference, $\DTep$,
which is critical for a correct estimate of the latitude temperature profile and of the surface habitability.
The results of the tests are shown
in Fig. \ref{figKS14validation}, where we compare the $\DTep$ values predicted by the 3D  model 
(diamonds) with those obtained from the \model\ (solid lines).
We also plot the predictions of
a ``dry'' transport  model (dashed lines) obtained by setting $\Lambda=0$ in Eq. (\ref{eq:DtermB}). 
Finally, for the sake of comparison with previous EBMs, we plot
the results obtained from a ``basic \model'' without moist term ($\Lambda=0$)
and without diabatic forcing term\footnote{
To ignore the diabatic forcing term we set 
$\frac{ \delta T }{T_w }  \left\{ \mathrm{ASR} \right\} =1$
in the scaling law (\ref{eq:SLdry}).}
(dotted lines).
In using this ``basic'' model, we test some alternative scaling laws
for the parameterization of the rotation rate,  surface pressure and radius, as we explain below. 
 
\subsubsection{Rotation rate}

In this experiment all parameters were fixed, with the exception of the
planet rotation rate, $\Omega$,
that was gradually increased from 1/10 to 10 times the Earth value, $\Omega_\oplus$. 
The results of this test are shown in the top-left panel of Fig. \ref{figKS14validation}.
One can see that the \model\ and GCM results show a similar trend, with a good quantitative agreement 
at $\Omega \ga 0.3 \, \Omega_\oplus$, but not at low rotation rate.
This result is expected since our parameterization is  appropriate
to simulate planets with horizontal transport dominated by mid-latitude eddies, 
i.e. planets with relatively high rotation rate 
 (see \S \ref{sectMeridionalTransport}). 
The dotted line in this figure shows the results of the basic model obtained by 
replacing the term $\Omega^{-4/5}$ in Eq. (\ref{eq:SLdry})
with the stronger dependence $\Omega^{-2}$ 
adopted in previous work \citep[e.g.][]{WK97,Vladilo13}. 
One can see that this strong dependence on rotation rate is not supported by the 3D model, 
while the more moderate dependence $D \propto \Omega^{-4/5}$  adopted in the \model\
yields a much better agreement  with the GCM experiments. 

\subsubsection{Stellar flux}

In the top-right panel of Fig. \ref{figKS14validation} we show the results obtained
by varying the insolation from 100 to 2000 W\,m$^{-2}$, 
i.e. from 0.07 to 1.47 times the present-day Earth's insolation.
The behavior predicted by the 3D model is bimodal, with a rise of $\DTep$
up to an insolation of $\simeq 800$ W\,m$^{-2}$ and a decline at higher values of stellar flux.
According to \citet{KS14}
the decline is triggered by the rise of the moist transport efficiency
resulting from the increase of temperature and water vapor content. 
The moist \model\ is able to capture this bimodal behavior,
even though a reasonable agreement with the 3D experiments
is only found in a range of insolation $\pm 25$\% around the present-Earth's value
(shaded area in the figure). 
The dry model (dashed line) is unable to capture the bimodal behavior of $\DTep$ versus flux.
The basic model
is even more discrepant (dotted line).

\subsubsection{Surface pressure or atmospheric columnar mass}

In the bottom-left panel of Fig. \ref{figKS14validation} we show the results obtained
by varying the surface pressure $p$ of the idealized aquaplanet from 0.2 to 20 bar. 
Since the surface gravity is not varied, this experiment is equivalent 
to vary the atmospheric columnar mass\footnote
{
In this experiment, \citet{KS14} adopted a constant optical depth of the atmosphere to focus on horizontal transport,
rather than vertical transfer effects. For the sake of comparison with their experiment, 
we used a constant value of atmospheric columnar mass in the OLR and TOA-albedo calculations,
while changing $p/g$  in the diffusion term.   
},
$p/g$, from 0.2 to 20 times that of the Earth.
Theoretical considerations indicate that  the efficiency of the horizontal transport must increase with increasing $p/g$
[e.g. Eq. (\ref{eq:fluxC})], and 
equator-pole temperature differences should decrease as a result. 
The 3D model predicts a monotonic decrease of $\DTep$, in line with this expectation.
However, the decrease is milder than expected by the basic model with
a simple law $D \propto p/g$ (dotted line).
The models with diabatic forcing (solid and dashed lines) 
predict a more moderate decrease, $D \propto (p/g)^{2/5}$ 
[Eq. (\ref{eq:SLdry})], 
and are in much better agreement with the results of the 3D experiments. 
The agreement of the moist \model\ (solid line) is  remarkable in the range of high columnar mass.  
 
\subsubsection{Planet radius or mass}

In this experiment all planetary parameters, including the columnar mass $p/g$, 
are fixed while changing the planet radius.
Assuming a constant mean density, $\rho=\rho_\oplus$, this is equivalent to scale
the planet mass  as $M \propto R^3$. The results are shown
in the bottom-right panel of Fig. \ref{figKS14validation}, where  
3D models predict an increase of $\DTep$ with increasing radius and mass,
indicating that the horizontal transport becomes less efficient in larger planets.
This is in line with theoretical expectations which suggest that the transport coefficient
decreases with increasing radius, possibly  with a quadratic law  [e.g. Eq. (\ref{eq:DtermA})]. 
However,  the increase of $\DTep$ appears to be too sharp
if we adopt the basic model with $D \propto R^{-2}$ (dotted line).
The models with diabatic forcing (solid and dashed lines)
predict a moderate decrease, $D \propto R^{-6/5}$ [Eq. (\ref{eq:Ddry})],
 in line with the 3D predictions. 
In the range of  masses typical of terrestrial planets
(shaded area in the bottom-right panel of Fig. \ref{figKS14validation})
the predictions of the \model\ are very similar to those obtained by \citet{KS14}.

\begin{figure*}
\begin{center}  
\includegraphics[width=7.5cm]{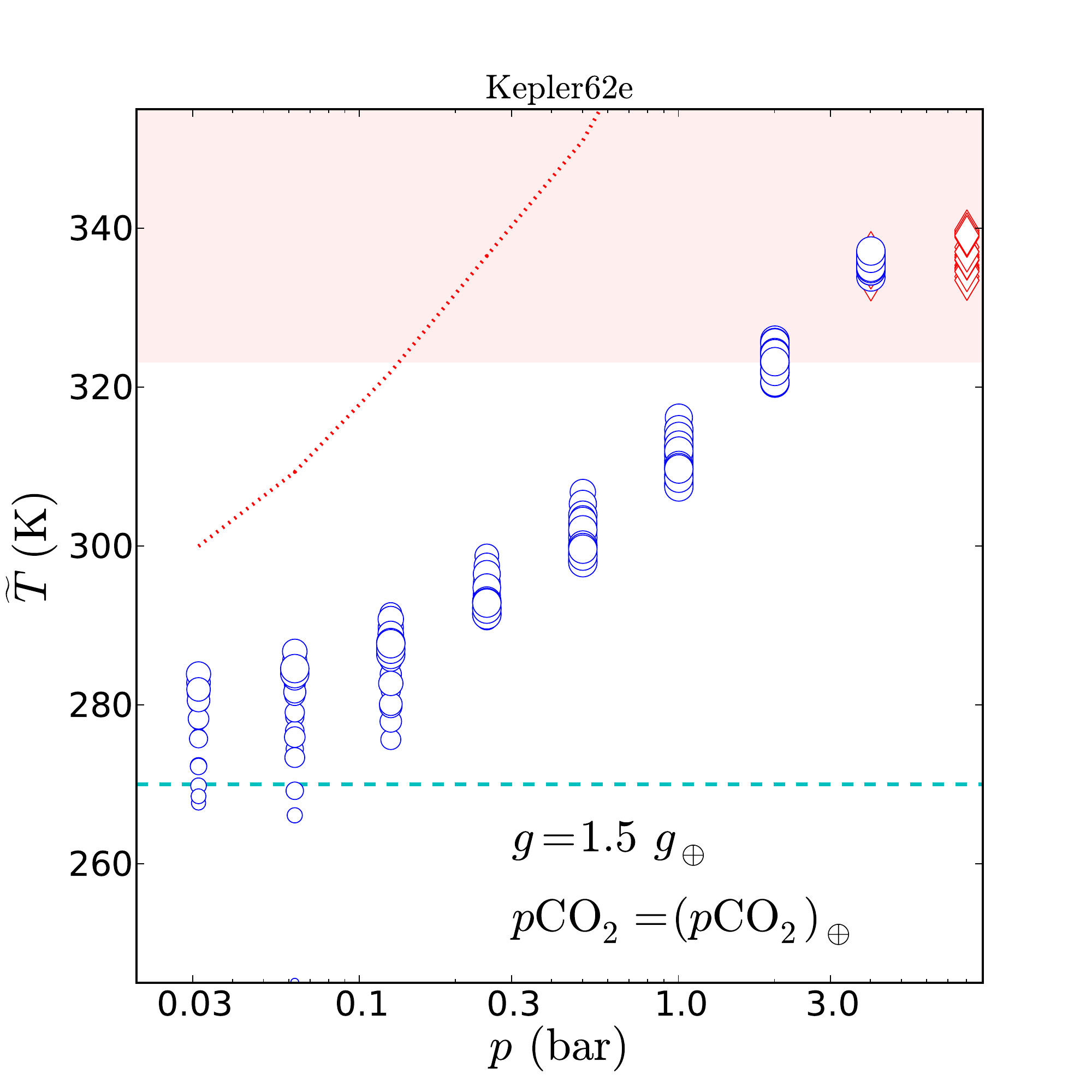} 
\includegraphics[width=7.5cm]{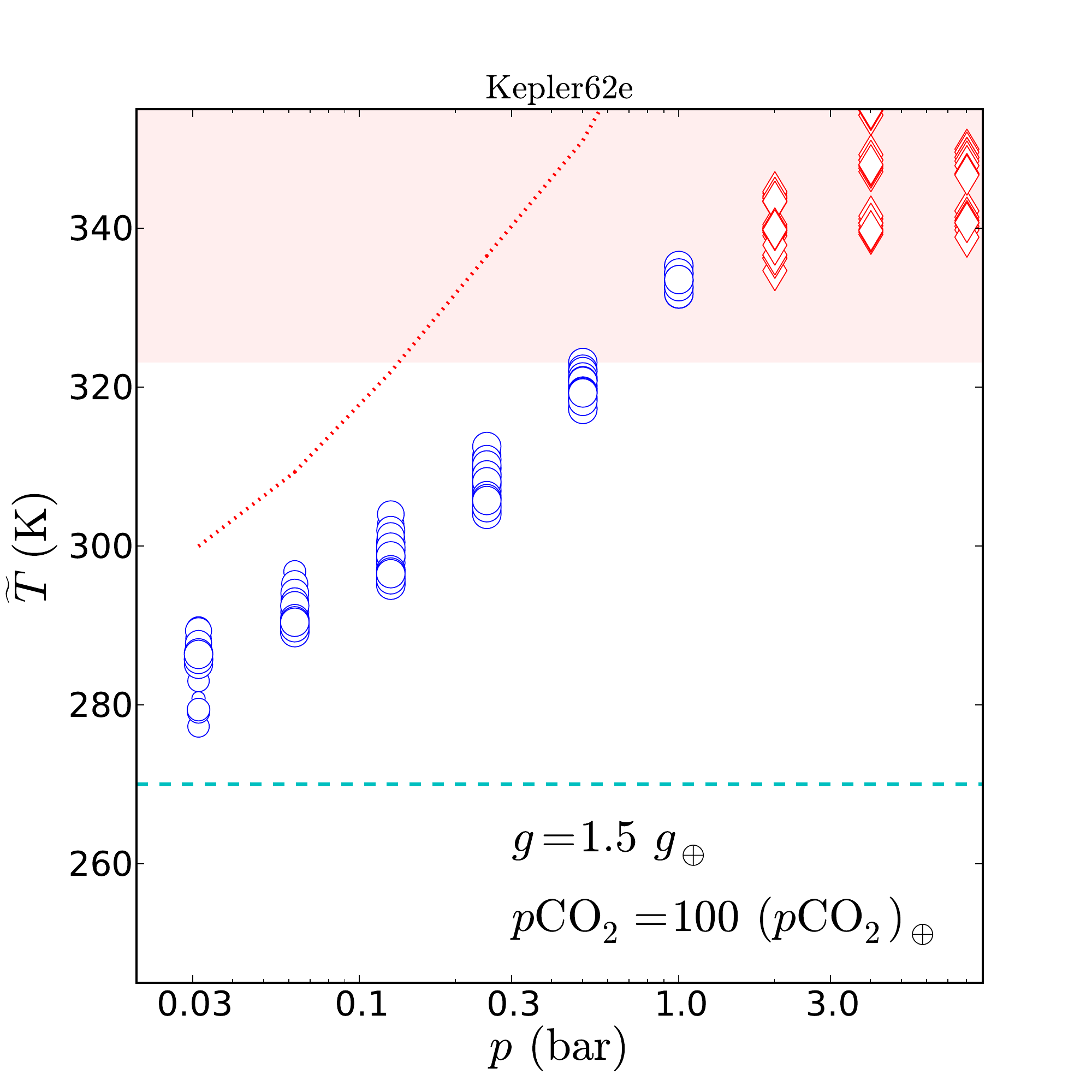} 
\caption{
Influence of  planetary parameters not constrained by observations
on the mean global surface temperature, $\Tm$, of Kepler62-e. 
Symbols: predictions of \model\ simulations  plotted as a function of surface pressure, $p$, 
for all possible combinations of rotation rate,
axis obliquity and ocean fraction described in \S \ref{sectKepler62e}. 
Circles: habitable solutions. Diamonds: solutions with 
water vapor column above the critical limit discussed in \S \ref{sectClimateSimulations}.  
The symbol size scales with the fractional habitability, $h_\text{lw}$.  
Left: Earth atmospheric composition.
Right: Earth-like atmospheric composition with a one hundred-fold increase of $p$CO$_2$.  
Dashed line: equilibrium temperature of Kepler62-e \citep{Borucki13K62}. 
Dotted line: water boiling point as a function of $p$. 
  }
\label{fig1Kepler62e}%
\end{center}
\end{figure*}

\section{Applications}

After the calibration and validation,  we apply the model to explore the dependence
of $T(\varphi,t)$ and the mean global surface temperature\footnote
{
The average in latitude is weighted in area and
the average in time is performed over one orbital period. }, $\Tm$,
on a variety of planet parameters.   
At variance with the  validation tests,  we now consider all the features of the model,  
including the ice-albedo feedback.
In Appendix C we present simulations  of idealized Earth-like planets.
Here we describe a test study of exoplanet habitability.

\subsection{Exoplanets}

The modelization of the surface temperature of exoplanets is severely constrained
by the limited amount of observational data. Typically, one can measure the stellar and orbital parameters and a few planetary quantities,
such as the radius and/or mass. From the stellar and orbital data one can estimate 
the planet insolation and its seasonal evolution. 
From the radius and mass one can estimate the surface gravity which enters
in the parameterization of the atmospheric columnar mass.
Unfortunately, many planet quantities  that are required for the modelization
are currently not observable. These  include the atmospheric composition\footnote
{At present time it is possible to measure the atmospheric composition of selected giant planets,
but not of Earth-size terrestrial planets.}, 
surface pressure, ocean/land distribution, axis obliquity and rotation period.
Taking advantage of the flexibility of the \model, we can perform a fast exploration
of the space of the unknown quantities, treating them as free parameters. 
From the application of this methodology we can
assess the relative importance in terms of climate impact of the planet quantities
that are not measurable. 
In addition, we can constrain the ranges of parameters values that yield habitable solutions.   

%
We show two examples of application of this methodology.
First we consider a specific exoplanet chosen as a test case,
then we introduce a statistical ranking of planetary habitability.   
We adopt an index of habitability, $\hm$, based on the liquid water criterion\footnote{
We define a function $H_\text{lw}(T)=1$ when $T(\varphi,t)$ is inside the
liquid-water temperature range; $H_\text{lw}(T)=0$ outside the range
\citep[][]{SMS08,Vladilo13}. The index $\hm$ is
the global and orbital average value of $H_\text{lw}(T)$.}.

\begin{figure}
\begin{center} 
\includegraphics[width=8.2cm]{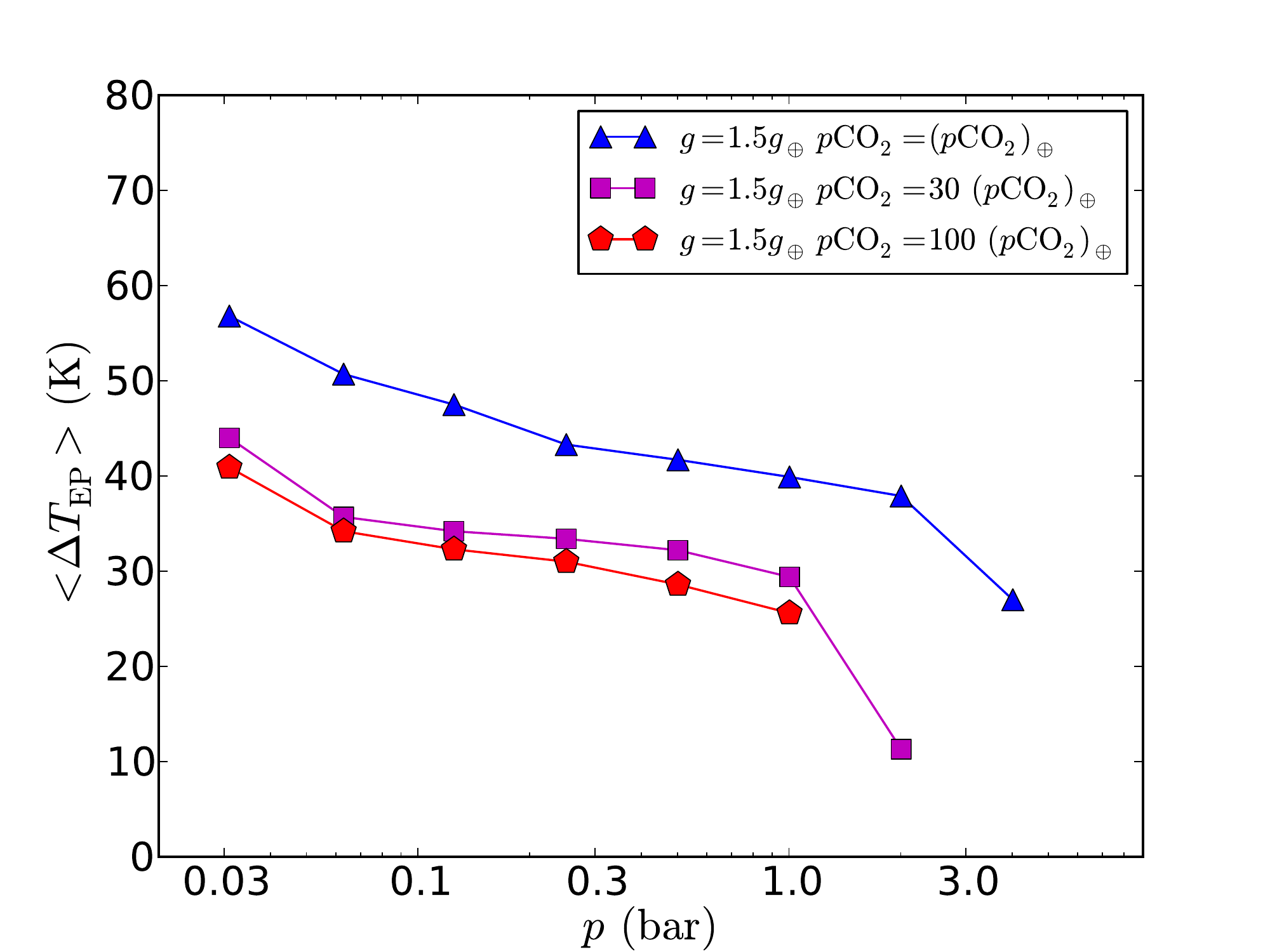} 
\caption{
Average equator-pole temperature difference, $\leftmean \DTep \rightmean$,
obtained from \model\ simulations of Kepler62-e   (\S \ref{sectKepler62e}),
plotted as a function of surface pressure, $p$. 
Each curve is calculated at constant $p$CO$_2$,  as specified in the legend. 
}
\label{fig2Kepler62e}%
\end{center}
\end{figure} 

\subsubsection{Kepler-62e as a test case \label{sectKepler62e}} 

The test-case exoplanet was chosen using three criteria. 
The first is that the planet should be of terrestrial type, i.e. rocky and without an extended atmosphere,
in order to be suitable for the application of the \model.  
We used the radius for a preliminary characterization of the planet,
since evidence is accumulating for the existence of a gradual transition,  correlated with radius, 
between planets of terrestrial type and planets with rocky cores but extended gas envelopes
\citep[e.g.][]{Wu13,Marcy14}.
We restricted our search to planets with $R \la 1.7 R_\oplus$,
the threshold  for terrestrial planets found in a statistical 
study of size, host-star metallicity and orbital period \citep{Buchhave14}.
As a second criterion, we required the orbital semimajor axis to be larger than the tidal lock radius, 
since the \model\  cannot be applied to tidally locked planets.
Finally, given the extreme dependence of habitability on insolation
(see e.g. Fig. \ref{mapsFlux}), 
we selected planets with an insolation within $\pm 50\%$ of the present-day Earth value.
By querying the Exoplanet Orbit Database \citep{Wright11}
at exoplanets.org, 
we found that only Kepler-62\,e \citep{Borucki13K62} satisfies the above criteria.  
The radius, $R=1.61 R_\oplus$, and orbital period, $P=122\,$d, 
suggest that Kepler-62\,e is probably of terrestrial type \citep[see][Fig. 2]{Buchhave14}. 
Its insolation is only 19\% higher than the Earth's value, and
its semimajor axis, $a=0.427$\,AU, is larger than the tidal lock radius\footnote
{
The tidal lock radius was calculated with the expression 
$r_\mathrm{TL}=0.027 (P_\circ t / Q)^{1/6} M_{\star}^{1/3}$
(Kasting 1993), where
$P_\circ$ is the original rotation period, $t$ is the amount of time during which the planet has been
slowed down, $Q^{-1}$ is the specific dissipation function,
and $M_\star$ is the stellar mass; 
we adopted 
$P_\circ=0.5$\,days, $t=10^9$\,yr, and $Q=100$. 
},
$r_\mathrm{tl}=0.31$\,AU.

To run the \model\ simulations of Kepler-62e we adopted at face value 
the radius, semimajor axis, eccentricity and stellar flux provided by the observations \citep{Borucki13K62}. 
Unfortunately, only a loose upper limit ($M<36\,M_\oplus$) is available for the mass, 
so that the surface gravity $g$ 
is poorly constrained at the present time. 
For illustrative purposes, we adopt
$g=1.5\,g_\oplus$ ($M=3.9\,M_\oplus$), 
corresponding to a mean density
$5.1$\,g\,cm$^{-3}$, similar to that of the Earth ($\rho_\oplus=5.5$\,g\,cm$^{-3}$).
As far as the atmosphere is concerned, we vary 
the surface pressure in the range $p \in (0.03,8)$\,bar
 and the CO$_2$ partial pressure in the range $p$CO$_2$/($p$CO$_2$)$_\oplus \in (1,100)$.
We adopt 3 representative values of rotation rate, $\Omega/\Omega_\oplus \in (0.5,1,2)$,
axis obliquity, $\epsilon \in (0^\circ,22.5^\circ,45^\circ)$, 
and ocean coverage, $f_o \in (0.5,0.75,1.0)$. 
For the remaining parameters we adopt the Earth's reference values. 
For each value of 
CO$_2$ partial pressure
we run simulations covering all possible combinations of background pressure, rotation rate,
axis obliquity and ocean coverage listed above. 
Part of the results of these simulations are shown in Figs. \ref{fig1Kepler62e} and \ref{fig2Kepler62e}.

In Fig. \ref{fig1Kepler62e} we plot the mean global temperature, $\Tm$, obtained for 
two different values of $p$CO$_2$, specified in the legends.
At each value of $p$, we show the values of $\Tm$ obtained from all possible combinations
of $\Omega$, $\epsilon$, and $f_o$.  
The typical scatter of $\Tm$ due a random combination of these 3 parameters
is $\simeq$ 10-20\,K. On top of this scatter,
the most remarkable feature is a positive trend of $\Tm$ versus $p$ extending over an interval of $\approx 60$\,K. Within the limits of application of the model,
these results indicate that an uncertainty on $p$
within the range expected for terrestrial planets\footnote{
Mars and Venus have a surface pressure of $\simeq 0.006$\,bar and $\simeq 90$\,bar, respectively.}
has stronger effects on $\Tm$ than uncertainties of rotation rate,
axis obliquity and ocean coverage.
In fact, variations of $p$  have strong effects both on $T(\varphi,t)$ and $\Tm$
because they are equivalent to
variations of atmospheric columnar mass, $p/g$, which affect both the latitudinal transport
(i.e. the surface temperature distribution), 
and the radiative transfer (i.e. the global energy budget). 

The results shown in Fig. \ref{fig1Kepler62e} constrain the 
interval of $p$ that allows Kepler-62e to be habitable. 
At high $p$, the habitability is limited by the rise of $\Tm$, which eventually
leads to a runaway greenhouse instability.
The red diamonds in Fig. \ref{fig1Kepler62e} indicate cases
where the water vapor column exceeds the critical value that we tentatively
adopt as a limit for the onset of such instability (\S \ref{sectClimateSimulations}). 
At low $p$, two factors combine to limit the habitability.
One is the onset of large temperature excursions and the other the 
decrease of the water boiling point  
(dotted lines in Fig. \ref{fig1Kepler62e}).  
As a result, the fraction of planet surface outside the liquid water temperature 
becomes larger at low $p$.
To evidentiate this effect, we have scaled the size of the symbols 
in Fig. \ref{fig1Kepler62e} according to the value of $\hm$. 
One can see that $\hm$ tends to become smaller at low $p$, especially when $\Tm$ approches the 
temperature regime
where the ice-albedo feedback becomes important. 
In some cases, not shown in the figure, the planet undergoes a complete snowball transition   
and the habitability becomes zero. 
The effect of temperature excursions is shown in   Fig. \ref{fig2Kepler62e}, where 
we plot as a function of $p$ the average value of 
$\DTep$ obtained from all possible combinations
of $\Omega$, $\epsilon$, and $f_o$. 
One can see that the temperature excursions become large with decreasing level of CO$_2$;
this happens because at low CO$_2$ the temperature is sufficiently low for the development of the ice-albedo feedback, and because the lowered IR optical depth of the atmosphere is less effective in reducing the effect of the geometrically-induced meridional insolation variation at the surface.

Fig. \ref{fig1Kepler62e} shows that the equilibrium temperature of Kepler-62e,  
$T_\text{eq}=270 \pm 15$\,K \citep[][dashed line]{Borucki13K62}, 
lies at the lower end of the predicted  $\Tm$ values. The difference $\Tm-T_\mathrm{eq}$
increases with  atmospheric columnar mass because the estimate of $T_\mathrm{eq}$
does not consider the greenhouse effect. 

\begin{figure}
\begin{center}  
\includegraphics[width=8.2cm]{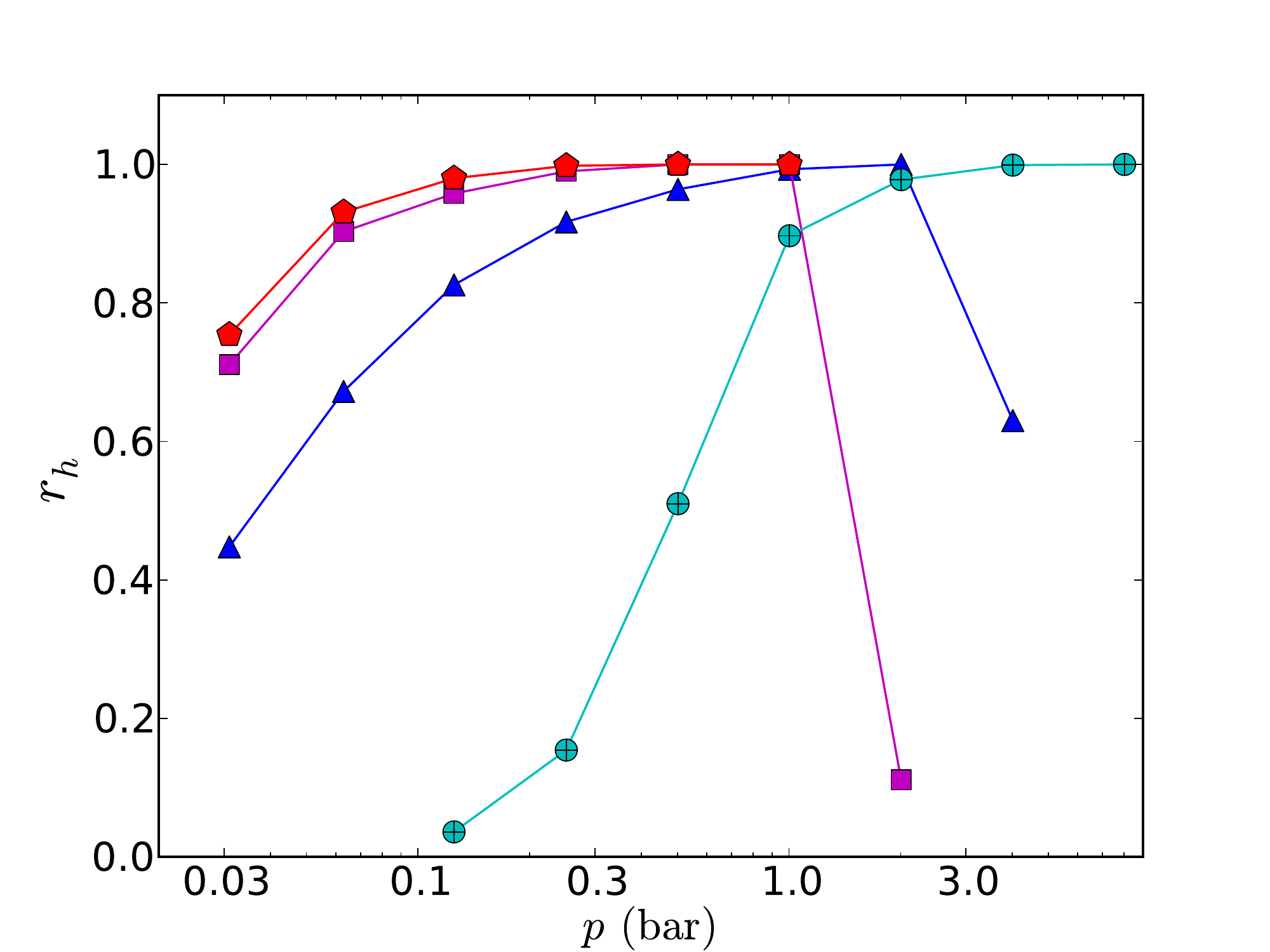}    %
\caption{ 
Ranking index of habitability, $r_h$,
obtained from \model\ simulations of Kepler-62e and an Earth twin (\S \ref{sectRanking})
plotted as a function of surface pressure, $p$.
Symbols for Kepler-62e as in Fig. \ref{fig2Kepler62e}.  
Crossed circles: Earth twin.
}
\label{figRanking}%
\end{center}
\end{figure}

\subsubsection{Statistical ranking of planetary habitability \label{sectRanking}}

By performing a large number of  simulations for
a wide combination of  parameters we can quantify the habitability in a statistical way. 
To illustrate this possibility with an example we consider again the test case of Kepler-62e.
We perform a statistical analysis of the results obtained from  
all combinations of $\Omega$, $\epsilon$, and $f_o$  values adopted at a given $p$.
We tag as ``non habitable''  the cases with a  snowball transition and those with a critical value of water vapor. 
For each set of parameters 
\{$\Omega$, $\epsilon$, $f_o$\} 
 we count the number of cases that are found to be habitable, $n_h$,
over the total number of simulations, $n_t$.
From this we calculate the fraction $\psi_h=n_h/n_t$,
which represents the probability for the planet to be habitable if 
the adopted parameter values are equally plausible a priori.
%
We then call $\leftmean \hm \rightmean =(1/n_h)\sum_{i=1}^{n_h} (\hm)_i$
the mean surface habitability of the $n_h$  sets
that yield a habitable solution. 
%
%
At this point we 
define a ``ranking index of habitability'' 
\begin{equation}
r_h \equiv \psi_h \times \leftmean \hm \rightmean = {1 \over n_t}\sum_{i=1}^{n_h}\, (\hm)_i  ~~.
\label{HabitabilityIndex}
\end{equation}
This index 
takes into account at the same time the probability that the planet is habitable
and the average fraction of habitable surface. As an example,
in Fig. \ref{figRanking} we plot $r_h$ versus $p$ for
the different values of $p$CO$_2$ 
considered in our simulations
of Kepler-62e.
One can see that $r_h \simeq 1$ only in a limited range of surface pressure.
At very low pressure, the index becomes lower  because
the fraction of habitable surface decreases  
and because $\psi_h$ drops below unity when a snowball transition is encountered.
At high pressure $\psi_h$ drops when the water vapor limit is encountered.
As an example of application, we can constrain the interval of $p$ suitable
for the habitability of Kepler62-e. For $p$CO$_2$= ($p$CO$_2$)$_\oplus$ (triangles)
the requirement of habitability yields the limit  $p \la 3$\,bar.
As $p$CO$_2$  increases (squares and pentagons), the upper limit becomes more stringent ($p \la 1$\,bar),
but the planet has a higher probability of being habitable at relatively low pressure. 

Clearly, the index $r_h$ does not have an absolute meaning since its value depends on the choice of 
the set of parameters.  However, by choosing a common set,
the index $r_h$ can be used to rank the relative habitability of different planets.
As an example, in Fig. \ref{figRanking} we plot $r_h$ versus $p$ obtained for an Earth twin\footnote{ 
Here an Earth twin is a planet with properties equal to those of the Earth
(including insolation, radius, gravity and atmospheric composition),
but with unknown values of $p$, $\Omega$, $\epsilon$ and $f_o$, which are treated as free parameters. 
}
with the same sets of parameters  \{$\Omega$, $\epsilon$, $f_o$\}
adopted for Kepler-62e. One can see that at $p \ga 1$\,bar
the Earth twin (crossed circles) is more habitable than Kepler-62e for the adopted set of parameters,
while at $p \la 1$\,bar Kepler-62 is more habitable than the Earth twin. 


%
%

\section{Summary and conclusions}

We have assembled the ESTM set of climate tools (Figure 1) to model
the latitudinal and seasonal variation of the surface temperature, $T(\varphi,t)$, on Earth-like planets. 
The motivation for building 
the ESTM is twofold.
From the general point of view of exoplanet research, Earth-size planets are expected to be rather common, 
but  difficult to characterize with experimental methods.
From the astrobiological point of view, Earth-like planets are excellent candidates in the
quest for habitable environments.
A fast simulation of $T(\varphi,t)$ enables us to characterize
the surface properties of these planets by sampling the wide parameter space
representative of the physical quantities not measured by observations.
The detailed modelization of the surface temperature is essential to estimate the habitability of these planets using the liquid water criterion or a proper set of thermal limits of life \citep[e.g.][]{Clarke14}. 
 
The \model\ consists of an upgraded type of 
EBM featuring a
multi-parameter description of the  physical quantities that dominate 
the vertical and horizontal energy transport (Fig.\,\ref{figScheme1}). 
The functional dependence of the physical quantities is derived using 
single-column atmospheric calculations and algorithms tested with 3D climate experiments.  
Special attention has been dedicated to improve (\S \ref{sectMeridionalTransport})
 and validate  (\S \ref{sectValidation})
the description of the meridional transport, a weak point of classic  EBMs. 
The functional dependence of the meridional transport 
on atmospheric columnar mass 
and rotation rate 
is significantly milder [see Eq. (\ref{eq:SLdry})] than the one adopted in previous EBMs. 
The reference Earth model obtained from the calibration process
is able to reproduce with accuracy the average surface temperature properties of the Earth
and to capture the main features of the Earth albedo and  meridional energy transport 
(Figs. \ref{annualLatProfiles} and \ref{TempLatMaps}).
Once calibrated, 
the \model\ is able 
to reproduce
the mean equator-pole temperature difference, $\DTep$, 
 predicted by 3D aquaplanet models (Fig. \ref{figKS14validation}). 

The \model\ simulations provide a fast ``snapshot'' of $T(\varphi,t)$ and temperature-based indices of habitability 
for any set of input parameters that yields a stationary solution.
The planet parameters that can be changed include 
radius, $R$, surface pressure, $p$, gravitational acceleration, $g$, 
rotation rate, $\Omega$, axis tilt, $\epsilon$, ocean/land coverage and
partial pressure of non-condensable greenhouse gases. 
The approximate limits of validity of the present version of the \model\ can be summarized as follows:
$0.5 \la R/R_\oplus \la 2$,
$p \la 10$\,bar,
$0.5 \la \Omega/\Omega_\oplus \la 5$,
$\epsilon \la 45^\circ$; $p$CO$_2$ and $p$CH$4$ can be changed, but should 
remain in trace abundances with respect to an Earth-like atmospheric composition.  
The requirement of a relatively high rotation rate is 
inherent to the simplified treatment of the horizontal transport. 
However, the ESTM can be applied to explore the early habitability 
of  slow rotating planets 
that had an initial fast rotation.  

We  have performed \model\ simulations of idealized Earth-like planets 
to evaluate the impact on the planet temperature and habitability
resulting from variations of 
rotation rate, insolation, atmospheric columnar mass, radius, axis obliquity, ocean/land distribution,
and  long-wavelength cloud forcing
(Figs. \ref{mapsRotation} - \ref{mapsOLRclouds}).
Most of these quantities can easily induce $\sim 30$-$40$\% changes of the mean annual habitability
for parameter variations well within the range expected for terrestrial planets. 
Variations of insolations within $\pm 10\%$ of the Earth value can impact the
habitability up to $100\%$. The land/ocean distribution  
mainly affects the seasonal habitability, rather than its mean annual value.
The impact of rotation rate is weaker than predicted by classic EBMs, without evidence
of a snowball transition at $\Omega/\Omega_\oplus \ga 3$.
A general result of these numerical experiments is that 
the ice-albedo feedback amplifies changes of $T(\varphi,t)$  
resulting from variations of planet parameters.   
  
We have tested the capability of the \model\ to explore
the habitability of a specific exoplanet in presence of a limited amount of observational data.  
For the exoplanet chosen for this test, Kepler-62e, 
we used the  
stellar flux, orbital parameters and planet radius provided by the observations \citep{Borucki13K62},
together with a surface gravity $g=1.5\,g_\oplus$ adopted for illustrative purposes. 
We treated the surface pressure, $p$CO$_2$, rotation rate, axis tilt and ocean coverage as free parameters.
We find that $\Tm$  increases from $\approx 280$\,K to $\approx 340$\,K
when the surface pressure increases between $p\simeq 0.03$\,bar and $3$\,bar;
this trend dominates the  scatter of $\simeq 10$-20\,K
resulting  from variations of rotation rate, axis tilt and ocean coverage at each value of $p$. 
We also find that the surface pressure of Kepler-62e
should lie above $p \approx 0.05$\,bar to avoid the presence of a significant ice cover and below $\approx 2$\,bar
to avoid the onset of a runaway greenhouse instability;
this upper limit is confirmed for different values of $p$CO$_2$ and surface gravity. 
These results demonstrate the \model\ capability  
to evaluate the climate impact of unknown planet quantities 
and to constrain the range of values that yield habitable solutions.   
The test case of Kepler-62e also shows that the equilibrium temperature commonly
published in exoplanet studies represents a sort of lower limit to the mean global temperature of more realistic models.   

We have shown that the flexibility of the \model\ makes it possible 
to quantify the habitability in a statistical way.
As an example, we have introduced a ranking index of habitability,
$r_h$, 
that can be used to compare the overall habitability of different planets
for a given set of reference parameters   (\S \ref{sectRanking}).
For instance, we find that at $p \la 1$\,bar Kepler-62e is more habitable than an Earth twin 
for the combination of rotation rates, axis tilts and ocean fractions considered in our test,
whereas the comparison favours the Earth twin at $p \ga 1$\,bar.
The index $r_h$ can be applied 
to select the best potential cases of habitable exoplanets for follow-up searches of biomarkers.

The results of this work indicate the level of accuracy required to estimate   
the surface habitability of terrestrial planets.
The quality of exoplanet orbital data and host star fluxes should be improved
to measure the insolation with  an accuracy of $\approx 1\%$.
In spite of the difficulty of characterizing terrestrial atmospheres \citep[e.g.,][]{Misra13},
an effort should be made to constrain the  atmospheric pressure, possibly within a factor of two.

\acknowledgements

We thank Yohai Kaspi for providing results in advance of publication. 
The comments and suggestions received 
from an anonymous referee have significantly improved the presentation of this work.
This research has made use of the Exoplanet Orbit Database and the Exoplanet Data Explorer at exoplanets.org.
We thank Rodrigo Caballero and Raymond Pierrehumbert for suggestions concerning the use of their climate utilities.

\appendix

\section{Running the  simulations\label{sectClimateSimulations}} 

The  \model\ simulations consist in a search for a stationary 
solution $T(\varphi,t)$ of Eq. (\ref{diffusionEq}). 
We solve the spatial derivates with the Euler method, with the boundary condition
that the flux of horizontal heat into the pole vanishes.
The temporal derivates are  solved with the Runge-Kutta method.
The solution is searched for by iterations,
starting with an assigned initial value of temperature equal in each zone,   
$T(\varphi,t_\circ) \equiv T_\mathrm{start}$. 
Every 10 orbits we calculate the mean global orbital
temperature, $\widetilde{T}$.  
The simulation is stopped when $\widetilde{T}$ converges within a prefixed accuracy. 
In practice, we calculate the increment $\delta \widetilde{T}$ every  
10 orbits and stop the simulation when  $| \delta \widetilde{T}| < 0.01$\,K.  
In most cases the convergence is achieved in fewer than 100 orbits.
After checking that the simulations converge to the same solution starting from
widely different values of $T_\mathrm{start}> 273$\,K,
we adopted $T_\mathrm{start}=275$\,K.  
The choice of this ``cold start'' allows us to study  atmospheres with
very low pressure, where the boiling point is just a few kelvins above the freezing point; 
in these cases, the adoption of a higher $T_\mathrm{start}$ 
would force most of the planet surface to evaporate at the very start of the simulation.
The adoption of a lower  $T_\mathrm{start}$, on the other hand, would
trigger artificial episodes of glaciation. 

In addition to the regular exit based on the convergence criterion, we interrupt the simulation  
in presence of water vapor effects that may lead to a condition of non-habitability.  
Specifically, two critical conditions are monitored in the course of the simulation.
The first takes place when $T(\varphi,t)$ exceeds the water boiling point, $T_\mathrm{b}$.
The second, when the columnar mass of water vapour\footnote{ 
The water columnar mass is  
${\cal M}_w \simeq (\mu_w/ \mu ) (q \, p^*_v(T)/g)$,
where $\mu_w$ and $\mu$ are the molecular weights of water and  air;
$q$ is the relative humidity and
$p^*_v(T)$ the saturated partial pressure of water vapor \citep[e.g.,][]{Pierrehumbert10}.
}
exceeds 1/10 of the total atmospheric columnar mass (see next
paragraph).
In the first case, the {\em long term} habitability might be compromised due to evaporation of the surface water.
The second condition might lead to the onset of a runaway greenhouse instability \citep{Hart78,Kasting88}
with a complete loss of water from the planet surface. 
The \model\ does not track variations of relative humidity
and is not suited to describe these two cases.  
By interrupting the simulation when one of these two conditions is met,
we limit the range of application of the simulations to cases 
with a modest content of water vapor that can be safely treated by the model.  

The limit of water vapor columnar mass that we adopt is inspired
by the results of a study of the Earth climate variation induced by a  rise of insolation
\citep{Leconte13}. %
When the insolation attains 1.1 times the present Earth value, 
the 3D moist model of \citet{Leconte13} predicts  
 an energy unbalance that would  lead  to a runaway greenhouse instability.
The  mixing ratio of water vapor over moist air predicted 
at the critical value of insolation is $\simeq 0.1$
\citep[][Fig. 3b]{Leconte13}.  
The limit of water columnar mass that we adopt is based on this mixing ratio.   


For the simulations presented in this work, we adopt a grid of $N=54$ latitude zones.
The orbital period is sampled at $N_s=48$ instants of time to investigate the seasonal evolution of the surface quantities
of interest (e.g. temperature, albedo, ice coverage).
With this set-up, the simulation of the Earth model presented below
attains a stationary solution after 50 orbital periods, with a CPU time of $\sim 70$\,s 
on a 2.3 GHz processor. 
This extremely high computational efficiency is the key
to perform the large number of simulations required
to cover the broad parameter space of exoplanets.

\begin{figure*}[ht]    
\begin{center} 
\includegraphics[width=7.5cm]{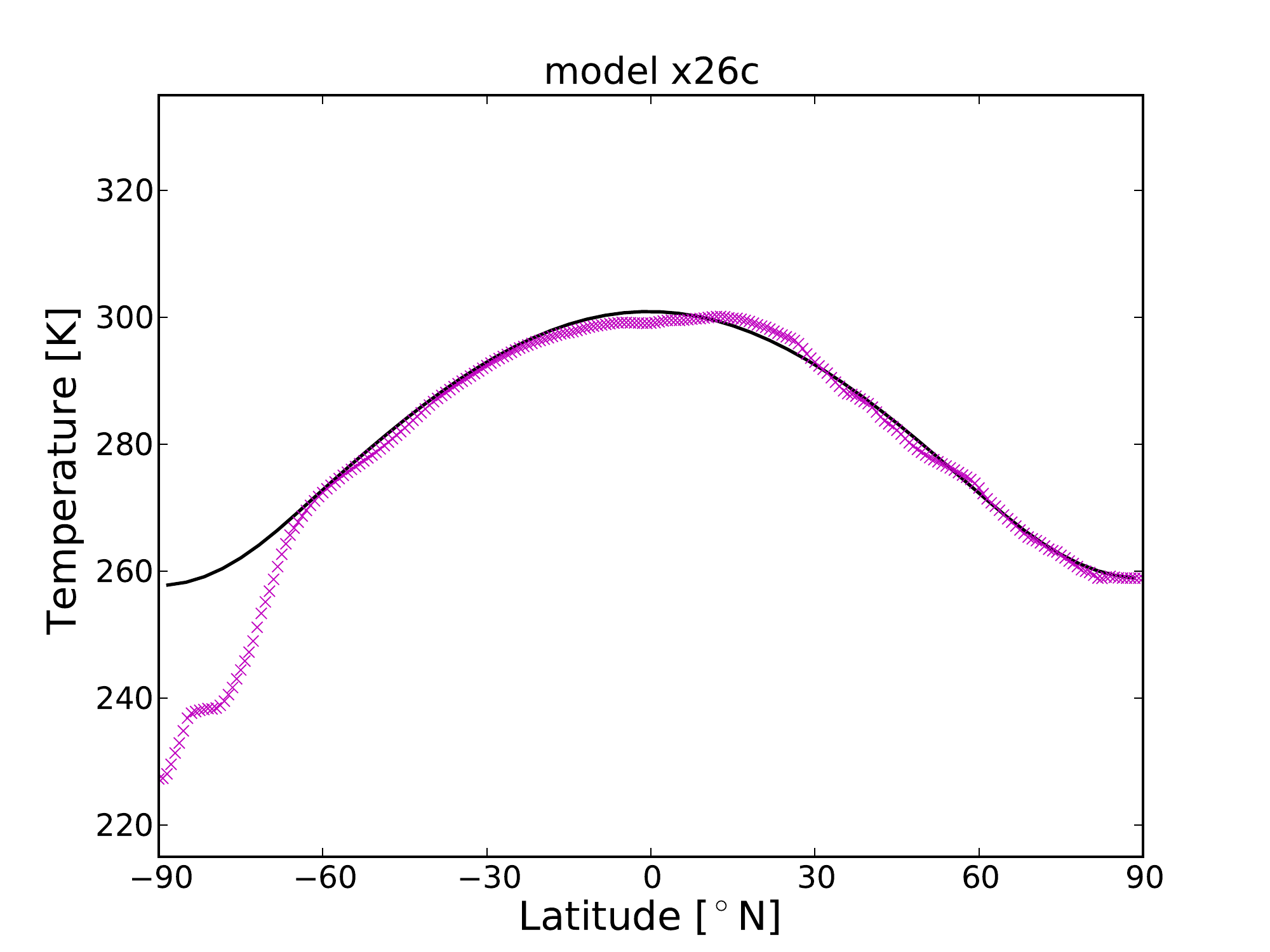}
\includegraphics[width=7.5cm]{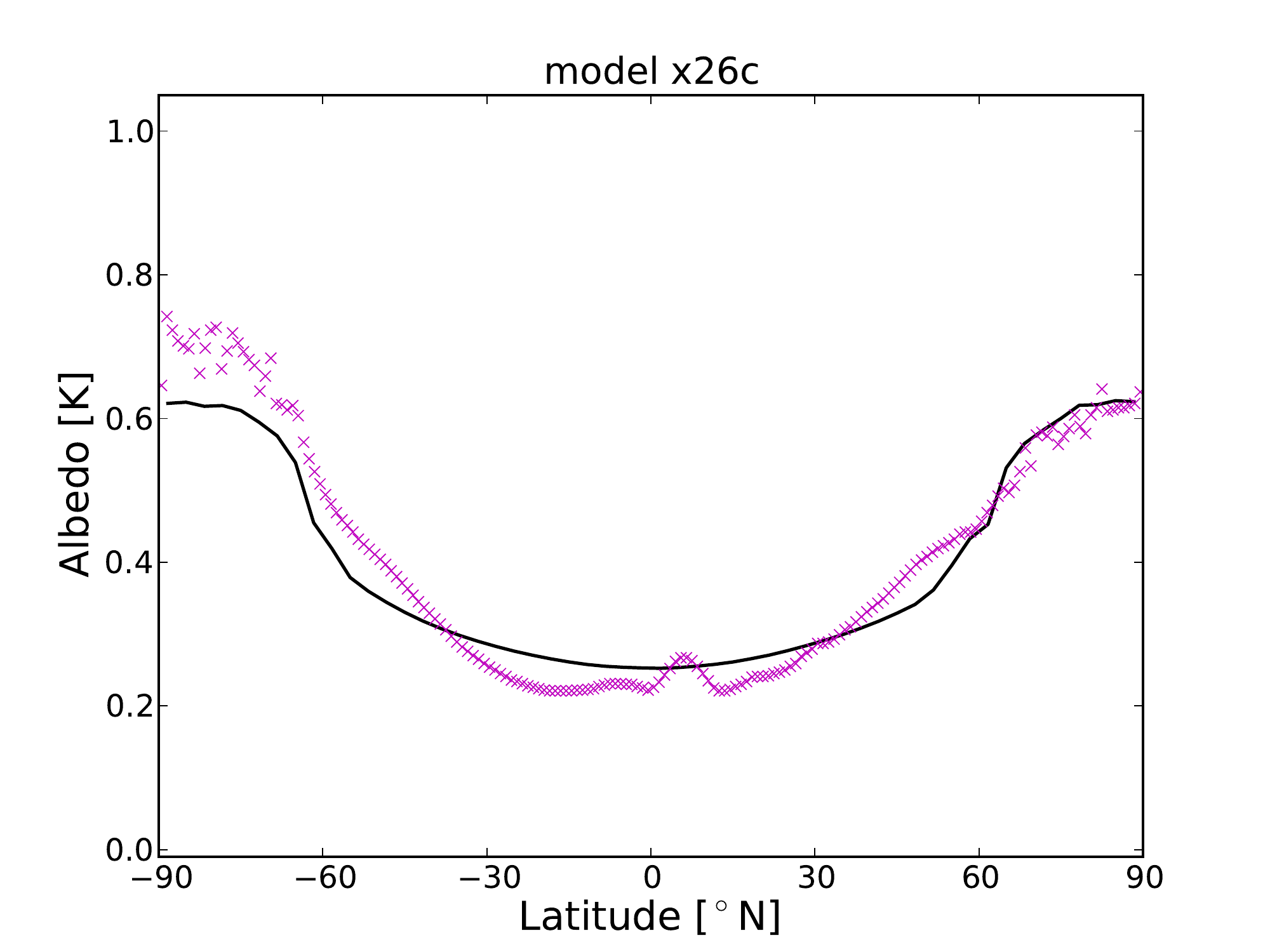} 
\includegraphics[width=7.5cm]{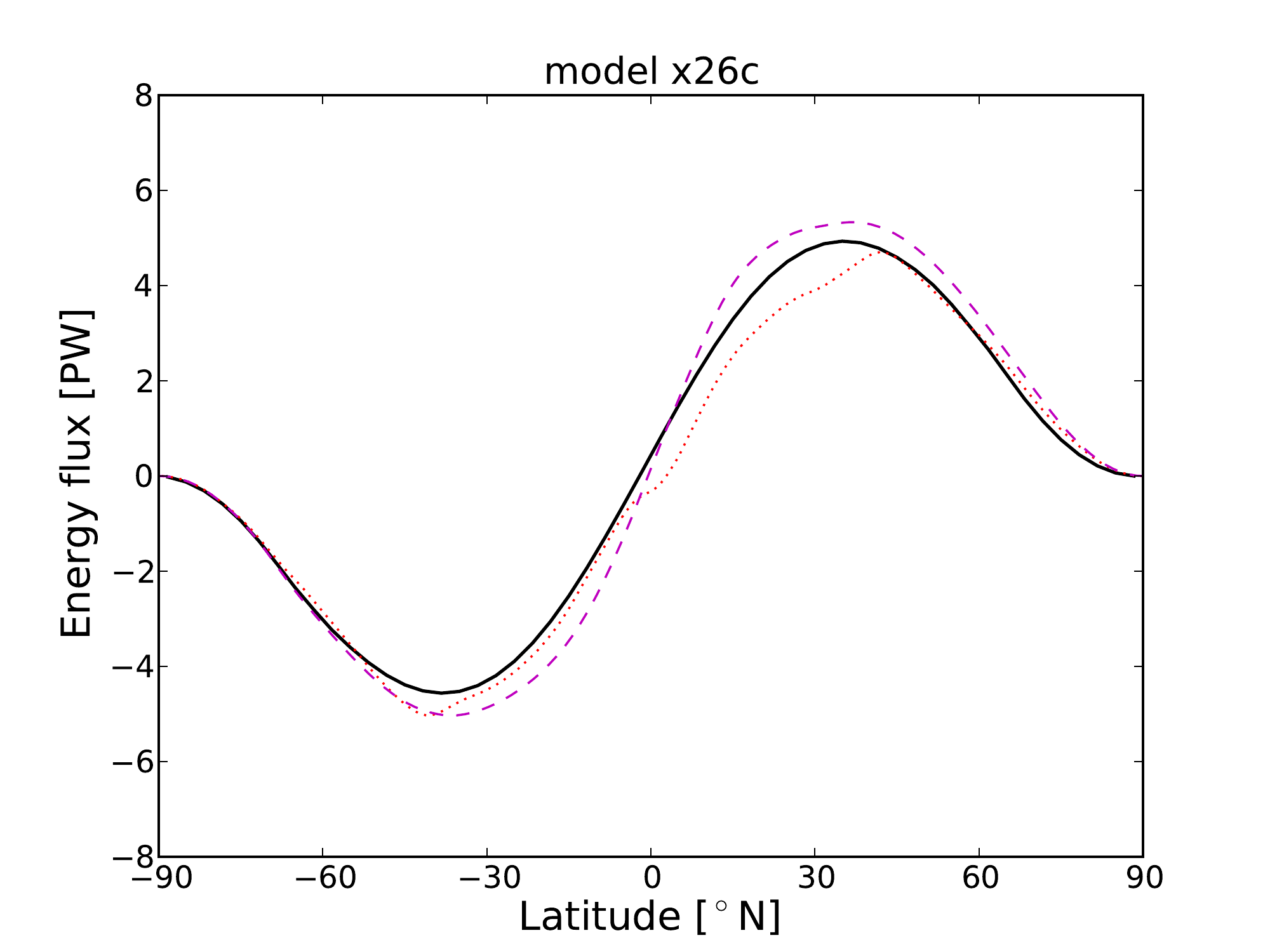} 
\caption{
Mean annual latitude profiles of surface temperature,  albedo
and meridional energy flux predicted by the reference Earth  model (solid line).
Top panel: the temperature profile is compared with  ERA Interim 2m temperatures \citep{Dee11}
averaged in the period 2001-2013  (crosses).
Middle panel: the albedo profile is compared with CERES short-wavelength albedo \citep{Loeb05,Loeb07}
averaged in the same period (crosses).
Bottom panel: the meridional energy flux profile is compared  
with the total (dashed line) and atmospheric (dotted line)
profiles obtained from the EC-Earth model \citep{Hazeleger10}.    }
\label{annualLatProfiles}%
\end{center}
\end{figure*}

\begin{figure*}[ht]    
\begin{center} 
\includegraphics[width=7.5cm]{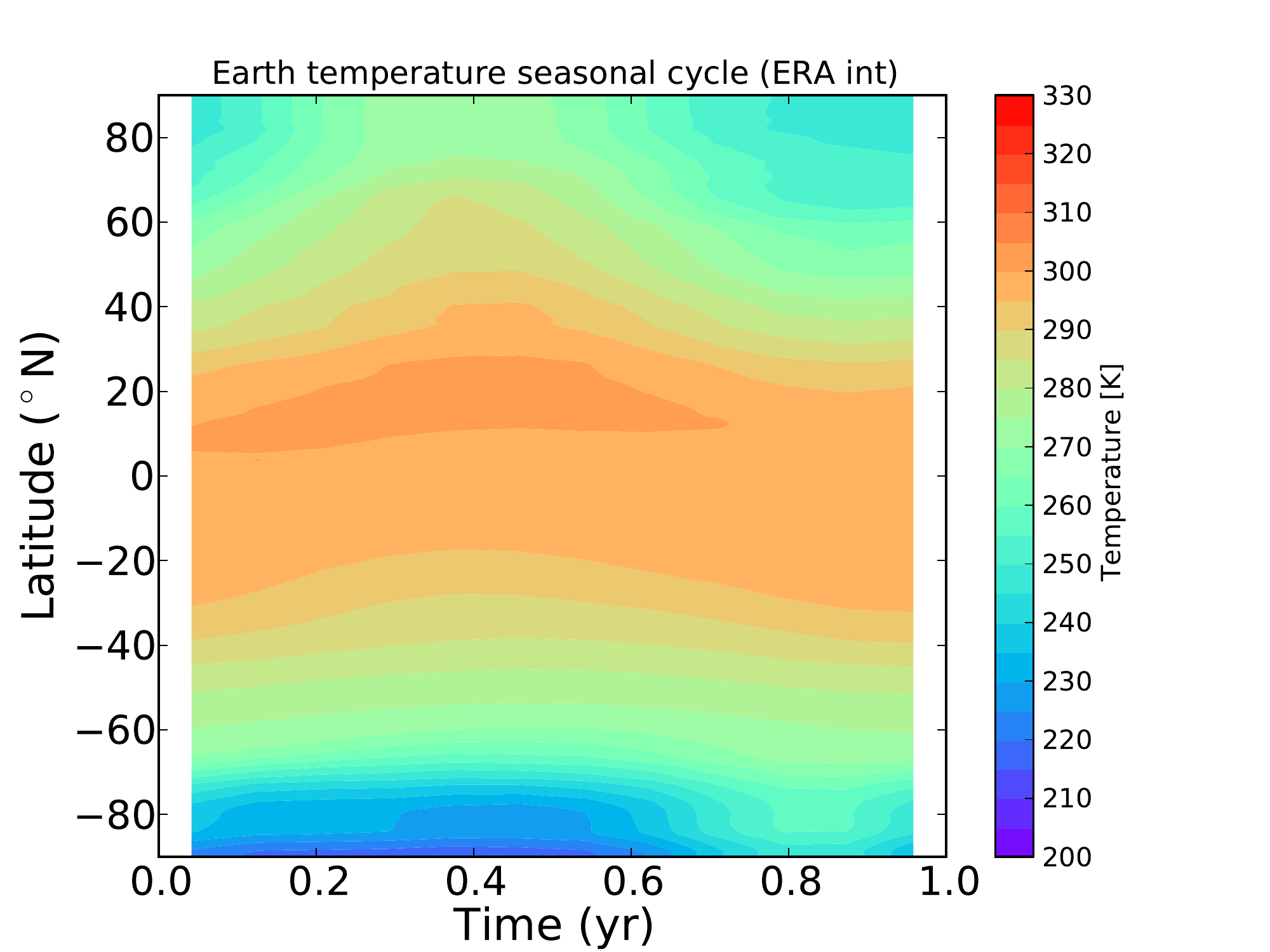}
\includegraphics[width=7.5cm]{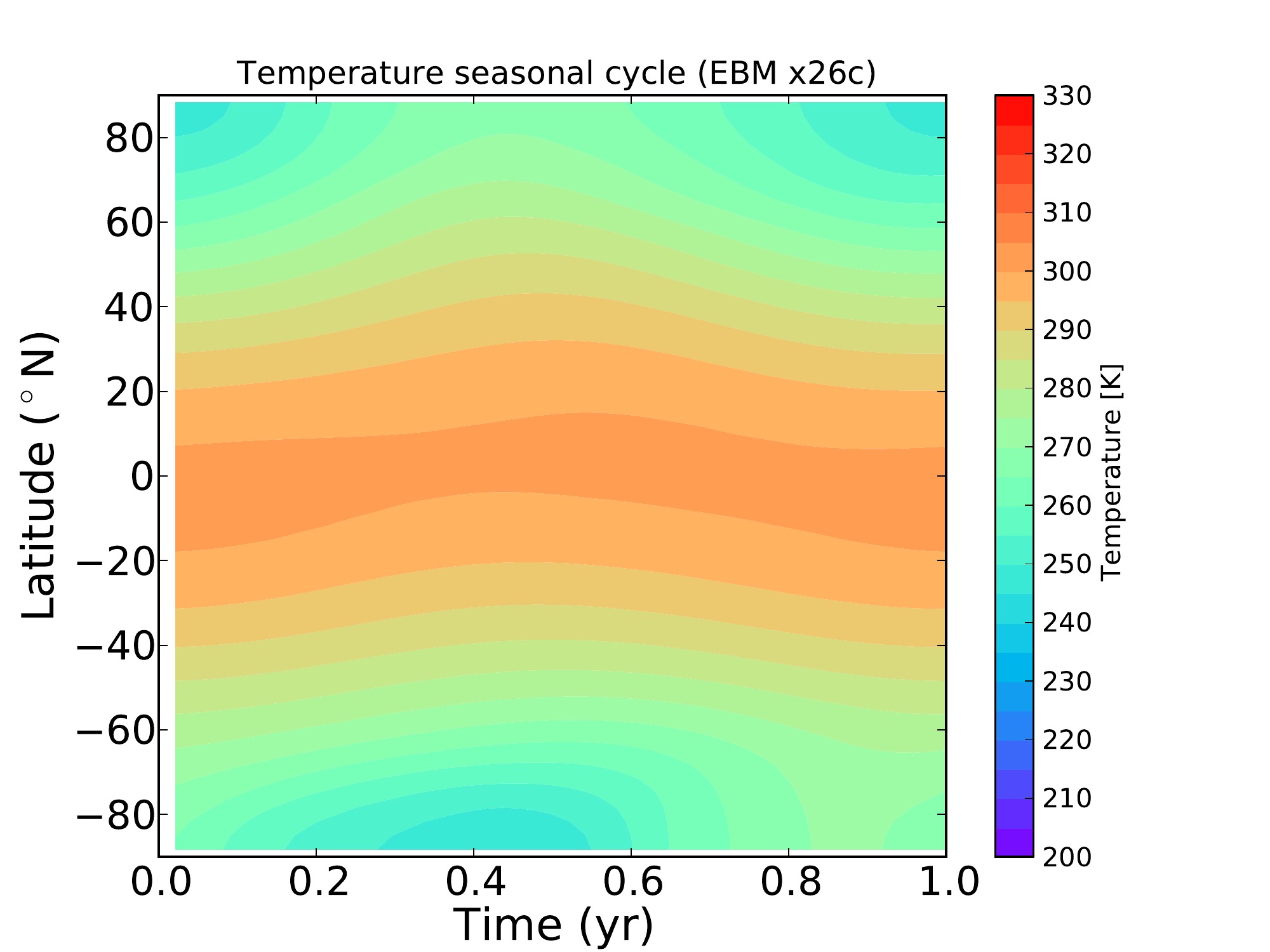}
\includegraphics[width=7.5cm]{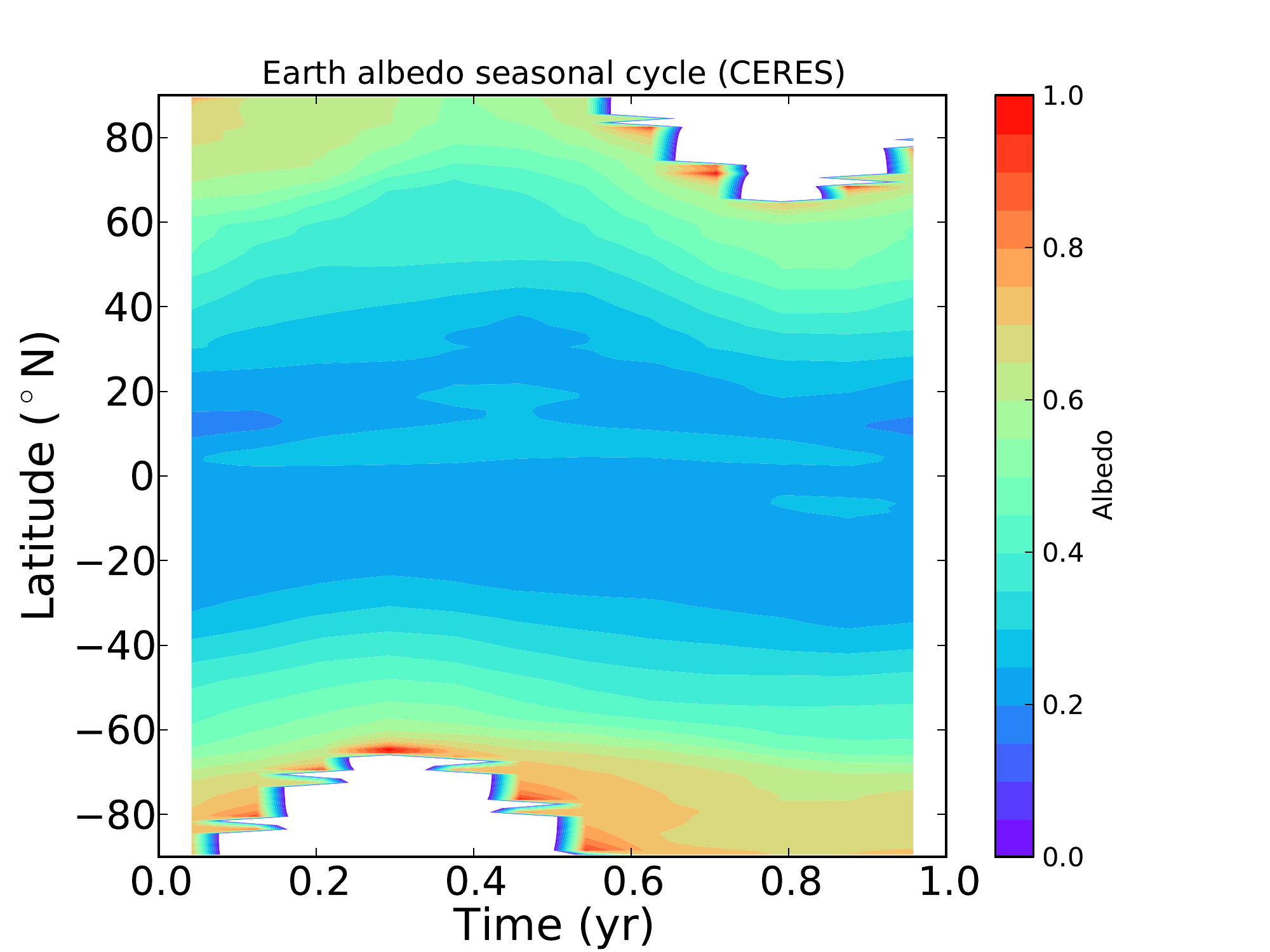}
\includegraphics[width=7.5cm]{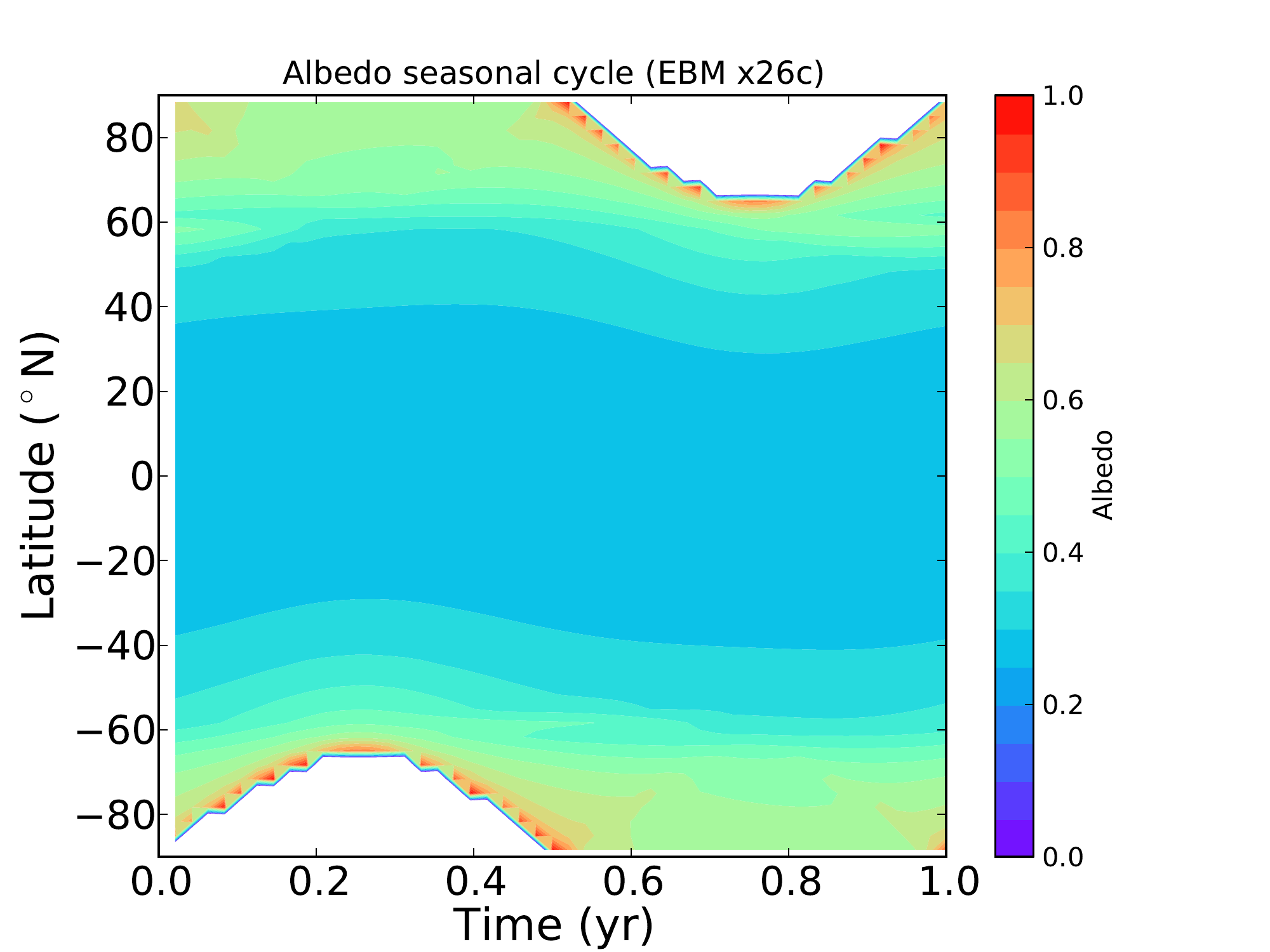} 
\caption{
Seasonal and latitudinal variations of surface temperature (top panels)
and top-of-atmosphere albedo (bottom panels) of the Earth.
Left panels: average ERA Interim 2m temperatures \citep{Dee11} and 
CERES short-wavelength albedo \citep{Loeb05,Loeb07}
in the years 2001-2013.
Right panels:  predictions of the reference Earth model. 
The zero point of the time scale is set at the spring equinox.
Blank areas in the albedo panels indicate regions of the seasonal-latitudinal space
without incident stellar flux. 
  }
\label{TempLatMaps}%
\end{center}
\end{figure*}

\section{The  reference Earth model \label{sectEarthModel}}

For the calibration of the reference Earth model we adopted 
orbital parameters, axis tilt, and rotation period from \citet{Allen00}.
For the solar constant we adopted $S_0 = 1360$ W\,m$^{-2}$ 
and $g=9.8$\,m\,s$^{-2}$ for the surface gravity acceleration. 
The zonal coverage of continents and oceans was taken from Table III in WK97. 
We adopted  a relative humidity $q= 0.6$ (see \S \ref{sectCRM}) and
volumetric mixing ratios of CO$_2$ and  CH$_4$  of 380 ppmV and 1.8 ppmV, respectively.
The surface pressure of dry air, $p_\mathrm{dry} = 1.0031 \times 10^5$\,Pa, was 
tuned to match the  moist surface pressure of the Earth, 
$p_\mathrm{tot} = 1.0132 \times 10^5$\,Pa. 
The remaining parameters of the model are shown in Table \ref{tabFiducialPar}. 
Some parameters were fine tuned to match 
the mean annual global quantities of the northern hemisphere of the Earth,
as specified in the Table. 
We avoided  using the southern hemisphere as a reference since its climate
is strongly affected by the altitude of Antartica,
while orography is  not included in the model. 
In Table \ref{GlobalEarthModel}, column 3,  we show the 
experimental data of the northern hemisphere used as a guideline to tune the model parameters.
In column 4 of the same table we show the corresponding predictions of the  Earth reference model.

In Fig. \ref{annualLatProfiles} we show the mean annual latitude profiles of surface temperature, top-of-atmosphere albedo
and meridional energy flux predicted by the reference model (solid line).
The temperature profile is compared with ERA Interim 2m temperatures \citep{Dee11} averaged in the period 2001-2013
(crosses in the top panel). 
Area-weighted temperature differences between observed and predicted profile 
have an rms of 1.1\,K in the northern hemisphere.
The albedo profiles are compared with the CERES short-wavelength albedo \citep{Loeb05,Loeb07}
averaged in the same period
(crosses in the middle panel). 
The \model\ is able to reproduce reasonably well the rise of albedo with increasing latitude.
This rise is due to two factors: the dependence of the atmosphere, ocean,  
and cloud albedo on zenith distance  (\S\S \ref{sectTOAalbedo},\ref{sectionAlbedo})
and the increasing coverage of ice at low  temperature (\S \ref{sectCoverage}). 
The meridional flux in the bottom panel is compared with the total flux (dashed line) and the atmospheric flux (dotted line)
obtained from EC-Earth model \citep{Hazeleger10}. 
In spite of the simplicity of the transport formalism intrinsic to Eq. (\ref{diffusionEq}), 
the model is able to capture remarkably well the latitude dependence of the meridional transport. 

The {\em seasonal} variations of the  temperature and  albedo latitudinal profiles
are compared with the experimental data in Fig. \ref{TempLatMaps}. One can see that
the reference model is able to capture the general patterns of seasonal evolution. 

Even if the reference model has been tuned using northern hemisphere data, 
the predictions shown in Figs. \ref{annualLatProfiles} and \ref{TempLatMaps} are in
general agreement with the data
also for most the southern hemisphere, with the exception of Antartica. 
It is  remarkable that the atmospheric transport in both hemispheres
is reproduced well, in spite of significant differences in the ocean contribution 
between the two hemispheres (see  \S \ref{sectOceanTransport}). 

Once the reference model is calibrated, 
some of the parameters that have been tuned to fit the present-day Earth's climate can be changed
for specific applications of the \model\ to exoplanets.
As an example, even if we adopt $a_s=0.18$ for the surface albedo of continents in the reference model, 
we may adopt lower values, typical of forests, or higher values, typical of sandy deserts, for specific applications. 
More information on the parameters that can be changed is given in Table \ref{tabFiducialPar}.

\section{Model simulations of idealized Earth-like planets \label{sectEarthLike}}

In this set of experiments we study the effects of varying a single planet quantity while
assigning Earth's values to all the remaining parameters. 
We consider variations of rotation period, insolation, atmospheric columnar mass, radius, obliquity, 
land distribution, and long-wavelength cloud forcing.

\subsection{Rotation rate \label{sectRotMap}}

\begin{figure*}
\begin{center} 
\includegraphics[width=7.5cm]{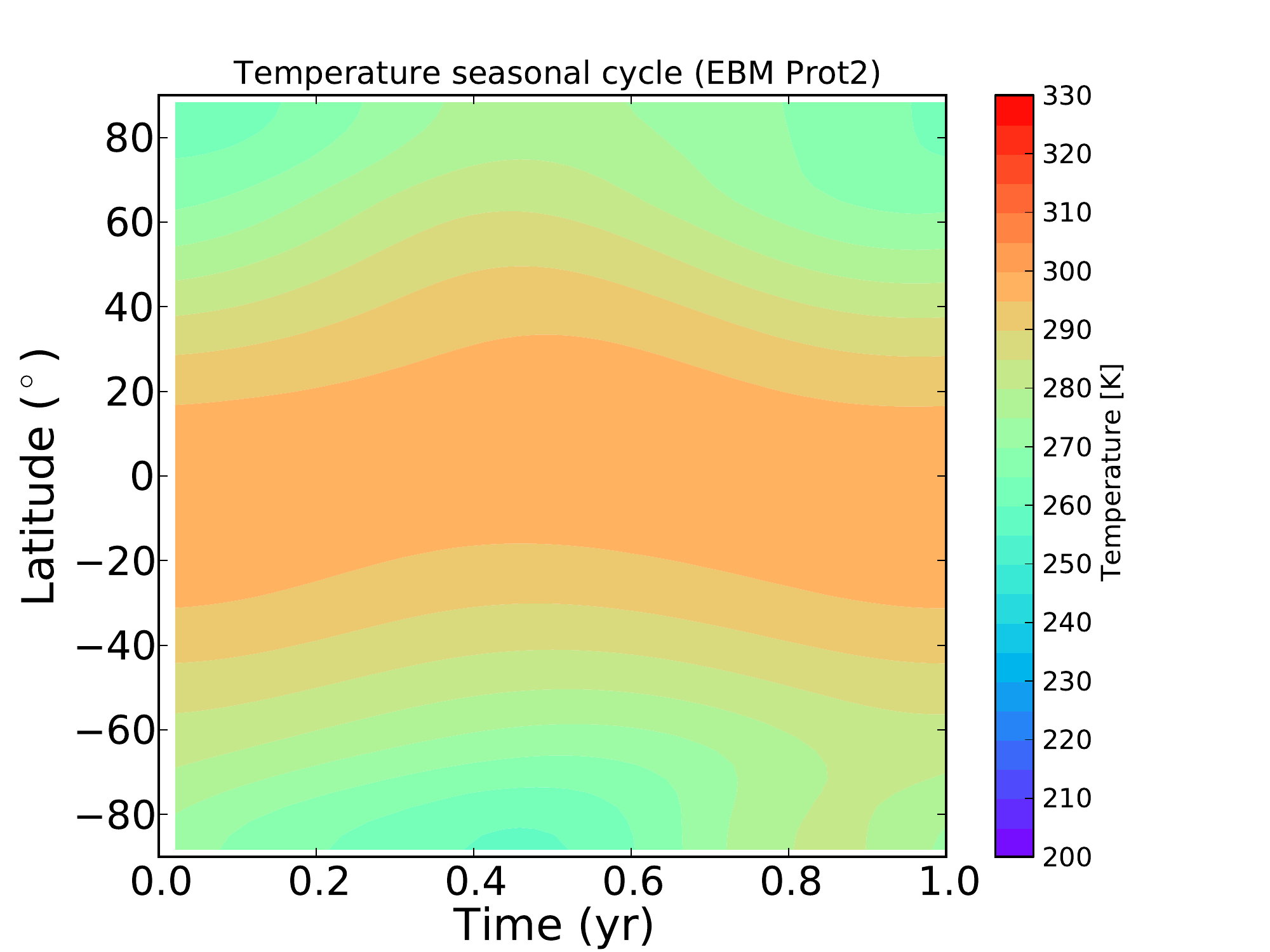} 
\includegraphics[width=7.5cm]{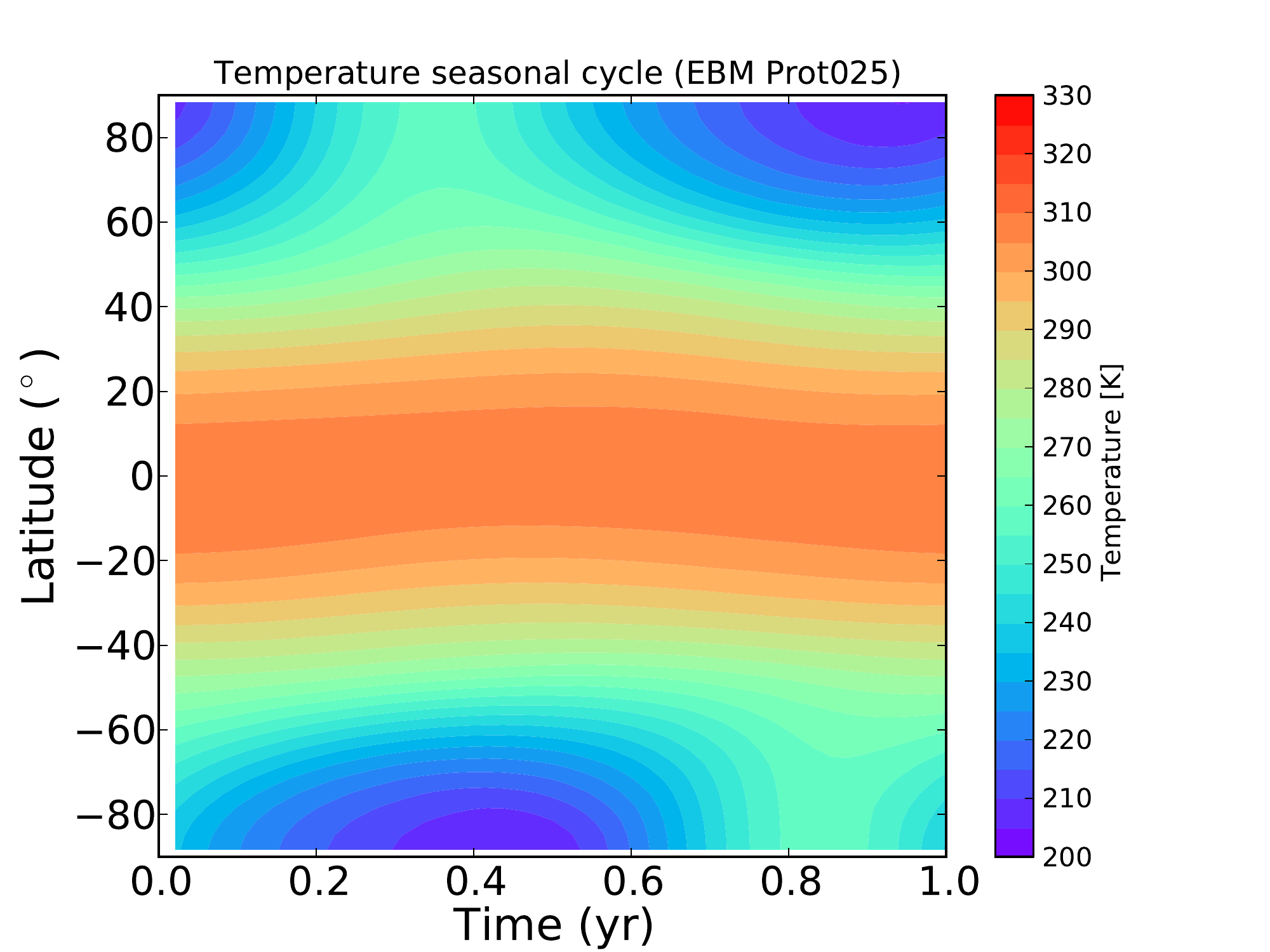}
\caption{
Variation of surface temperature as a function of latitude and orbital phase 
for an Earth-like planet with rotation rate $\Omega=0.5 \,\Omega_\oplus$ (left)
and $\Omega=4 \, \Omega_\oplus$ (right). 
All the remaining parameters are those used for the Earth reference model (Appendix B).}
\label{mapsRotation}%
\end{center}
\end{figure*}

In Fig. \ref{mapsRotation} we show how $T(\varphi,t)$ is affected by variations of rotation rate,
the left and right panel corresponding to the cases 
$\Omega=0.5 \,\Omega_\oplus$ and $\Omega=4 \, \Omega_\oplus$, respectively. 
One can see that the
change of the surface temperature distribution is quite dramatic
in spite of the modest dependence of the transport coefficient
on rotation rate that we adopt, $D \propto \Omega^{-4/5}$ [see Eq.(\ref{eq:Ddry})]. 
The mean global habitability changes from $\hm=0.94$ in the slow-rotating case
to $\hm=0.71$ in the fast-rotating case. The corresponding change of mean global temperature
is relatively small, from $\widetilde{T}=284$\,K to 290\,K. 
These results show the importance of estimating $T(\varphi,t)$, 
rather than $\widetilde{T}$, in order to quantify the habitability.

The behavior of the mean equator-pole difference, $\DTep$,
is useful to interpret the results of this test. 
We find $\DTep = 28$\,K in the slow-rotating case
and $\DTep = 80$\,K in the fast-rotating case.
This variation of $\DTep$ is much higher than
that found for the same change of rotation rate 
in the case of the KS14 aquaplanet  
(top-left panel in Fig. \ref{figKS14validation}). We interpret this
strong variation of  $\DTep$ in terms of the ice-albedo feedback, 
which is positive and tends to amplify variations of the surface temperature.
This feedback is accounted for
in the present experiment, but not in the case of the aquaplanet.
These results illustrate the importance of using climate models with 
latitude temperature distribution and ice-albedo feedback in order to estimate the  
fraction of habitable surface. 
  
The analysis of the ice cover evidentiates another important difference
between the \model\ and classic EBMs. 
With the \model\ the ice cover increases from $\simeq 3\%$ at $\Omega=0.5 \,\Omega_\oplus$ 
to  $\simeq 23\%$ at $\Omega=4 \, \Omega_\oplus$. 
This increase is less dramatic than the transition 
to a complete ``snowball'' state (i.e. ice cover $\simeq 100\%$)
found with classic EBMs at $\Omega=3 \, \Omega_\oplus$
\citep[e.g.][]{SMS08}. 
This difference is due to two factors. One is the strong dependence of the transport on rotation rate
adopted in most EBMs ($D \propto \Omega^{-2}$), 
which is not supported by the validation tests discussed above  (top-left panel of Fig. \ref{figKS14validation}).
Another factor is the algorithm adopted for the albedo, which in classic EBMs is a simple analytical function $A=A(T)$,
while in the \model\ is a multi-parameter function $A=A(T,p,g,p\mathrm{CO_2},a_\mathrm{s},Z)$
that takes into account the vertical transport of stellar radiation (\S \ref{sectTOAalbedo}). 
These results show the importance of adopting algorithms calibrated with 3D experiments 
and atmospheric column calculations.

\begin{figure*}
\begin{center} 
\includegraphics[width=7.5cm]{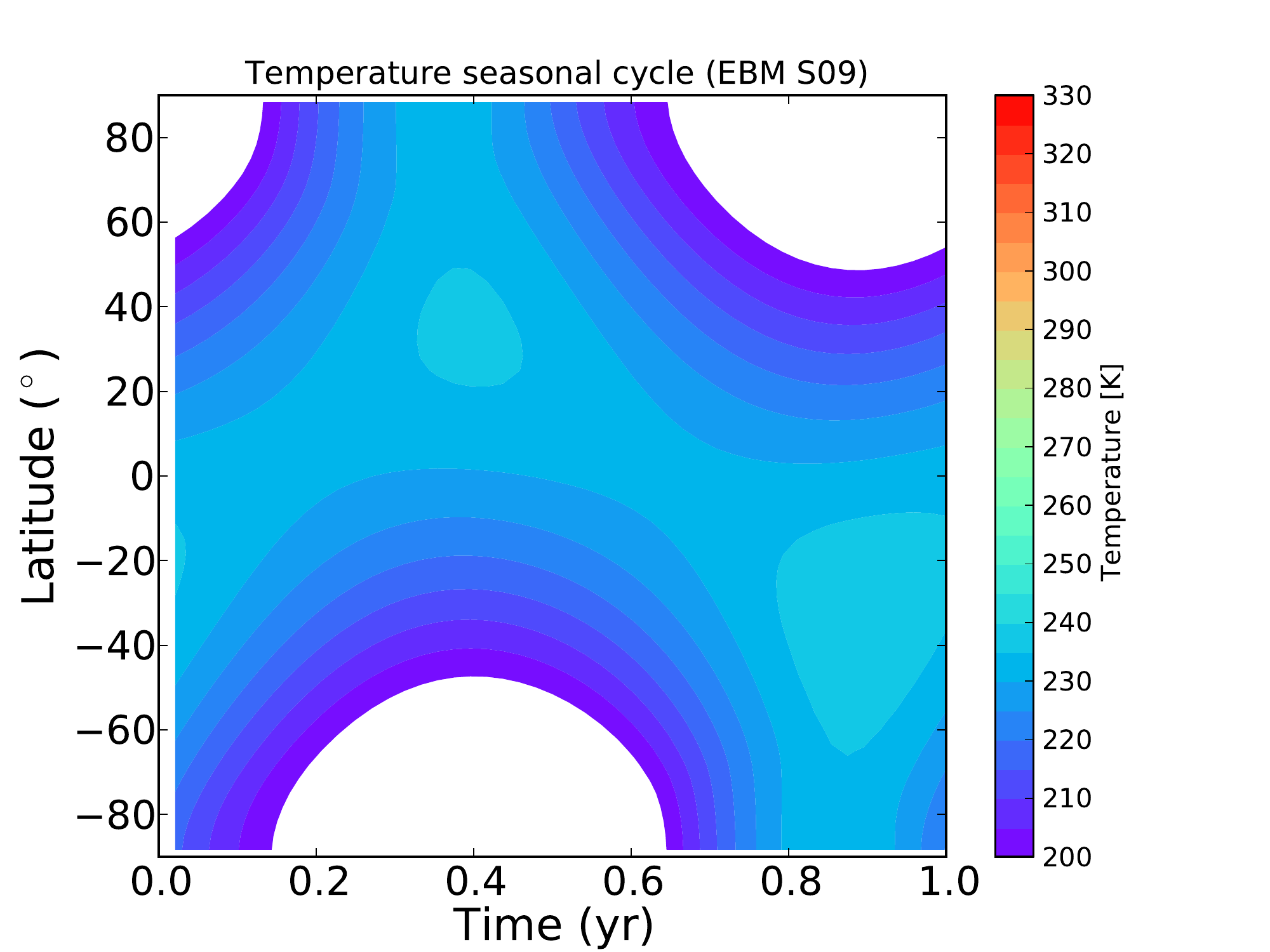}
\includegraphics[width=7.5cm]{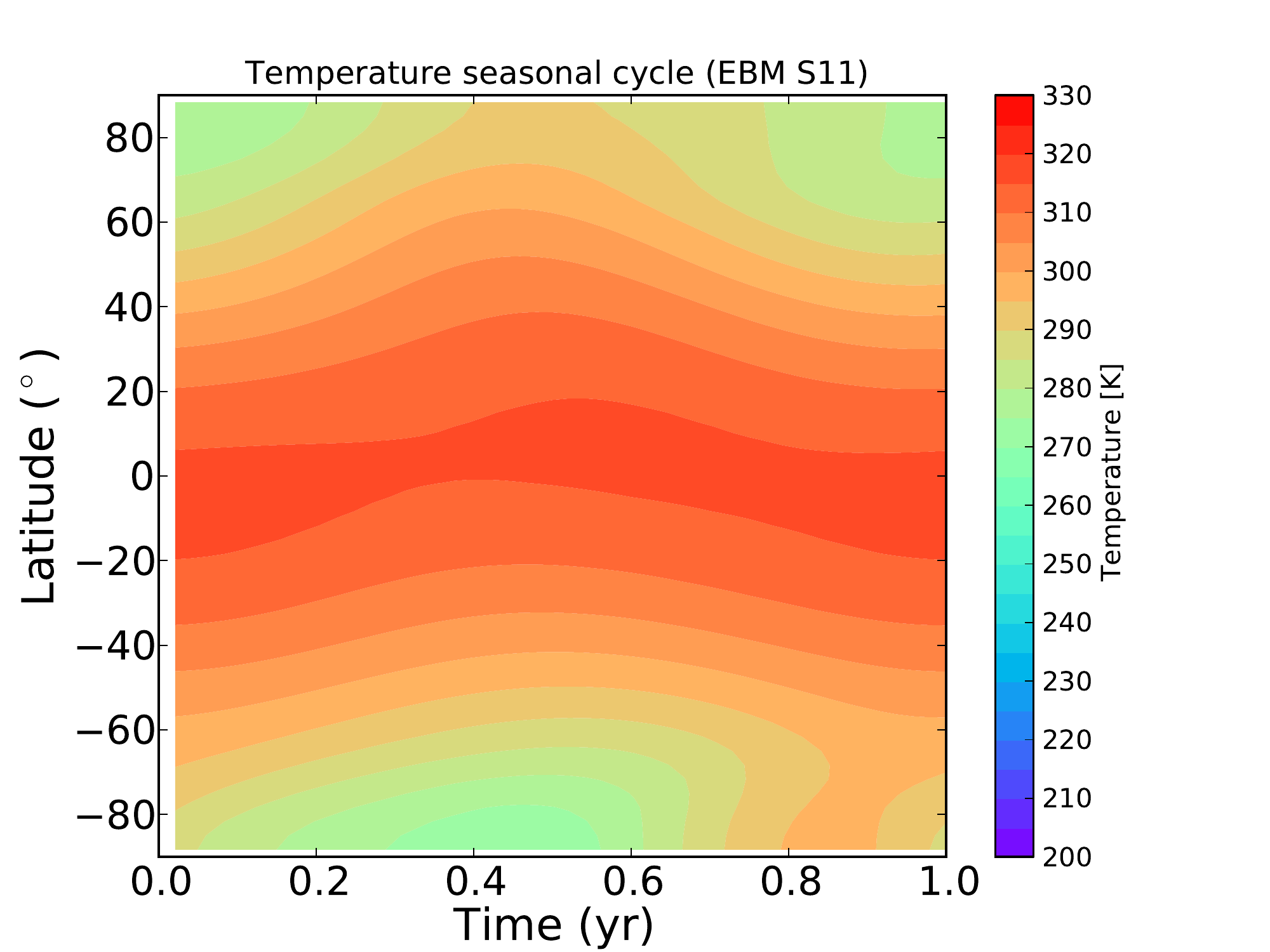} 
\caption{
Variation of surface temperature as a function of latitude and orbital phase 
for an Earth-like planet with insolation 10\% lower (left) and 10\% higher (right)
than the present-day value. 
All the remaining parameters are those used for the Earth reference model (Appendix B).
Blank areas in the left panel indicate regions of the $(\varphi,t)$ space with surface temperature below 200\,K.
  }
\label{mapsFlux}%
\end{center}
\end{figure*}

\subsection{Insolation \label{sectFluxMap}}

In Fig. \ref{mapsFlux} we show how $T(\varphi,t)$ is affected by variations of stellar insolation,
the left and right panel corresponding to an insolation of 0.9 and 1.1 times
the present-day Earth's insolation ($S/4\simeq 341$\,W\,m$^{-2}$), respectively.
In the first case the \model\ finds a complete snowball with $\Tm=223$\,K and $\hm=0$,
while in the latter  $\Tm=306$\,K and $\hm=1$ with no ice cover.
These results demonstrate the extreme sensitivity of the surface temperature to variations of insolation
and the need of incorporating  feedbacks in global climate models in order to  
define the limits of insolation of a habitable planet.

Our model is able to capture the ice-albedo feedback, but is not suited for treating
hot atmospheres and the runaway greenhouse instability.
To test the limits of the \model\ we have gradually increased the insolation and compared
our results with those obtained by the 3D model of \citet{Leconte13}.
We found that 
the \model\ tracks the rise of $\Tm$ with insolation predicted by the 3D model
up to $S/4 \simeq 365$\,W\,m$^{-2}$.  
At higher insolation, the 3D model predicts a faster rise of $\Tm$ due to an increase of the radiative cloud forcing. 

By decreasing the insolation with respect to the Earth's value,
the \model\ finds solutions characterized by increasing ice cover. 
At $S/4 \simeq 310$\,W\,m$^{-2}$ the simulation displays a runaway ice-albedo feedback  
that leads to a complete snowball configuration. 
This result sets the \model\ limit of minimum insolation for the liquid-water habitability 
of an Earth-twin planet. 

\begin{figure*}
\begin{center} 
\includegraphics[width=7.5cm]{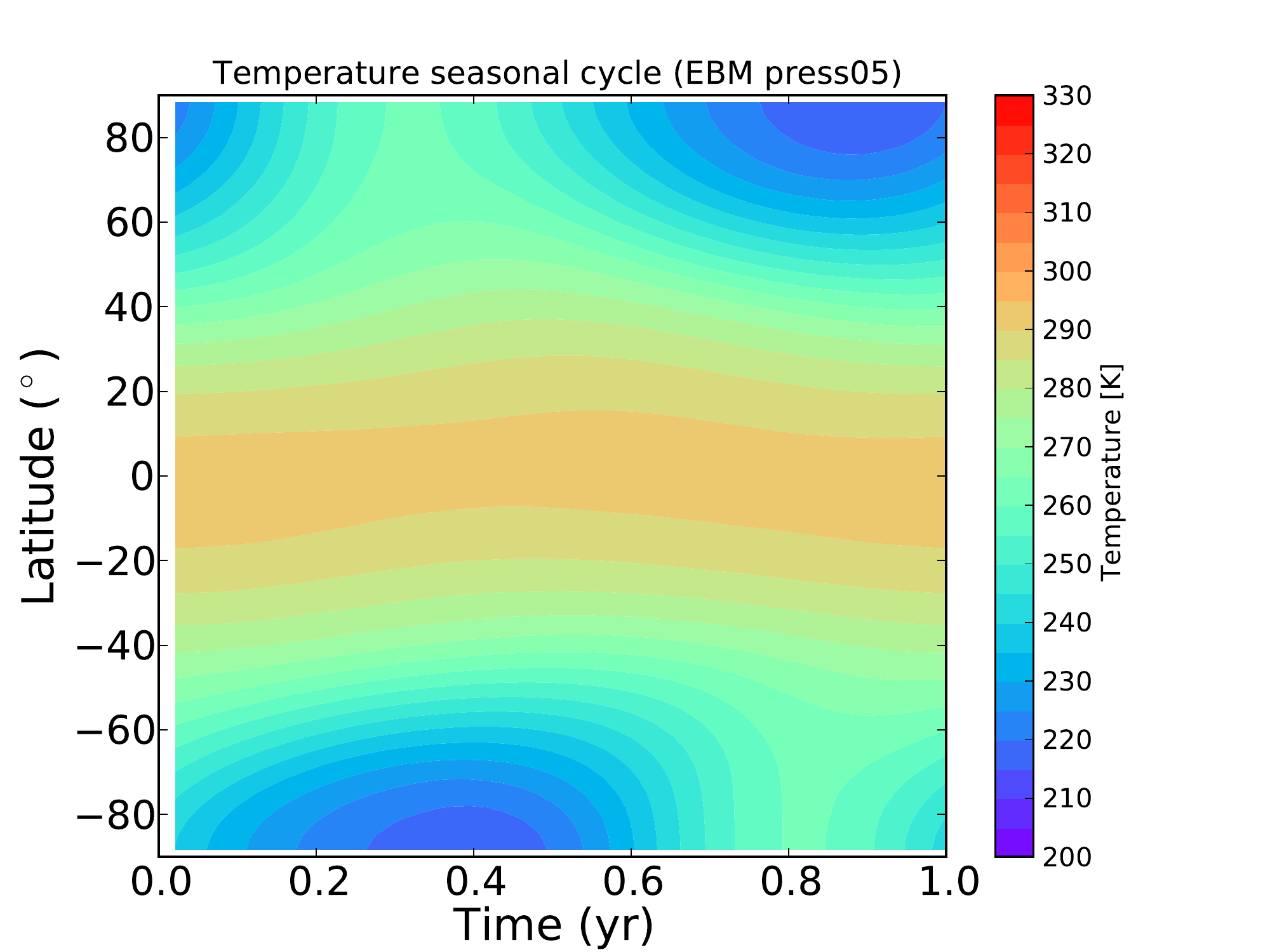}
\includegraphics[width=7.5cm]{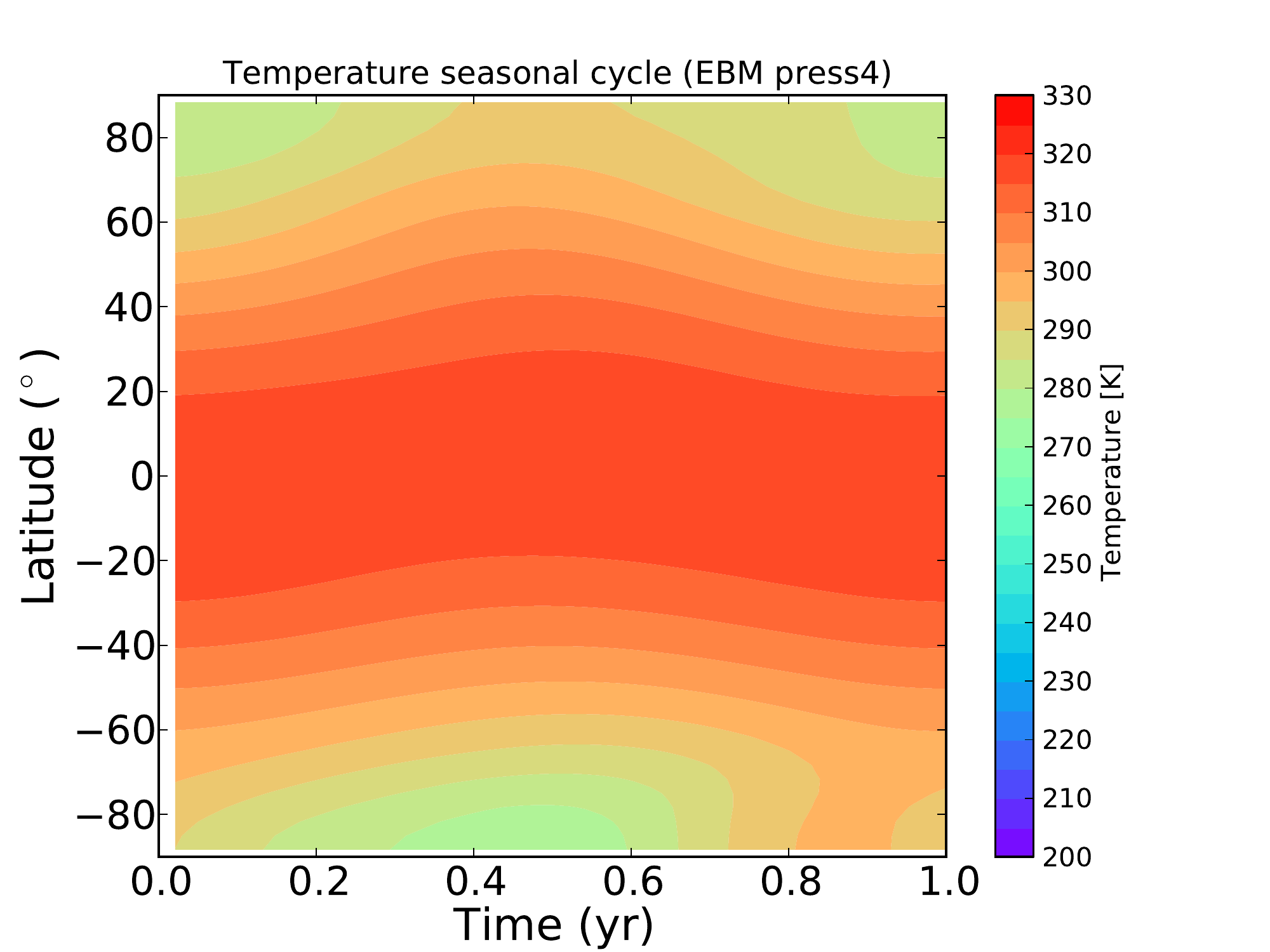} 
\caption{
Variation of surface temperature as a function of latitude and orbital phase 
for an Earth-like planet with surface pressure $p=0.5$\,bar (left) and $4$\,bar (right).
All the remaining parameters are those used for the Earth reference model (Appendix B).}
\label{mapsPressure}%
\end{center}
\end{figure*}

\subsection{Atmospheric columnar mass \label{sectColumnarMass}}

In Fig. \ref{mapsPressure} we show how $T(\varphi,t)$ is affected by variations of surface pressure,
the left and right panel corresponding to the cases $p=0.5$\,bar and $4$\,bar, respectively.
Since the surface gravity is kept fixed at $g=g_\oplus$, this experiment also investigates
the climate impact of variations of atmospheric columnar mass, $p/g$. 
We find a significant difference in mean temperature and habitability between the two cases,
with $\Tm=274$\,K and $\hm=0.62$ in the low-pressure case and
$\Tm=310$\,K and $\hm=1.00$ in the high-pressure case. 
The mean equator-pole difference, 
decreases from $\DTep=$55\,K to 33\,K between the two cases.
The rise of mean temperature is due to the existence of a positive correlation 
between  columnar mass and intensity of the greenhouse effect.
The decrease of temperature gradient results from the correlation
between $p/g$ and the efficiency of the horizontal transport.  
Both effects have been already discussed in Paper I. 
Here we find a more moderate trend with $p/g$, as a result of the new formulation of $D$ that we adopt.

\begin{figure*}
\begin{center} 
\includegraphics[width=7.5cm]{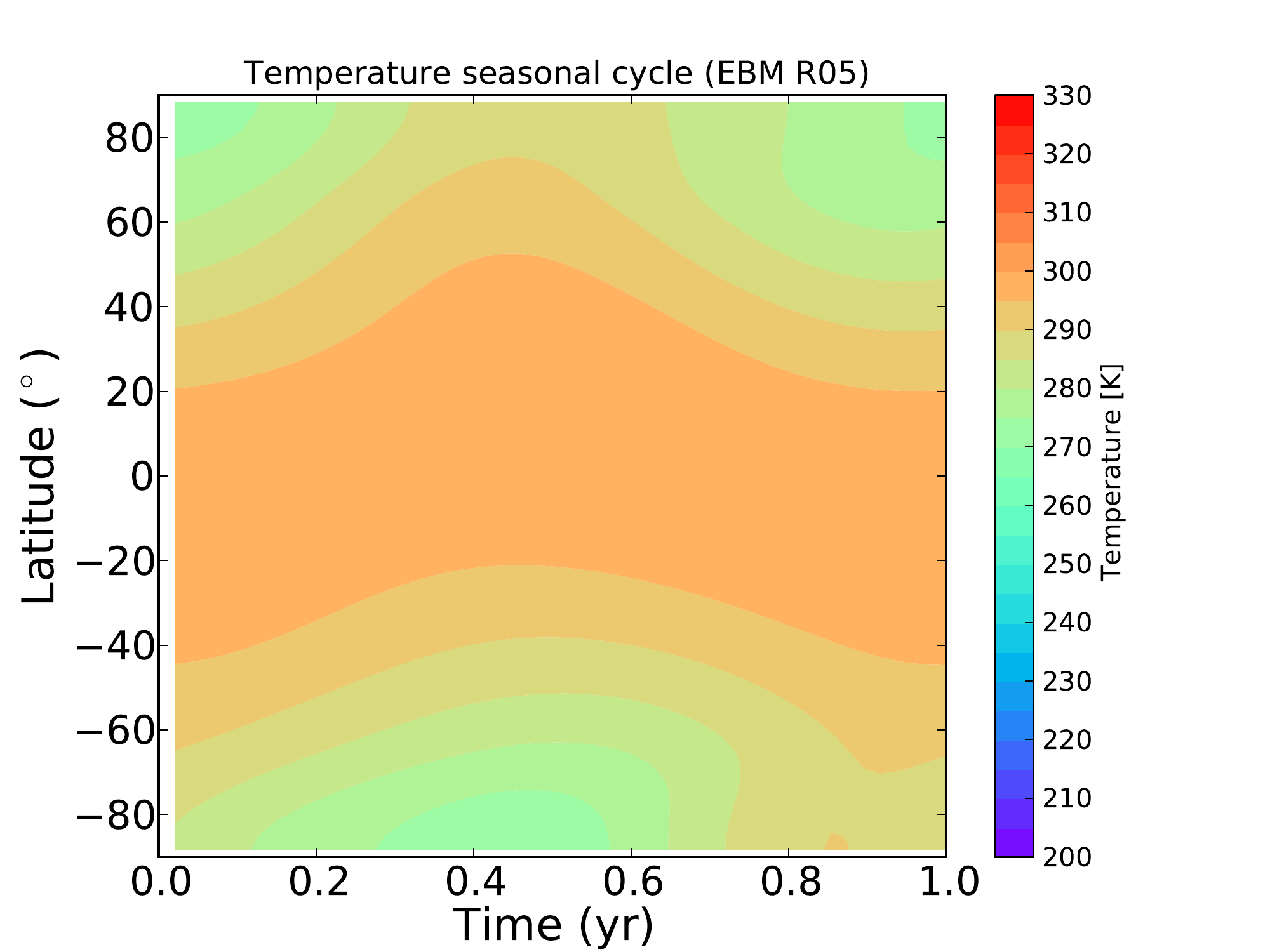}
\includegraphics[width=7.5cm]{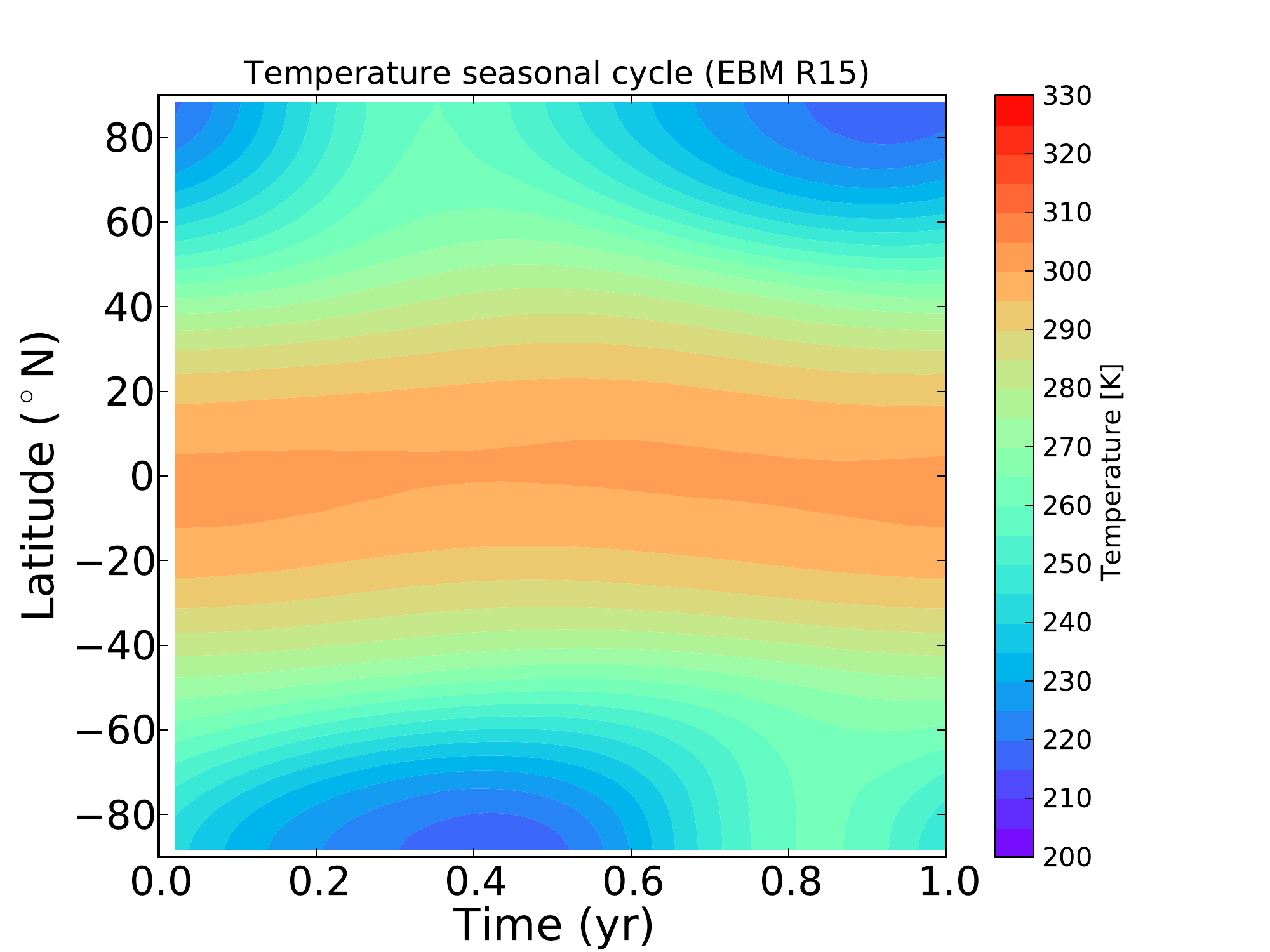} 
\caption{
Variation of surface temperature as a function of latitude and orbital phase 
for an Earth-like planet with radius $R=0.5 \,R_\oplus$ (left) and $1.5 \,R_\oplus$ (right).
The mean density of the planet is assumed to be constant, $\rho=\rho_\oplus$, so that
the same model predictions are also appropriate for an Earth-like planet with mass $M=0.125\,M_\oplus$ (left panel)
and $3.375\,M_\oplus$ (right panel).  
The atmospheric columnar was kept constant by scaling $p$ and $g$ with $R$.   
All the remaining parameters are those used for the Earth reference model (Appendix B).}
\label{mapsRadius}%
\end{center}
\end{figure*}

\subsection{Planet radius or mass \label{sectRadius}}

In Fig. \ref{mapsRadius} we show how $T(\varphi,t)$ is affected by variations of planet radius,
the left and right panel corresponding to the cases $R=0.5\,R_\oplus$ and $1.5\,R_\oplus$, respectively.
In this ideal experiment we keep fixed the planet mean density, $\rho=\rho_\oplus$, so that the planet mass
and gravity scale as $M \propto R^3$ and $g \propto R$, respectively. We also keep fixed the columnar mass,
$p/g=(p/g)_\oplus$, by scaling $p$ and $g$ with $R$. With this experimental setup, the radius
is the only parameter that varies in the transport coefficient $D$ [Eqs. (\ref{eq:SLdry}),  (\ref{eq:SLdm}), and  (\ref{eq:DtermF})]. 
The left panel corresponds to the case $M=0.125\,M_\oplus$, $p=0.5$\,bar and $g=0.5\,g_\oplus$
and the right panel to the case $M=3.375\,M_\oplus$, $p=1.5$\,bar and $g=1.5\,g_\oplus$.

We find that the planet cools significantly when the radius, mass and gravity increase,
with a variation of the mean global temperature from $\Tm=294$\,K to 281\,K.
The equator-to-pole temperature differences increases dramatically, from
$\DTep=$18\,K to 64\,K, respectively.
This  change of $\Delta T_\mathrm{PE}$ is much higher than
that found for the same change of radius in the case of the aquaplanet 
(bottom-right panel in Fig. \ref{figKS14validation}).
The inclusion of the ice-albedo feedback in the present experiment 
amplifies variations  of  $\Delta T_\mathrm{PE}$.
In fact, the ice cover increases from 0\% to 21\% between  $R=0.5\,R_\oplus$ and $1.5\,R_\oplus$.
As a result of the variations of $\Tm$ and $\DTep$, the mean global habitability is significantly affected, 
changing from $\hm=1.00$ to 0.71 with increasing planet size.

\begin{figure*}
\begin{center} 
\includegraphics[width=7.5cm]{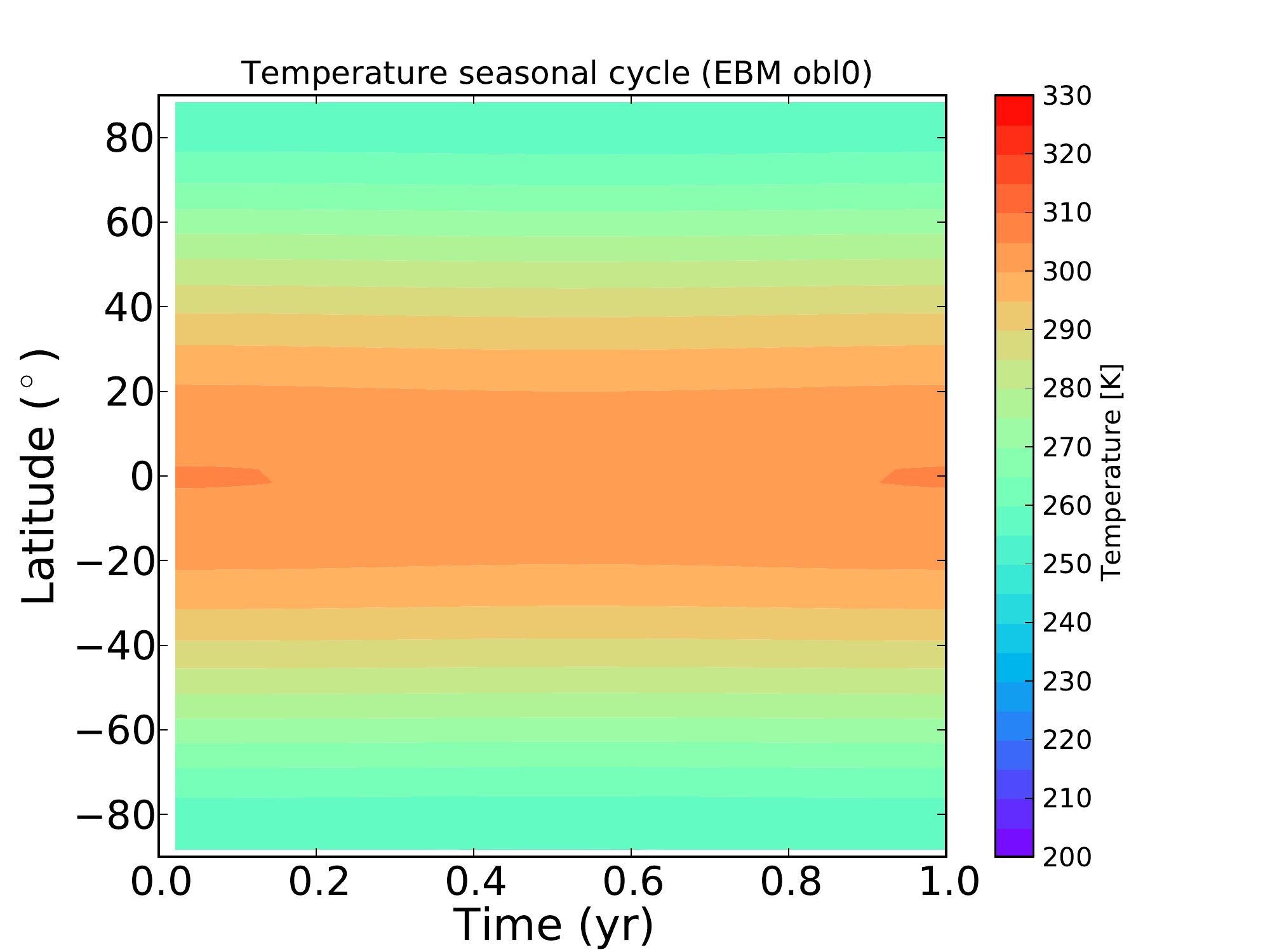}
\includegraphics[width=7.5cm]{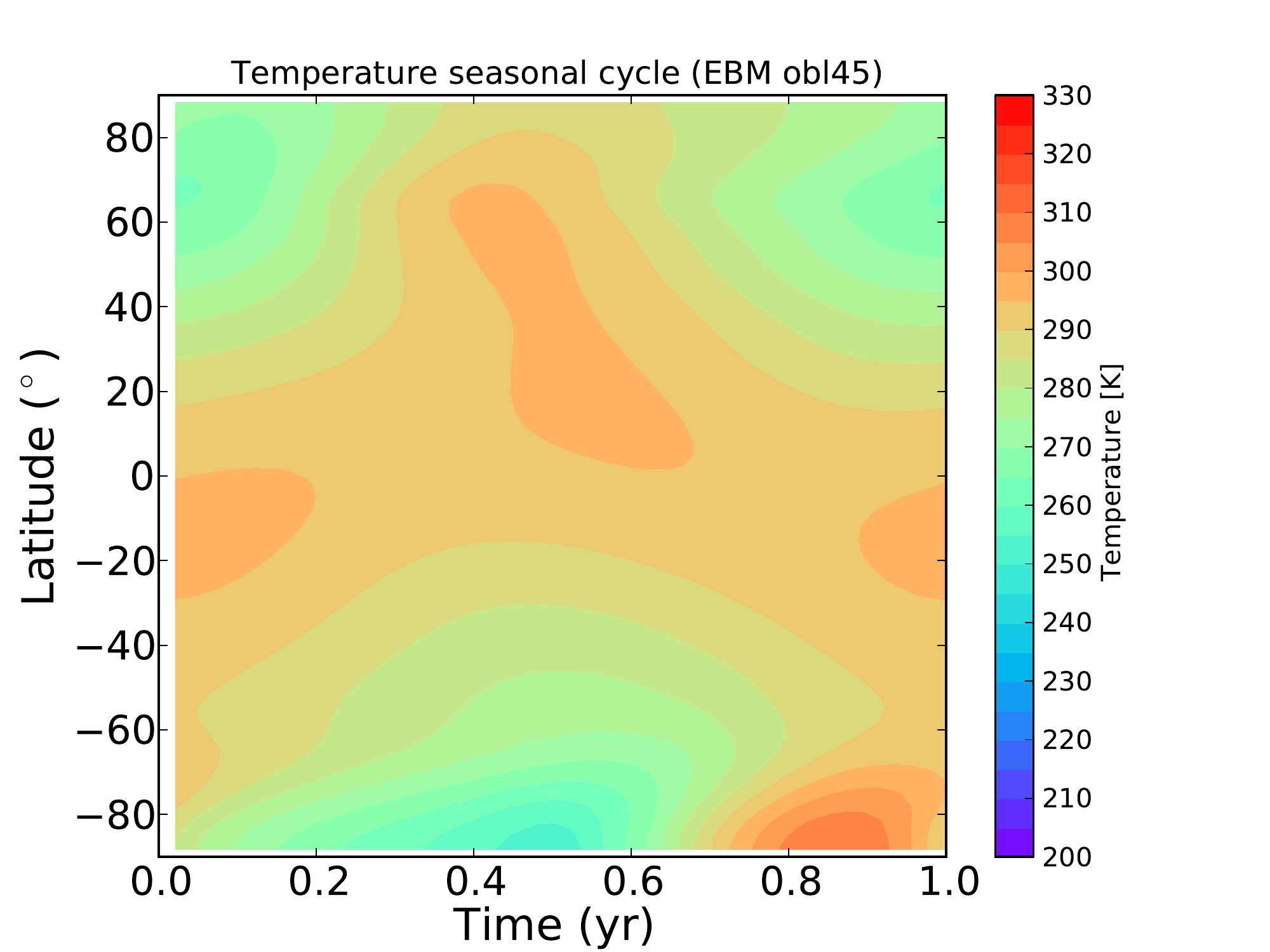} 
\caption{
Variation of surface temperature as a function of latitude and orbital phase 
for an Earth-like planet with axis obliquity $\epsilon=0^\circ$ (left) and $45^\circ$\,d (right).
All the remaining parameters are those used for the Earth reference model (Appendix B).}
\label{mapsObliquity}%
\end{center}
\end{figure*}

\subsection{Axis obliquity \label{sectObliquity}}

In Fig. \ref{mapsObliquity} we show how $T(\varphi,t)$ is affected by variations of axis obliquity,
the left and right panel corresponding to the cases $\epsilon=0^\circ$ and $45^\circ$, respectively.
We find a modest decrease of $\Tm$, 
from $291$\,K to 289\,K, and a significant decrease of $\DTep$,
from 42\,K to 15\,K. 
The mean annual temperature at the poles is lower at $\epsilon=0^\circ$ than at $\epsilon=45^\circ$, 
because in the first case the poles have a constant, low temperature, while in the latter 
they alternate cool and warm seasons. As a result, the ice cover decreases
and the habitability increases in the range from $\epsilon=0^\circ$ to $\epsilon=45^\circ$
For the conditions considered in this experiment, the initial ice cover is relatively small
($\simeq 7\%$) and the increase of habitability relatively modest (from $0.87$ to 0.95). 
Larger variations of habitability 
are found starting from a higher ice cover. 
%
These results confirm the necessity of determining  $T(\varphi,t)$ 
and accounting for the ice-albedo feedback in order to estimate the habitability.

At $\epsilon > 45^\circ$  
EBM studies predict a stronger climate impact of obliquity,
with  possible formation of equatorial ice belts  
\citep{WK97,SMS09,Vladilo13}.  The physically-based derivation of the  coefficient $D$
prevents using the \model\ when the equatorial-polar gradient is negative, 
because Eqs. (\ref{eq:Ddry}) and (\ref{eq:SLdry}) require $\delta T>0$.
In the Earth model this condition is satisfied when $\epsilon \leq 52^\circ$. 
Clearly, 
the climate behavior at high obliquity should be tested
with 3D climate experiments \citep[e.g.][and refs. therein]{Williams03,Ferreira14},
being  cautious with  predictions obtained with EBMs.

\begin{figure*}
\begin{center} 
\includegraphics[width=7.5cm]{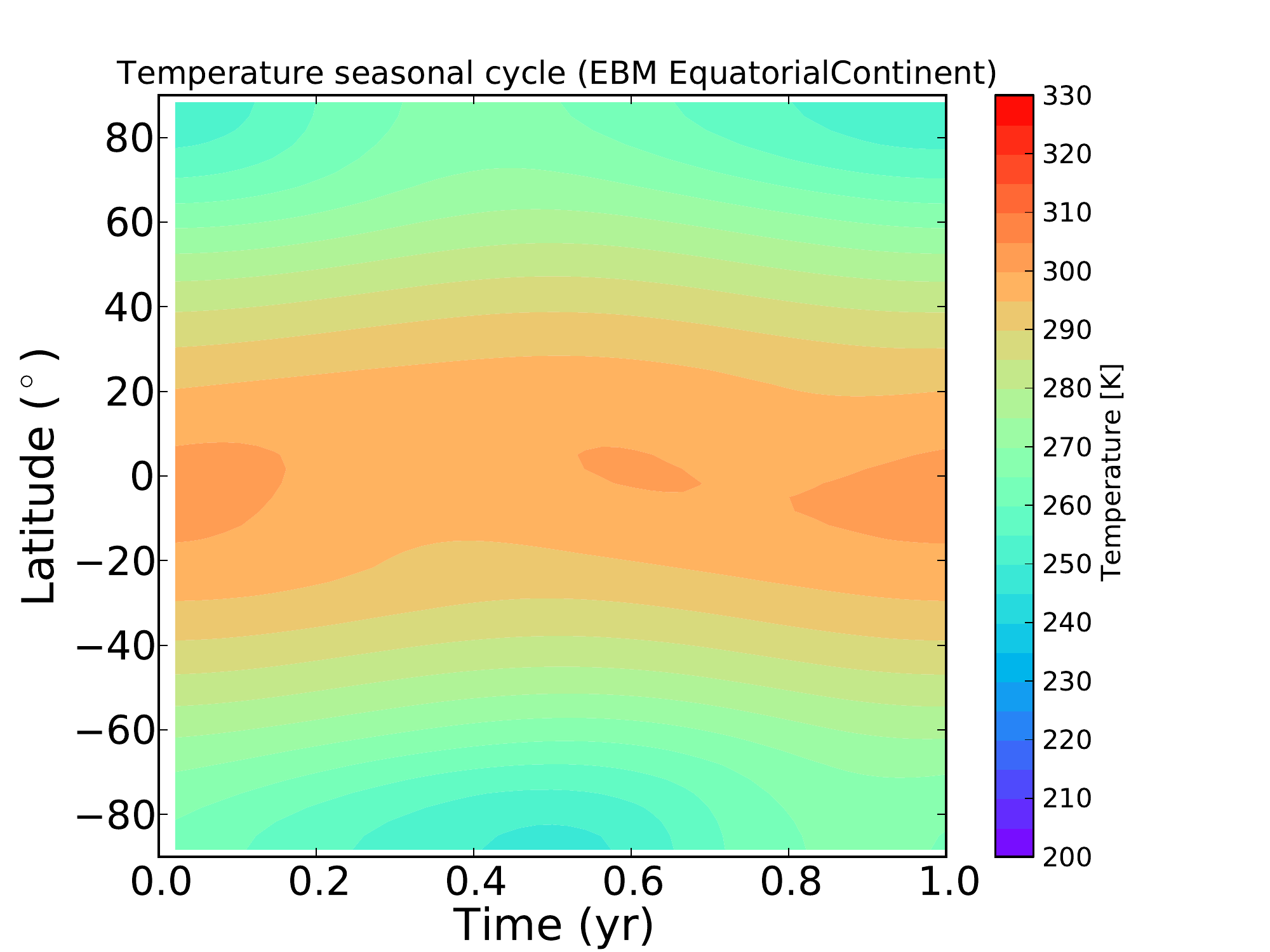}
\includegraphics[width=7.5cm]{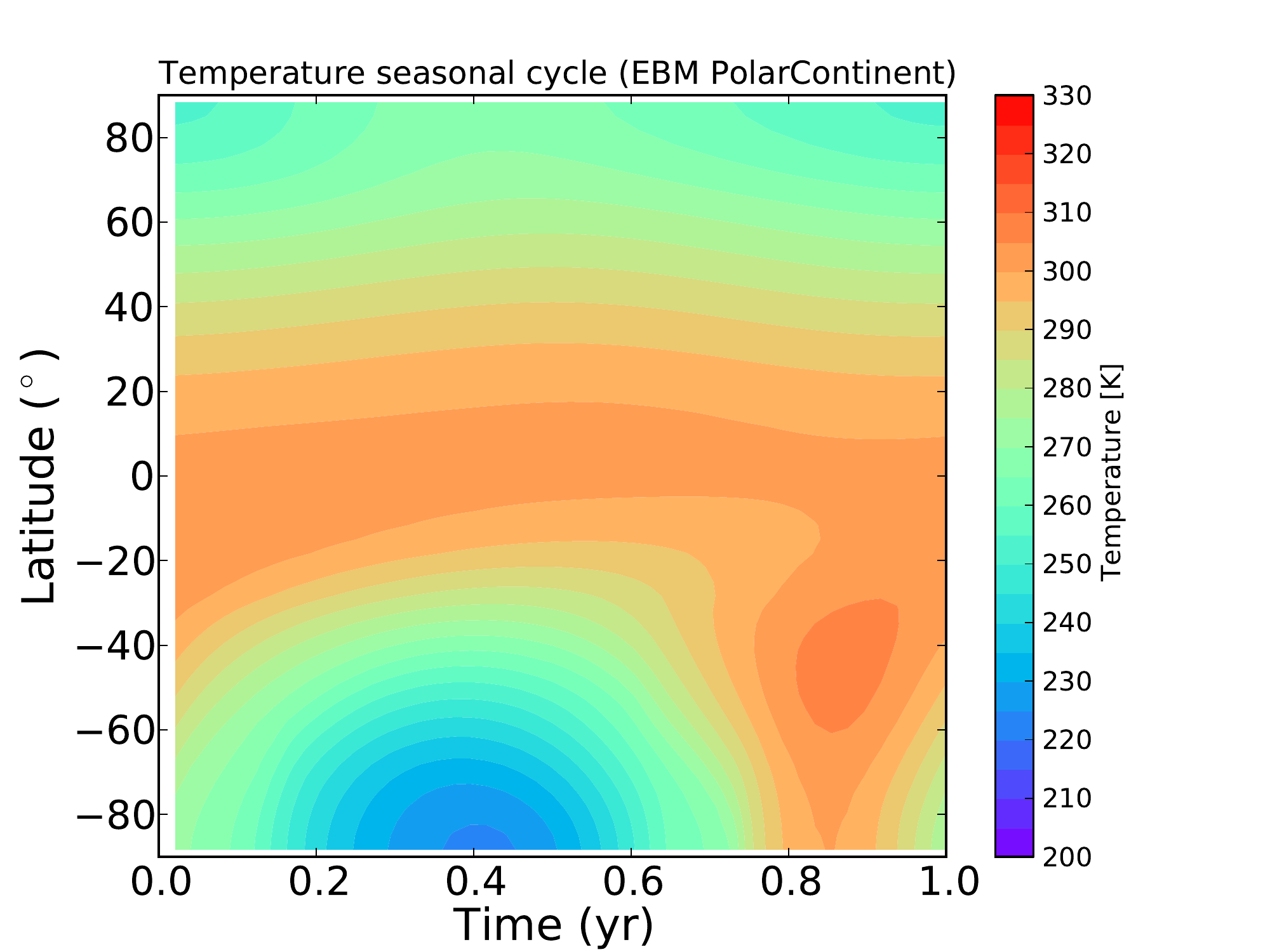} 
\caption{
Variation of surface temperature as a function of latitude and orbital phase 
for an Earth-like planet with global ocean coverage of 0.7 and
a single continent covering the remaining surface. Left panel: equatorial continent spread over all longitudes.
Right panel: polar continent centered at latitude $-90^\circ$.  
All the remaining parameters are those used for the Earth reference model (Appendix B).}
\label{mapsContinents}%
\end{center}
\end{figure*}

\subsection{Ocean/land distribution \label{sectOceanLand}}

In Fig. \ref{mapsContinents} we show how $T(\varphi,t)$ is affected by variations of 
the geographical distribution of the continents. 
In these experiments we consider a single continent covering all longitudes,
but located at different latitudes in each case. 
The global ocean coverage is fixed at 0.7, as in the case of the Earth.
In the left panel we show the the case of a continent centered on the equator, 
while in the right panel a continent centered on the southern pole.
The variation of $T(\varphi,t)$ is quite remarkable given the little change of
 mean global annual temperature ($\Tm=288$\,K and 289\,K for the equatorial and polar case, respectively).
 The mean annual habitability is almost the same in the two continental configurations  ($0.86$ and $0.85$),
 but  in the case of the polar continent the fraction of habitable surface shows
 strong seasonal oscillations. This behaviour is due to
 the low thermal capacity of the continents and the large
 excursions of polar insolation.

\begin{figure*}
\begin{center} 
\includegraphics[width=7.5cm]{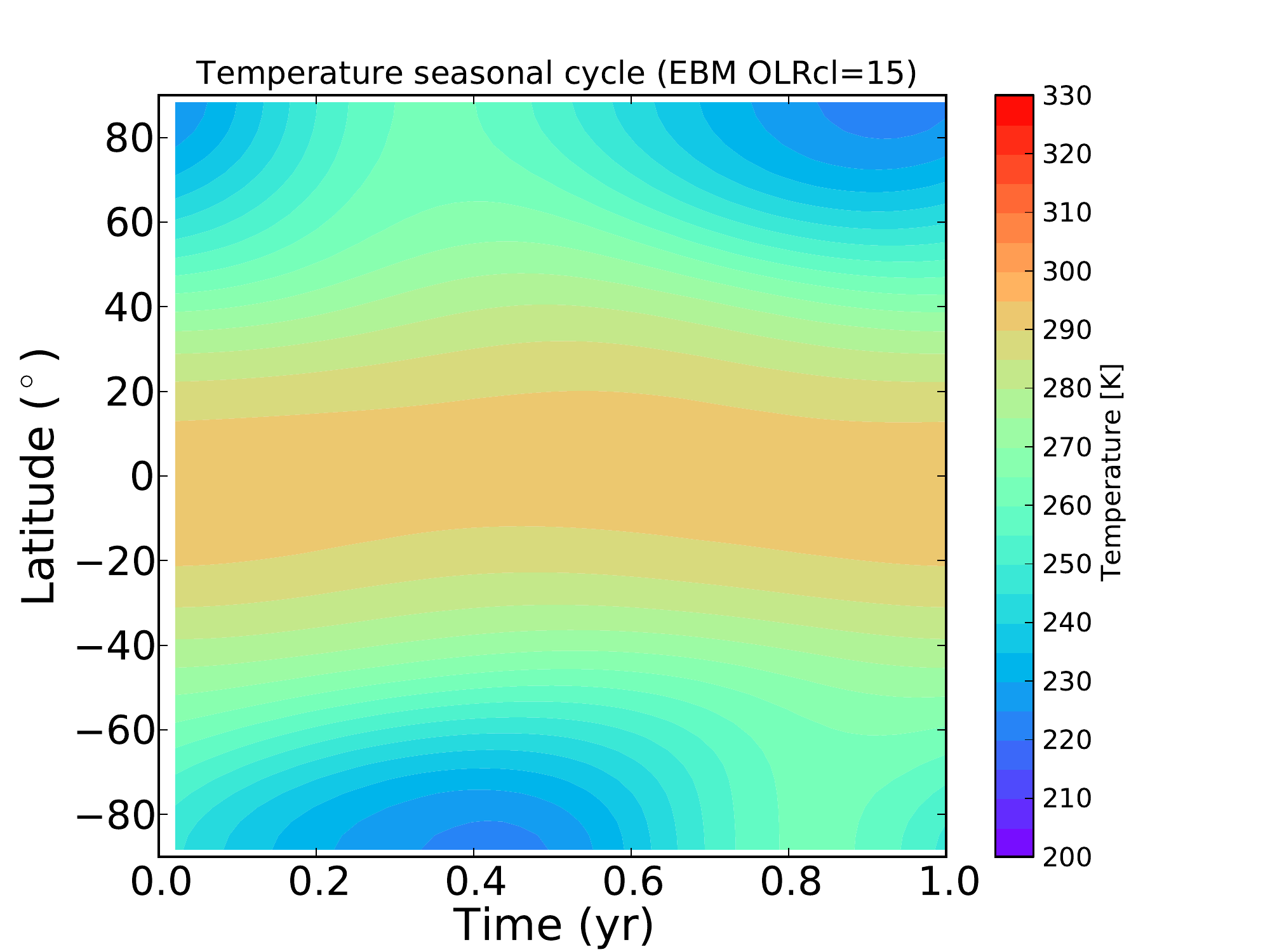}
\includegraphics[width=7.5cm]{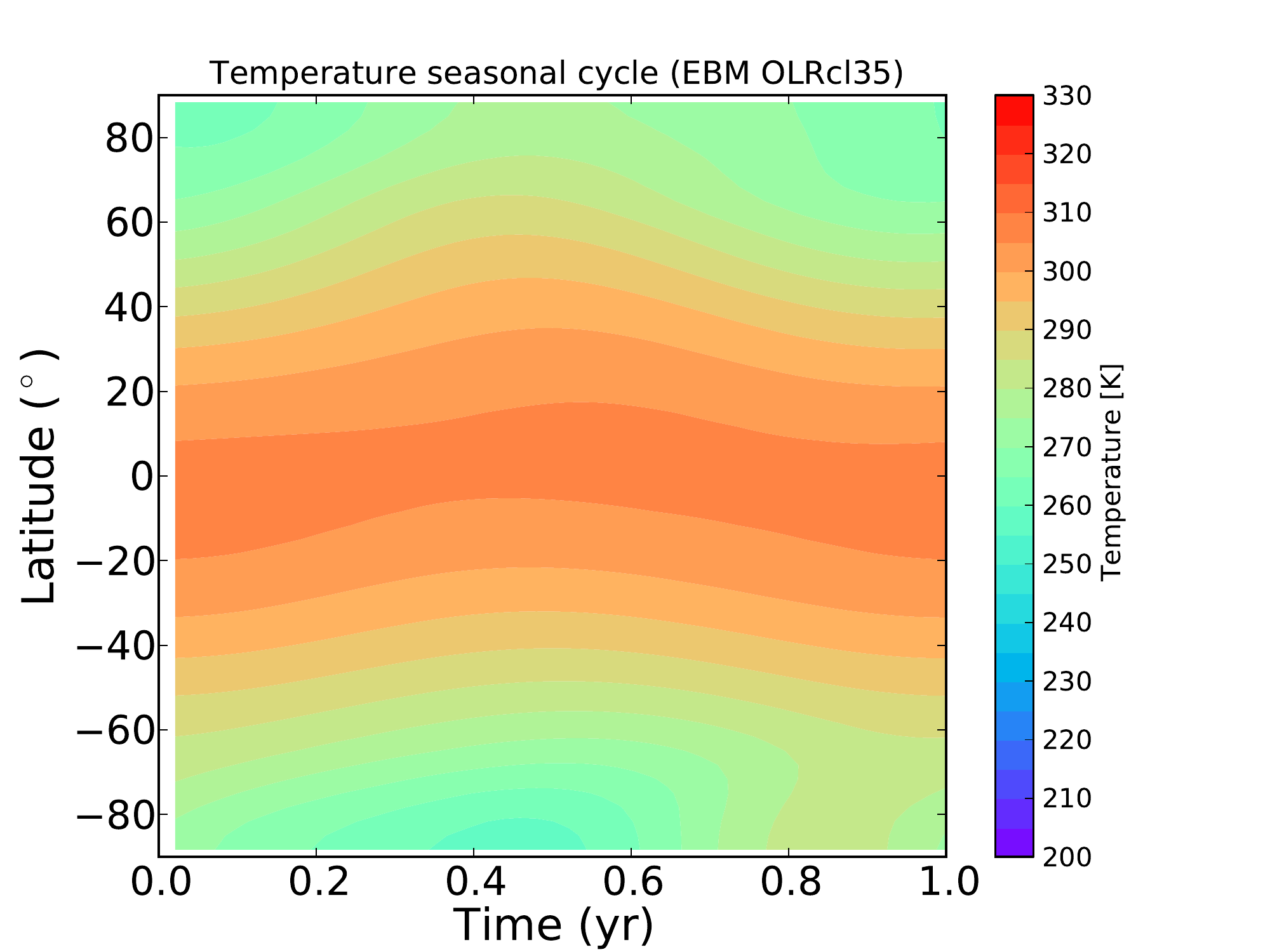} 
\caption{
Variation of surface temperature as a function of latitude and orbital phase 
for an Earth-like planet with long-wavelength cloud forcing 15\,W/m$^2$ (left) and  35\,W/m$^2$ (right). 
All the remaining parameters are those used for the Earth reference model (Appendix B).} 
\label{mapsOLRclouds}%
\end{center}
\end{figure*}

\subsection{Long-wavelength cloud forcing \label{sectOLRclouds}}

In Fig. \ref{mapsOLRclouds} we show how $T(\varphi,t)$ is affected by variations of 
the long-wavelength forcing of clouds. Analysis of Earth data indicates a mean value 
26.4\,W/m$^2$ \citep{Stephens12}, but with large excursions \citep[e.g.][]{Hartmann92}. 
To illustrate the impact of this quantity on the surface temperature 
we have adopted $15$\,W/m$^2$ (left panel) and $35$\,W/m$^2$ (right panel)
since this range brackets most of the Earth values.  
The impact on the mean global temperature is relatively high, 
with a rise from $\Tm =277$\,K to 295\,K between the two cases. 
The ice cover correspondingly decreases from 24\% to 2\%, while
the habitability increases from $\hm=0.68$ to 0.95. 
The mean equator-pole temperature difference shows a decrease 
from $\DTep= 52$\,K to 36\,K.
This decrease of $\DTep$
is more moderate than found for variations of rotation rate, radius and axis tilt.






\bibliographystyle{apj}
\bibliography{exoclimates}  

\clearpage


\begin{deluxetable}{clll}
\tabletypesize{\scriptsize}
\tablecaption{Fiducial parameters of the  \model}
\tablewidth{0pt}
\tablehead{\colhead{Parameter} & \colhead{Fiducial value} & \colhead{Description}  & \colhead{References/Comments} }
\startdata
$C_\text{ml50}$ & $210 \times  10^6$ J m$^{-2}$ K$^{-1}$ 
& Thermal inertia of the oceans$^\text{a}$ (\S \ref{sectThermalCapacity}) &  \citet{Pierrehumbert10} \\ 
$C_{\text{atm},\circ}$ & $10.1 \times 10^6$ J m$^{-2}$ K$^{-1}$
 & Thermal inertia of the  atmosphere$^\text{a}$ (\S \ref{sectThermalCapacity})& \citet{Pierrehumbert10} \\
$C_\text{solid}$ & $1 \times 10^6$ J m$^{-2}$ K$^{-1}$ 
& Thermal inertia of the solid surface(\S \ref{sectThermalCapacity}) & \citet{Vladilo13} \\ 
$D_\circ$ & 0.66 W m$^{-2}$ K$^{-1}$ & Coefficient of latitudinal transport   (\S \ref{sectEddiesTransport})
& Tuned$^b$ to match $T$-latitude profile (Fig.\,\ref{annualLatProfiles}) \\
$\mathcal{R}$ & 2.2 & Modulation of latitudinal transport (\S \ref{sectHadleyCell}) 
&  Tuned to match $T$-latitude profile  (Fig.\,\ref{annualLatProfiles}) \\
$a_l$ & 0.18      & Albedo of lands$^\text{a}$ &   Tuned to match albedo-latitude profile  (Fig.\,\ref{annualLatProfiles}) \\
$a_{il}$ & 0.70  & Albedo of frozen surfaces and overlooking clouds  &  Tuned to match albedo-latitude profile  (Fig.\,\ref{annualLatProfiles}) \\  
$\alpha$ & $-0.11$ & Cloud albedo [Eq. (\ref{cloudAlbedo})] &  Tuned$^\text{c}$ using Fig. 2 in \citet{Cess76} \\
$\beta$ & $7.98 \times 10^{-3}$ $(^\circ)^{-1}$ &  Cloud albedo [Eq. (\ref{cloudAlbedo})]  &  Tuned using Fig. 2 in \citet{Cess76} \\
$\leftmean$OLR$\rightmean_\mathrm{cl,\circ}$ & 26.4 W m$^{-2}$ & Long wavelength forcing of clouds$^\text{a}$ &  \citet{Stephens12} \\
$f_{cw}$ & 0.70       &  Cloud coverage on water &  \citet{Sanroma12,Stubenrauch13}\\ 
$f_{cl}$ & 0.60         & Cloud coverage on land and frozen surface &  \citet{Sanroma12,Stubenrauch13} \\ 
$\Lambda_\circ$ & 0.7 & Ratio of moist over dry eddie transport &   \citet[][Fig.\,2]{KS14} \\
\enddata 
\tablenotetext{a}{Representative Earth's value that can be changed 
 to model exoplanets with different types of surfaces or cloud properties.}
\tablenotetext{b}{$D_\circ$ is also tuned to match the Earth's peak of atmospheric transport at mid latitudes,
$\Phi_\mathrm{max}$ (Table \ref{GlobalEarthModel}).}
\tablenotetext{c}{The parameter $\alpha$ is also tuned to match the minimum value of the albedo-latitude profile.}
\label{tabFiducialPar}
\end{deluxetable}

\begin{deluxetable}{llccl}
\tabletypesize{\scriptsize}
\tablecaption{Northern hemisphere data used to calibrate the Earth model} 
\tablewidth{0pt}
\tablehead{\colhead{Quantity} & \colhead{Description} &  \colhead{Earth value} & \colhead{Model} &\colhead{Units} }
\startdata 
$\leftmean T \rightmean_\mathrm{NH}$ & Surface temperature & 288.61$^a$ & 288.60 &  K   \\
$\Delta T_\mathrm{PE}$ & Pole-equator temperature difference & 40.3$^a$ & 41.7 &  K \\
$\leftmean h\rightmean_\mathrm{NH}$ & Fraction of habitable surface & 0.851$^b$ & 0.858 &  ... \\
$\leftmean A \rightmean_\mathrm{NH}$ & Top-of-atmosphere albedo  &  0.322$^c$ & 0.323 & ...  \\
$\leftmean OLR\rightmean_\mathrm{NH}$ & Outgoing longwave radiation &  240.3$^c$ & 237.6 & W\,m$^{-2}$ \\ 
$\Phi_\mathrm{max}$ & Peak atmospheric transport at mid latitude & 5.0$^d$ & 4.9 &  PW \\
\enddata
\tablenotetext{a}
 {Average ERA Interim 2m temperatures \citep{Dee11} in the period 2001-2013.}
\tablenotetext{b}
{Average fraction of planet surface with temperature satisfying the liquid water criterion.}
\tablenotetext{c}
{Average CERES data \citep{Loeb05,Loeb07} in the period 2001-2013.}
\tablenotetext{d}
{\citet{Trenberth01}}
\label{GlobalEarthModel}
\end{deluxetable}

\end{document}